%% file: 14743.tex
\documentclass[structabstrac]{aa}
\usepackage{natbib}
\usepackage[T1]{fontenc}
\usepackage[version=3]{mhchem}
\usepackage{times}
\usepackage{amssymb}
\usepackage{amsmath}
\usepackage{graphicx}

\bibpunct{(}{)}{;}{a}{}{,} 

\makeatletter
\newcounter{reaction}
\renewcommand\thereaction{C\,\arabic{reaction}}
\newcommand\reactiontag{\refstepcounter{reaction}\tag{\thereaction}}
\newcommand\reaction@[2][]{\begin{equation}\ce{#2}%
\ifx\@empty#1\@empty\else\label{#1}\fi%
\reactiontag\end{equation}}
\newcommand\reaction@nonumber[1]{\begin{equation*}\ce{#1}%
\end{equation*}}
\newcommand\reaction{\@ifstar{\reaction@nonumber}{\reaction@}}
\makeatother

\begin{document}
\input{journals.tex}

\title{Refit to numerically problematic UMIST reaction rate coefficients}
\titlerunning{Refit to numerically problematic UMIST reaction rate coefficients}
\author{M. R\"ollig\inst{1}}
\institute{I. Physikalisches Institut, Universit\"at zu K\"oln, Z\"ulpicher Str. 77, D-50937 K\"oln,
Germany}
\offprints{M.R\"ollig,\\
 \email{roellig@ph1.uni-koeln.de}}

\date{Received <date> / Accepted <date>}

\abstract
{}
{Chemical databases such as the UMIST Database for Astrochemistry (UDFA) are indispensable in the numerical modeling of astrochemical networks. Several of the listed reactions in the UDFA have properties that are problematic in numerical computations: Some are parametrized in a way that leads to extremely divergent behavior for low kinetic temperatures. Other reactions possess multiple entries that are each valid in a different temperature regime, but have no smooth transition when switching from one to another. Numerically, this introduces many difficulties. We present corrected parametrizations for these sets of reactions in the UDFA06 database.} 
{From the tabulated parametrization in UDFA, we created artificial data points and used a Levenberg-Marquardt algorithm to find a set of improved fit parameters without divergent behavior for low temperatures. For reactions with multiple entries in the database that each possess a different temperature regime, we present one joint parametrization that is designed to be valid over the whole cumulative temperature range of all individual reactions.} 
{We show that it is possible to parametrize numerically problematic reactions from UDFA in a form that avoids low temperature divergence. Additionally, we demonstrate that it is possible to give a collective parametrization for reaction rate coefficients of reactions with multiple entries in UDFA. We present these new fitted values in tabulated form.}
{}

\keywords{Astrochemistry -- Astronomical data bases: miscellaneous -- Methods: numerical -- ISM: general}

\maketitle

\section{Introduction}
Astrochemistry is a field of increasing scientific attention. Over the last three decades, a lot of effort went into understanding how interstellar chemical species are formed, destroyed, and how interstellar chemistry works in detail. For a review of recent astrochemical progress see for instance \cite{lis05,herbst06rma,combes07,floris08}.  
To model interstellar chemistry, reliable laboratory data on the respective reaction rate coefficients are of utmost importance. A few comprehensive databases of astrochemical reactions are available today, such as  UDFA\footnote{html://www.udfa.net} \citep{udfa06}, the Ohio database OSU\footnote{http://www.physics.ohio-state.edu/~eric/research.html} \citep[e.g.,][]{garrod08}, and NIST Chemistry Webbook\footnote{http://webbook.nist.gov/chemistry/}. There are also efforts to pool all available reaction data into a unified database (KIDA: KInetic Database for Astrochemistry)\footnote{http://kida.obs.u-bordeaux1.fr/} \citep{kida}. Results from astrochemical laboratory experiments are collected in these databases in an indispensable effort to make these results known to the wider community. They represent the interface between chemistry and astrophysical modeling.  In the following we will mainly concentrate on UDFA06, which contains parametrized reaction rate coefficients for more than 4500 reactions, involving 13 different elements and 420 different chemical species -- far more than have actually been observed in interstellar space.

\begin{table*}[htb]
\begin{center}
\caption{All reactions with negative $\gamma$ coefficients that were refitted. The coefficients on the left are the original UDFA values, the new fits to the coefficients are given on the right side (numbers in parentheses mean: times ten raised to that power). The last column gives the relative error brackets across the given temperature range (given in K) with respect to the original fit.}\label{tab1}
\begin{tiny}
\begin{tabular}{lclclclrrrrr|rrrr}
\hline\hline
\multicolumn{3}{c}{Educts}&&\multicolumn{3}{c}{Products}&{\footnotesize $\alpha_\mathrm{old}$}&{\footnotesize $\beta_\mathrm{old}$}&{\footnotesize $\gamma_\mathrm{old}$}&T$_\mathrm{min}$&T$_\mathrm{max}$ &{\footnotesize $\alpha_\mathrm{new}$}& {\footnotesize $\beta_\mathrm{new}$}& {\footnotesize $\gamma_\mathrm{new}$}&{\footnotesize [error]}\\ \hline
 {C} & + & O$_2$ &   $\longrightarrow$ & {CO} & + & {O} & 2.48(-12) & 1.54 & -613 & 295 & 8000 & 9.94(-12)
& 1.05 & -42.48 & -0.42, 0.26 \\
 {N} & + & {OH} &   $\longrightarrow$ & {NO} & + & {H} & 4.06(-11) & 0.05 & -78 & 103 & 2500
& 5.73(-11) & -0.15 & 1.34 & -0.19, 0.09 \\
 {NH} & + & {NH} &   $\longrightarrow$ & {NH}$_2$ & + & {N} & 1.03(-14) & 3.07 & -1123 & 300 &
3000 & 1.81(-13) & 1.80 & -70.03 & -0.47, 0.36 \\
 {NH} & + & {NO}$_2$ &   $\longrightarrow$ & N$_2$O & + & {OH} & 1.44(-13) & 0 & -1140 & 200 & 300 & 1.06(-11)
& -5.36 & 168 & -0.07, 0.04 \\
 {CH}$_3$ & + & {NH}$_2$ &   $\longrightarrow$ & {CH}$_2$ & + & {NH}$_3$ & 4.76(-17) & 5.77 & -151 &
300 & 2000 & 4.07(-17) & 5.85 & -205 & -0.03, 0.03 \\
 {O} & + & {NH}$_2$ &   $\longrightarrow$ & {HNO} & + & {H} & 4.56(-11) & 0 & -10 & 200 & 3000
& 4.72(-11) & -0.02 & 0.38 & -0.01, 0.01 \\
 {O} & + & {OH} &   $\longrightarrow$ & O$_2$ & + & {H} & 1.77(-11) & 0 & -178 & 158 & 5000 & 3.35(-11)
& -0.28 & 4.30 & -0.29, 0.15 \\
 {O} & + & O$_2$H &   $\longrightarrow$ & O$_2$ & + & {OH} & 3.17(-11) & 0 & -174 & 200 & 2500 & 5.76(-11)
& -0.30 & 7.48 & -0.17, 0.09 \\
 {O} & + & {HS} &   $\longrightarrow$ & {SO} & + & {H} & 8.25(-11) & 0.17 & -254 & 298 & 2000
& 1.74(-10) & -0.20 & 5.70 & -0.12, 0.06 \\
 {O} & + & {NO}$_2$ &   $\longrightarrow$ & O$_2$ & + & {NO} & 6.5(-12) & 0 & -120 & 200 & 2500 & 9.82(-12)
& -0.21 & 5.16 & -0.12, 0.06 \\
 {NH}$_2$ & + & {OH} &   $\longrightarrow$ & H$_2$O & + & {NH} & 7.78(-13) & 1.50 & -230 & 250 & 3000
& 1.35(-12) & 1.25 & -43.45 & -0.14, 0.07 \\
 {OH} & + & C$_2$H$_2$ &   $\longrightarrow$ & {CO} & + & {CH}$_3$ & 6.51(-18) & 4.00 & -1006 & 500 & 2500
& 4.75(-17) & 3.16 & -128 & -0.18, 0.10 \\
 {OH} & + & H$_2${CO} &   $\longrightarrow$ & {HCO} & + & H$_2$O & 2.22(-12) & 1.42 & -416 & 200 & 3000
& 7.76(-12) & 0.82 & -30.62 & -0.35, 0.21 \\
 {OH} & + & {HNO} &   $\longrightarrow$ & {NO} & + & H$_2$O & 4.44(-12) & 1.37 & -169 & 298 & 4000 &
6.17(-12) & 1.23 & -44.29 & -0.09, 0.04 \\
 {OH} & + & O$_2$H &   $\longrightarrow$ & O$_2$ & + & H$_2$O & 3.66(-11) & -0.13 & -244 & 200 & 2500 & 8.58(-11)
& -0.56 & 14.76 & -0.23, 0.13 \\
 {NH}$_3$ & + & {CN} &   $\longrightarrow$ & {HCN} & + & {NH}$_2$ & 3.41(-11) & -0.90 & -9.90 & 25
& 761 & 3.73(-11) & -1.08 & 10.00 & -0.23, 0.09 \\
 C$_2$H & + & C$_2$H$_2$ &   $\longrightarrow$ & H$_2${CCCC} & + & {H} & 1.31(-10) & 0 & -25 & 143 & 3400 &
1.44(-10) & -0.05 & 0.80 & -0.05, 0.02 \\
 {CN} & + & {CH}$_3${CH}$_3$ &   $\longrightarrow$ & C$_2$H$_5$ & + & {HCN} & 4.8(-12) & 2.08 & -484
& 185 & 1140 & 2.34(-11) & 1.02 & -34.95 & -0.28, 0.17 \\
 {CN} & + & O$_2$ &   $\longrightarrow$ & {NO} & + & {CO} & 5.01(-12) & -0.46 & -8 & 13 & 1565 &
5.12(-12) & -0.49 & -5.16 & -0.11, 0.02 \\
 {CN} & + & O$_2$ &   $\longrightarrow$ & {OCN} & + & {O} & 1.86(-11) & -0.13 & -40 & 13 & 4526 &
2.02(-11) & -0.19 & -31.91 & -0.30, 0.07 \\
 {CN} & + & {NO}$_2$ &   $\longrightarrow$ & {NO} & + & {OCN} & 3.93(-11) & 0 & -199 & 297 & 2500
& 7.02(-11) & -0.27 & 8.27 & -0.11, 0.06 \\
 C$_2$H$_3$ & + & O$_2$ &   $\longrightarrow$ & H$_2${CO} & + & {HCO} & 4.62(-12) & 0 & -171 & 200 & 362 & 8.87(-12)
& -0.73 & 22.67 & -0.02, 0.01 \\
 C$_2$H$_3$ & + & O$_2$ &   $\longrightarrow$ & O$_2$H & + & C$_2$H$_2$ & 2.16(-14) & 1.61 & -193 & 300 & 3500 & 3.15(-14) &
1.45 & -51.97 & -0.09, 0.04 \\
 {HCO} & + & O$_2$ &   $\longrightarrow$ & O$_2$H & + & {CO} & 1.58(-12) & 1.24 & -353 & 200 & 2500 & 4.64(-12)
& 0.70 & -25.61 & -0.29, 0.17 \\
{O$_2$} & + & S &   $\longrightarrow$ & SO &+& O & 4.74(-13)&1.41&-439&200&3460&1.76(-12)&0.81&-30.75&-0.38,0.24\\ \hline  

\end{tabular}
\end{tiny}
\end{center}
\end{table*}
\begin{figure*}
\centering
\includegraphics[width=8.5cm]{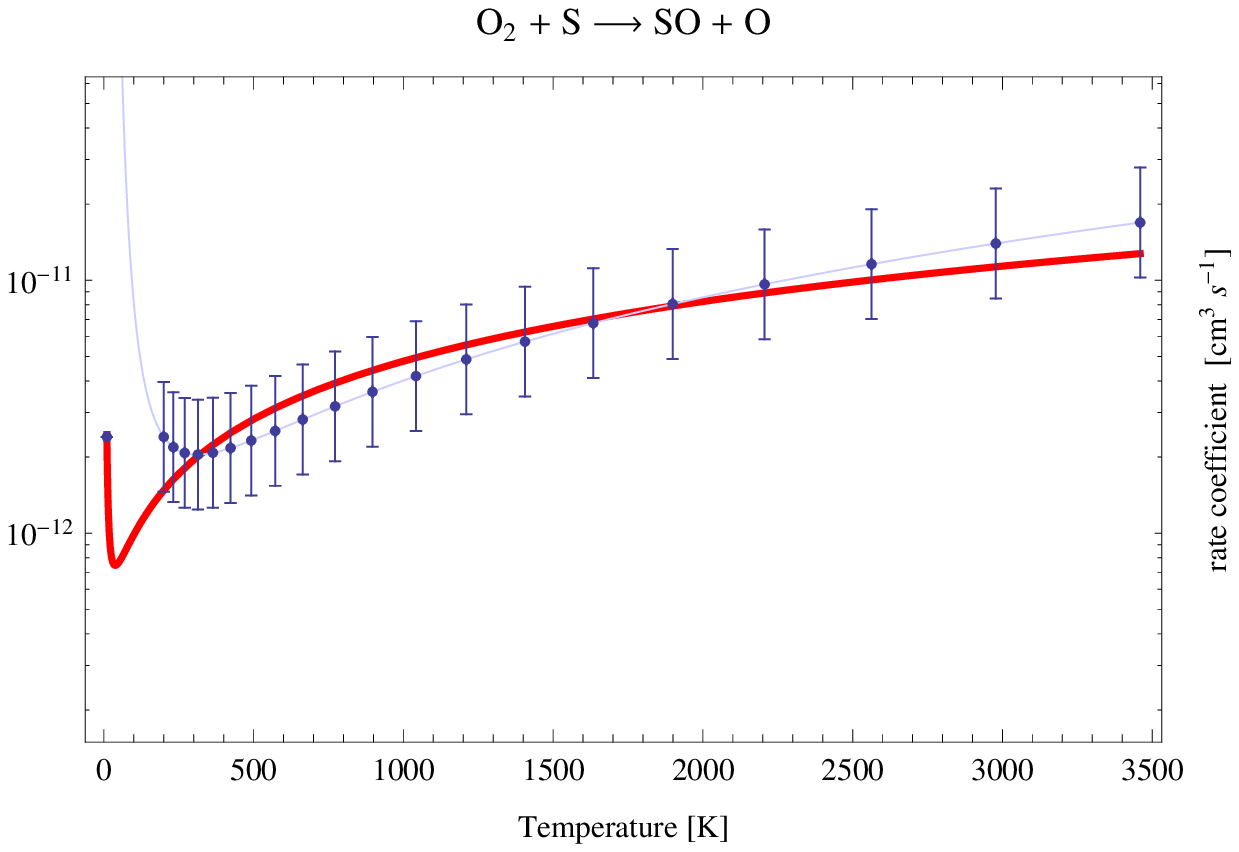}
\hfill
\includegraphics[width=8.5cm]{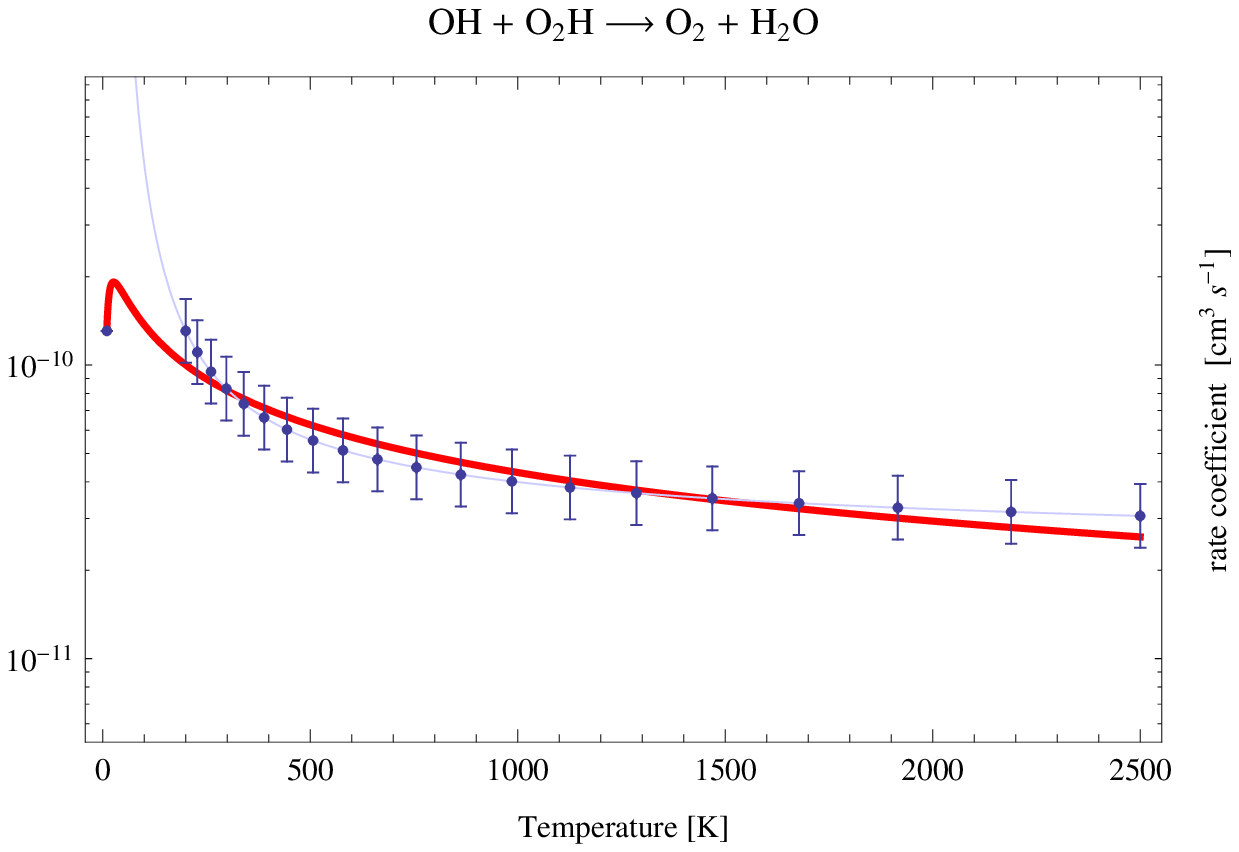}
\caption{Two example reactions to demonstrate how extremely reactions might diverge from physically reasonable numbers if $\gamma\ll 0$. The final fit is shown in red, the UDFA rates are shown in light blue. {\bf Left} The UDFA extrapolation to $k(10\mathrm{K})=4.5\times 10^4$ compared to $2.5\times 10^{-12}$ in our new fit. {\bf Right}The UDFA extrapolation to $k(10\mathrm{K})=2.3$ compared to $1.3\times 10^{-10}$ in our new fit.}
\label{neg1}
\end{figure*}
UDFA (and most other databases as well) tabulate two-body reactions together with a set of fit parameters $\alpha, \beta, \gamma$ in the form $k(T)=\alpha(T/300~K)^\beta \exp(-\gamma/T)$ and a temperature range at which the rate coefficients hold. For some reactions, more than one entry with non-overlapping temperature ranges and respective fit parameters is present because of multiple sets of experimental data. In the given fit form, $\gamma$ is equivalent to the reaction's activation energy, the energy barrier that has to be overcome for the reaction to start. When fitting the measured reaction rates over a particular range of temperatures, a negative $\gamma$  may sometimes give the best fit. Special care has to be taken when using these rate coefficients outside their given range. Especially at very low temperatures, which are relevant for many astrophysical situations, negative $\gamma$ may lead to unphysically high reaction rates. 

To prevent unreasonable results, \citet{udfa06} give a list of reactions whose rate coefficients should be set to zero at 10~K. This may be feasible in chemical models, where the gas temperature is a constant, given parameter. However, in the framework of, e.g., numerical models of photon-dominated regions (PDRs), where the temperature is determined from detailed energy balance and may cover a wide range ($\sim$5~K-5000~K), it is preferable to use reaction rate coefficients, which do not show discontinuities across the applied temperature range \citep[see, e.g.,][]{ht99,comparison07}.

In Figure~\ref{neg1} we show two example reactions that have a diverging reaction rate coefficient at low temperatures because of $\gamma\ll 0$. Using these extrapolated values in a chemical network calculation may lead to artificial and very unrealistic results! 
\begin{figure}[hbt]
\resizebox{\hsize}{!}{\includegraphics{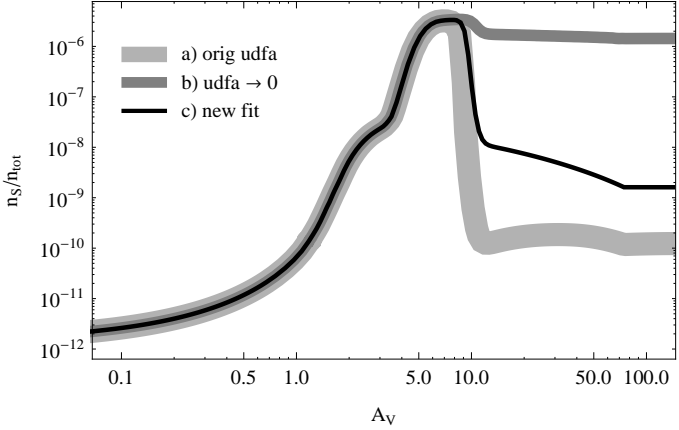}}
\caption{Relative sulfur abundance as function of A$_\mathrm{V}$ for a spherical cloud. The model parameters are mass: $M=1$~M$_\odot$, surface density $n=10^6$~cm$^{-3}$, and FUV intensity $\chi=10^6$ in units of Draine. Three different chemical networks were applied: a) the original UDFA rates, b) the original UDFA rates with all reactions possessing $\gamma<0$ set to zero, and c) all reactions with $\gamma<0$  replaced by the new fits from Table~\ref{tab1}. }
\label{sulfur}
\end{figure}


To illustrate the problem, we compare the results of three PDR model calculations using the KOSMA-$\tau$ PDR model \citep{roellig06}. We calculated spherical model clouds with mass $M=1\,\mathrm{M}_\odot$, surface gas density $n=10^6$~cm$^{-3}$, and FUV illumination $\chi_\mathrm{Draine}=10^6$. The chemical network consists of 198 species and 3237 reactions. Except for the surface formation of molecular hydrogen, a pure gas-phase chemistry was calculated. The only difference between the models is the applied set of reaction rate coefficients.
In the first model we used the original parametrization of the reaction rate coefficients\footnote{Using the original UDFA rates may lead to non-converging models depending on the given parameters. The model shown in Figure~\ref{sulfur} is one of the few that converged at all within our parameter grid.}. In the second model we deactivated these reactions by setting their reaction rates to zero. In a third model we used our new fit to the reaction rate coefficients from Table~\ref{tab1}.
In Figure~\ref{sulfur} we show the relative sulfur abundance versus A$_\mathrm{V}$ for all three cases. For A$_\mathrm{V} > 7$ the models start to deviate significantly. The relevant reaction for this comparison is
\begin{equation}
\mathrm{O}_2 + \mathrm{S} \longrightarrow \mathrm{SO} + \mathrm{O} \label{reaction}\;\;\;.
\end{equation}
The original UDFA rates are valid for $T>200$~K (see Fig.~\ref{neg1}, left). The model temperature goes down to 40~K at A$_\mathrm{V}=3.5$. At this temperature, the UDFA reaction rate coefficient is a factor 670 higher than at 200~K, leading to a very efficient destruction of S. At A$_\mathrm{V}=20$ reaction \ref{reaction} is three orders of magnitudes stronger than the next important destruction reaction. Switching off this reaction allows the atomic sulfur to remain in the gas up to large A$_\mathrm{V}$. Our new rate coefficients give a result between these two extremes. In our comparison, the choice between the scenarios a), b), and c) strongly affects the whole sulfur chemistry at large  A$_\mathrm{V}$.


The UDFA lists the accuracies of the given fits, usually ranging between 25-50\%. However, outside their given temperature ranges, negative $\gamma$ may lead to errors many orders of magnitude larger. For these reactions we calculated new fit parameters to allow for continuous rate coefficients even when $T$ becomes very low. Furthermore, we calculated a single set of fit parameters for those reactions that have multiple entries in the database. These multiple sets may or may not have overlapping temperature ranges. Our final fit is designed to be valid over the total temperature interval with only small deviations. 

It is well possible that the results obtained with our new fits have large errors, especially outside the original temperature regime. The peculiar shape of the fit in Figure~\ref{neg1} below $T_\mathrm{min}$ is a direct result of the choice of the parametrization function in UDFA and it is unlikely that the low temperature behavior of this reaction follows that particular shape. However, the divergent low temperature rates of the original UDFA parametrization are unphysical, and the new fit is our best attempt at improving this behavior while minimizing any new uncertainties.

\section{Fit procedure}
Below we describe the applied fit strategy. Generally, we are willing to sacrifice some of the originally given accuracy to gain continuity down to very low temperatures. To be precise, we did not calculate a new fit to previously attained experimental data. We calculated a new fit to rate coefficients calculated from the original UDFA fits over the given range. The steps are the following:
\begin{enumerate}
\item tabulate the reaction rate coefficients over the given temperature range, using the given fit-accuracy as error. These are our artificial data.
\item if $\gamma<0$, fix $k(T=10K)$
\item calculate new fit parameters  $\alpha, \beta, \gamma$\footnote{The fitting was made with Mathematica from Wolfram Research, using a Levenberg-Marquardt algorithm. The fitting procedure is available upon request.}
\item calculate the error of the new fit with respect to the original UDFA parametrization.
\end{enumerate}
In the appendix we show plots of all new fits presented in this paper.
\subsection{Discussion of the fit strategy}

The details of the fit strategy may have considerable influence on the final fit result. Particularly point 2 prevents the reaction rates from becoming unphysically high. Of course, this procedure introduces new, possibly very large errors to the rate coefficients. But these errors are usually of the same order as those from the UDFA fits. We now consider some major influences on the fit results.
\subsubsection{Placement of 'data' points}
The number and placement of `data` points has a direct influence on the quality of the fit because it implicitly introduces a weighting of the respective temperatures. Using the Levenberg-Marquardt algorithm is in effect applying a sequence of linear least-squares fits to a nonlinear function. Regions with more 'data' points contribute more to the total sum of squares and force the fit function closer to these 'data' points. For instance, placing temperature points equidistant in log-space will place more grid points at lower temperatures compared to equidistant linear spacing, weighting lower temperatures stronger than higher temperatures.
Optimal fit results could be achieved by adaptively changing the number and positions of grid points until the error is minimal, but to keep a consistent error estimate, we used the same temperature gridding for all fits.
To fit reactions with  $\gamma<0$, we employed an equidistant temperature grid in log-space with 20 grid points, while we used linear spacing for fitting reactions with multiple entries, using 20 grid points per temperature interval.
\subsubsection{Fixing  $k(T=10K)$}
\begin{figure}[hbt]
\resizebox{\hsize}{!}{\includegraphics{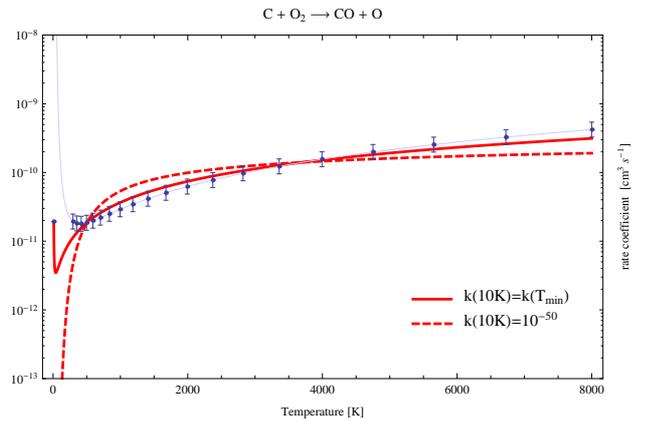}}
\caption{\ Comparison of two fit results for different choices of $k(T=10K)$. The UDFA rate coefficient, valid between 295 and 8000~K, is shown in light blue. The best fit for $k(10K)=10^-50$ is shown as red, dashed line ($\alpha=1.01(-11), \beta=0.23, \gamma=913$). The best fit for $k(10K)=k(\mathrm{T}_\mathrm{min})$ is shown as red, solid line ($\alpha=9.95(-12), \beta=1.05, \gamma=-42.5$). }
\label{rate10kinfluence}
\end{figure}
Choosing a fixed $k(T=10K)$ prevents the reaction rates from becoming unphysically high. However, the exact choice of $k(T=10K)$ is completely arbitrary since no real measurement is available. Moreover, it poses a major constraint to the fit and an unfortunate choice will lead to very large deviations from the true reaction rate coefficient.  \citet{udfa06} suggest to set $k(10K)=0$. It is possible that  $\lim_{T\rightarrow 10}k(T)=0$. It is also possible that $k(10K)>>0$, we simply do not know. To minimize the deviations from the only available 'data', we chose  to set $k(10K)=k(T_\mathrm{min})$ if $\gamma<0$ with $T_\mathrm{min}$ the minimal valid temperature from the UDFA parametrization. This also reduces any over/under-swinging behavior of the new fit.
 To illustrate how the choice of $k(10K)$ influences the form of the final fit function we compare in Figure~\ref{rate10kinfluence} the fit for $k(10K)=k(T_\mathrm{min})$ (red, solid line) and for $k(10K)=10^{-50}$ (red, dashed line). In this particular case, the latter leads to significantly stronger deviations from the UDFA parametrization than for $k(10K)=k(T_\mathrm{min})$. 
\subsubsection{Calculating new fit parameters}
\begin{figure}[hbt]
\resizebox{\hsize}{!}{\includegraphics{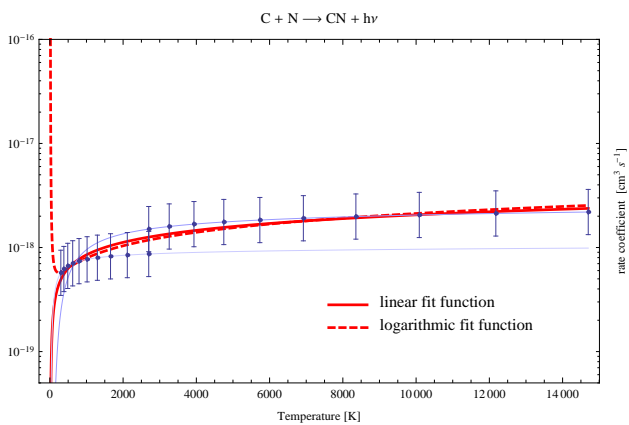}}
\caption{Comparison of fit results for a linear fit function $k(T)$, shown as solid, red line, and a logarithmic fit function $\log(k(T))$, shown as dashed, red line.}
\label{logFitinfluence}
\end{figure}
We fitted the fit function $k(T)=\alpha\left(\frac{T}{300}\right)^{\beta } e^{-\frac{\gamma }{T}}$ to the list of 'data' points created in steps 1 and 2. In most cases it is better to perform the fit in log-space, i.e. use $\log(k(T))$ instead of $k(T)$ as fit function. Otherwise, the Levenberg-Marquardt fit would have the tendency to fit $k(T)$ to absolutely higher temperatures, while we are interested in the lower temperature regime. In some cases however, a linear fit function is more appropriate. This case is shown in Figure~\ref{logFitinfluence}. We try to find a simultaneous fit across the multiple temperature regimes of the reaction C + N $\rightarrow$ CN.  The linear fit produces a continuously declining result for $T\rightarrow0$, while a logarithmic fit function produces the opposite behavior, which results in large deviations of the 10~K rate predicted by the UMIST parametrization and our fit result. Again, the true reaction rate behavior at very low temperatures is unknown, hence, our aim is to minimize any deviation from the UDFA fit. The logarithmic fit differs from the 10~K UDFA rate by 8 orders of magnitude, while the linear fit differs only by a factor of 40. For this reaction we chose a linear fit function, for all other fits in this paper a logarithmic fit function was used. 

\subsubsection{Error calculation}
Strictly speaking, the term error is misleading in our context, since it suggest model deviations from real measurements\footnote{The error bars in some of the figures show the uncertainties of the UDFA reaction rates. These errors are not taken into account by the Levenberg-Marquardt algorithm when fitting a new parametrization.}. Instead, whenever we are speaking of fit errors,  we actually compare how two different models deviate from each other. Nevertheless, because our aim is to change the original parametrization as little as possible, we will use this deviation to quantify the quality of our fits. 

In Table~\ref{tab1} and \ref{tab2} we give the relative error brackets of our new fit, i.e. $[\min(e_i),\max(e_i)]$, with $e_i=\left(k^f(T_i)-k^U(T_i)\right)/k^U(T_i)$ for all $ T_i \in [ T_\mathrm{min}, T_\mathrm{max} ]$. Here, $k^U$ is the UDFA parametrization of the reaction, and $k^f$ is our new fit.

\subsection{Fitting reactions with negative $\gamma$.}
\begin{figure}[hbt]
\resizebox{\hsize}{!}{\includegraphics{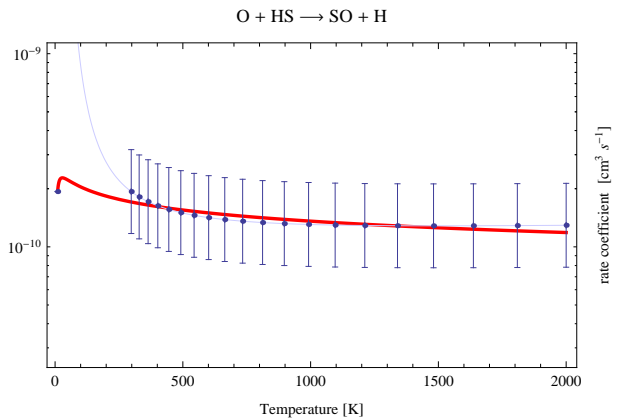}}
\caption{Example reaction that can be fitted very well (maximum additional error is 11\%), but has a less extreme low temperature behavior. The final fit is shown in red, the UDFA rates for each individual temperature regime are shown in light blue. The difference in k(10~K) is 10 orders of magnitudes between the UDFA rates and our new fitted rates.}
\label{neg2}
\end{figure}
 
\begin{figure}
\resizebox{\hsize}{!}{\includegraphics{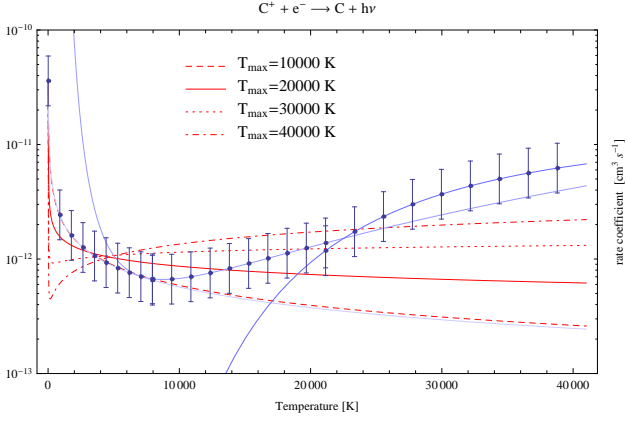}}
\caption{UDFA lists three C$^+$ recombination reactions for temperatures from 10 to 41000~K. The two reaction rates for the higher temperatures show a very different behavior compared to the reaction that is valid from 10 to 8000~K. Applying different values of $T_\mathrm{max}$ during the fit leads to significantly different results.}
\label{tmaxinfluence}
\end{figure}
\begin{figure}
 \resizebox{\hsize}{!}{\includegraphics{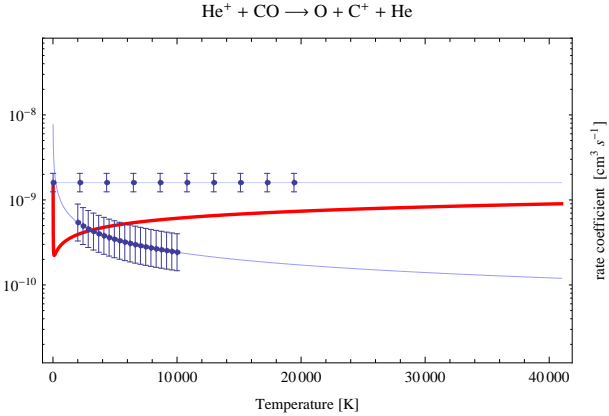}}
 \caption{Example reaction with apparently mutually exclusive reaction rates listed in UDFA. The final fit is shown in red, the UDFA rates for each individual temperature regime are shown in light blue. The two temperature ranges overlap and the respective fits to the reaction rates, here shown in light blue, display a very different behavior.}
 \label{multi1}
 \end{figure}

To obtain a new fit for reactions with $\gamma\ll 0$, we used a log-spaced temperature grid with 20 grid points and a logarithmic fit function. Collider reactions, i.e. reactions of type ``CL'' in UDFA, were excluded from the fit.
Apparently ist is possible to find a very good fit for some reactions without any divergence at low temperatures and without introducing any significant additional error. The prime example for this is shown in Figure~\ref{neg2}. Using the UDFA reaction rates gives an unreasonably high reaction rate coefficient at 10~K of $5$! The new fitted rate coefficient at 10~K is $2\times 10^{-10}$. Even though we do not know the real reaction strength at 10~K, we consider our new result to be much more reasonable compared to the extrapolated numbers from UDFA.

Some of our new fits even have negative $\gamma$, which shows that $\gamma < 0$ not always has to lead to divergent behavior. If $\gamma \gtrsim -100$, the rate inclination for low temperatures is moderate and should not lead to unreasonable results\footnote{UDFA list more reactions with $\gamma<0$ than given in Table~\ref{tab1}. We omitted those reactions because they showed no significant divergence at low temperatures.}. In Table~\ref{tab1} we present the fit results for reactions that are listed in UDFA with $\gamma < 0$. We give the original and the new fit parameters as well as the relative error brackets of the new fit. 

Some of the reactions given in Table~\ref{tab1} also appear in Table~\ref{tab2} because they have counterparts with different temperature regimes. Some of the numerical problems described above can be circumvented by choosing an alternative reaction rate coefficient with a more appropriate temperature regime. Nevertheless, we list these reactions for cases where that is not possible.  
\subsection{Fitting reactions with multiple entries.}

\begin{figure*}
\centering
\includegraphics[width=8.5cm]{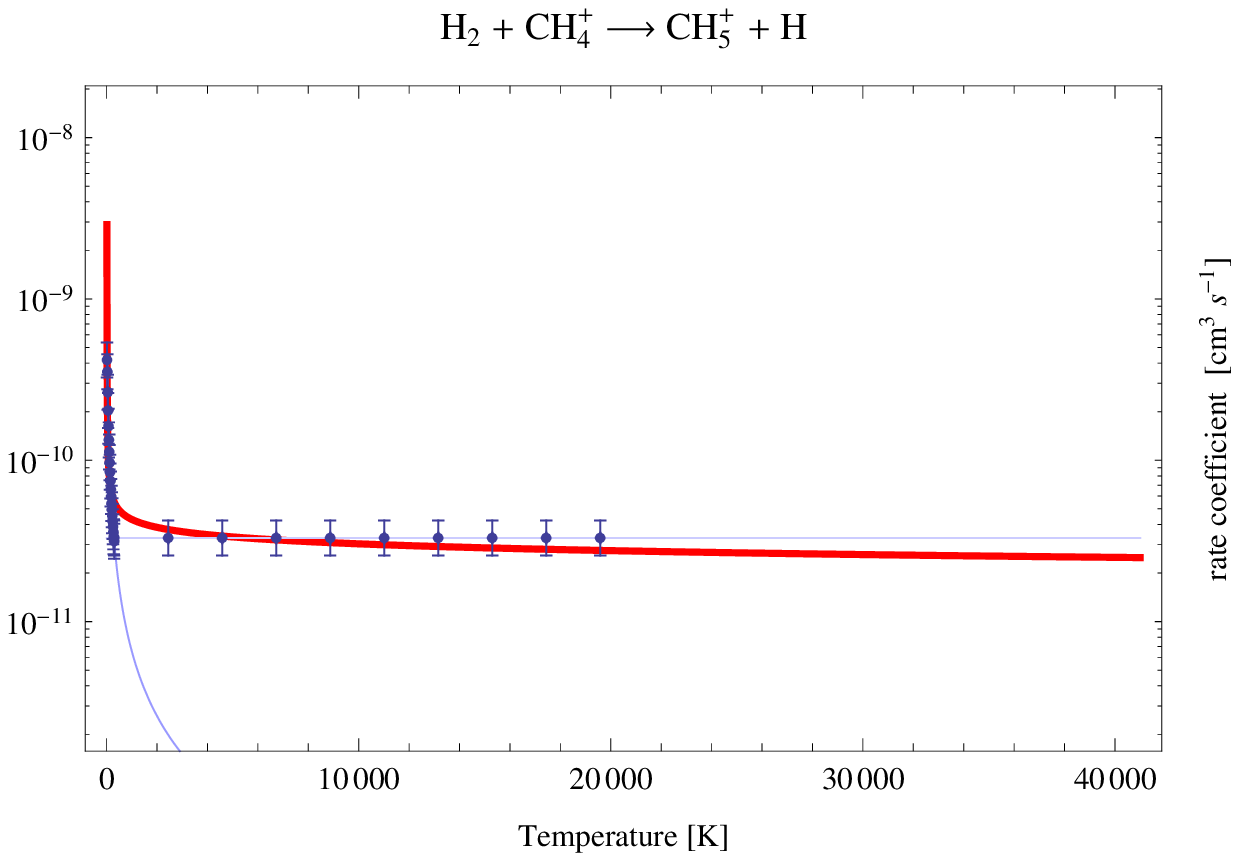}
\hfill
\includegraphics[width=8.5cm]{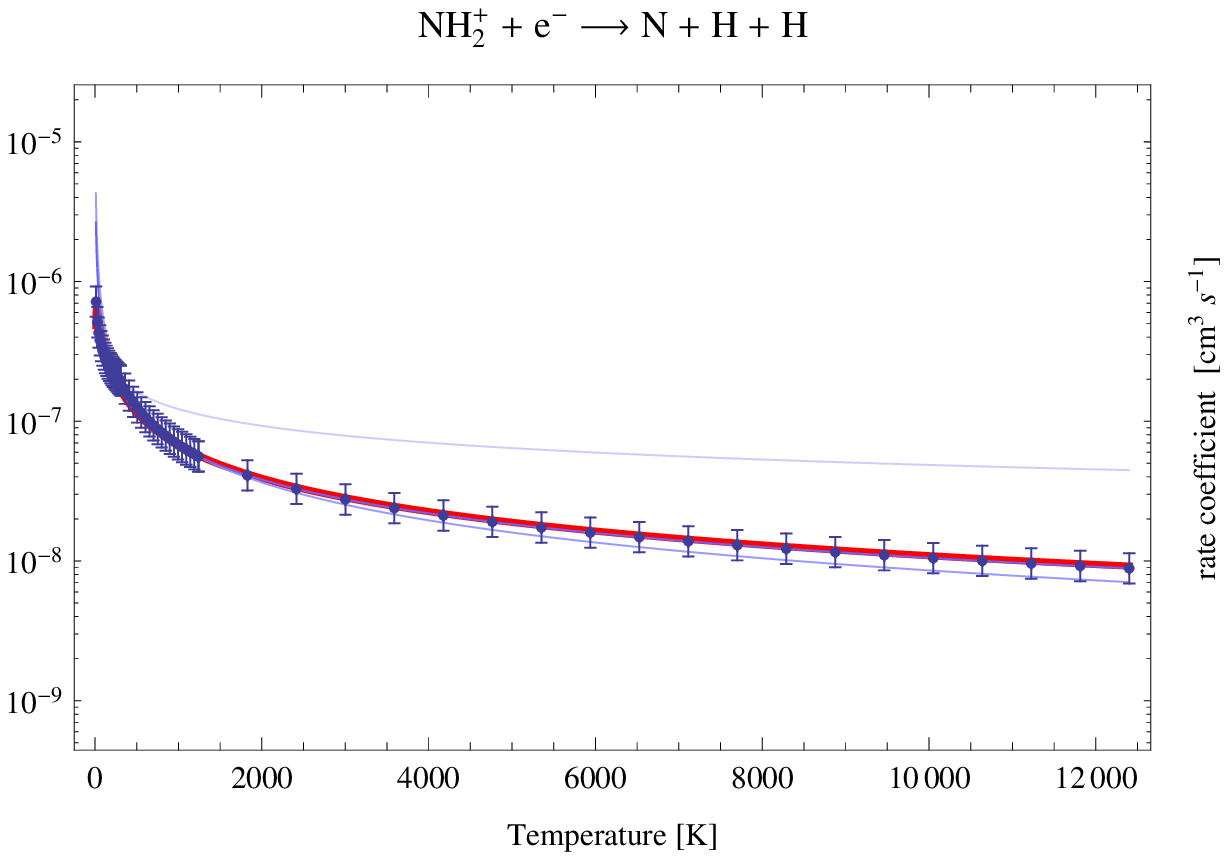}
\caption{Two example reactions to demonstrate the wide range of error values. The final fit is shown in red, the UDFA rates for each individual temperature regime are shown in light blue. {\bf Left} This reaction is listed with two very differently behaving reaction rates. Consequently, the simultaneous fit has relatively large errors especially at the interface between bow temperature regimes. Still, we consider the fit to be very good because of its overall capability to reproduce the general behavior in both regimes. {\bf Right} This reaction is listed with three distinct temperature regimes. The simultaneous fit is nevertheless a very good approximation to all three UDFA rates in the relevant temperature ranges.}
\label{multi2}
\end{figure*}
The fits to reactions with multiple database entries are, in most cases, much less sensitive to the details of the fit strategy, but, overall, the errors of the simultaneous fits are larger compared to the fits to reactions with $\gamma < 0$.  Here, the partly very inconsistent entries in UDFA for the different temperature ranges contribute the most to the errors in the final, simultaneous fit. For cases thta span a very large temperature range, the choice of $T_\mathrm{max}$ when setting up the artificial 'data' points poses a major constraint on the fit result. This is demonstrated in Figure~\ref{tmaxinfluence}.

In Table~\ref{tab2} we present the fit results for reactions with multiple entries in UDFA. The valid temperature ranges of these reaction rates may or may not overlap. Even worse, some of the multiple entries appear to be mutually exclusive. Especially the high temperature rate coefficients for the reaction He$^+$+CO$\rightarrow$O + C$^+$ + He appear to be at odds with both theory and experiment and it may be advisable to only use the low temperature rate coefficients (see Figure~\ref{multi1}). However, this assessment is far beyond the scope of this paper. We apply our fit routine to all reactions in UDFA with multiple entries and leave it to the informed reader to decide whether to use the simultaneous fit or whether to select a reaction with an appropriate temperature regime.


  For the fitting, we limited the upper temperature range to 20000~K where T$_\mathrm{max}$ is higher, thus neglecting higher temperatures. The errors are explicitly calculated for this range and may be significantly larger for higher temperatures. Despite the apparently large errors we consider the fits relatively good in most cases. Remember that we fit multiple, sometimes very disparate sets of `data` points, and therefore, large errors are unavoidable. This is demonstrated in Fig.~\ref{multi2} for two cases from Table~\ref{tab2}.

\begin{figure}
\resizebox{\hsize}{!}{\includegraphics{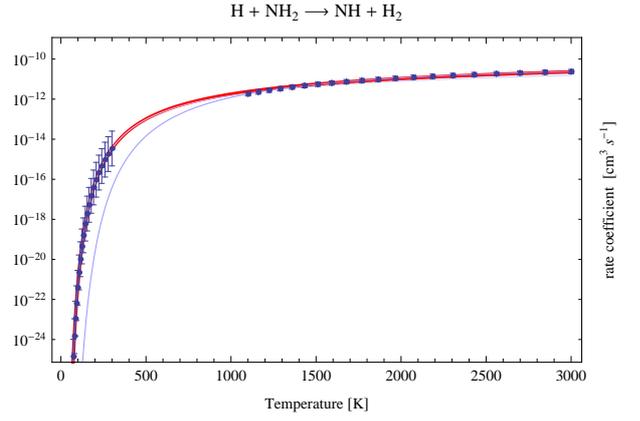}}
\caption{Simultaneously fitted reaction rates in this figure have a maximum  relative error of 33\%. The final fit is shown in red, the UDFA rates for each individual temperature regime are shown in light blue. }
\label{multi3}
\end{figure}

The reader should also recall that the reaction rates from UDFA usually have errors on the order of 25-200\%. One could be tempted not to use the UDFA rates to prevent the errors to rise even more because of our simultaneous fitting. However, by doing so one might implicitly accept much larger errors  when moving from one temperature regime to the other. This is demonstrated in Figure~\ref{multi3}. 

The simultaneously fitted reaction rates in Figure~\ref{multi3} have a maximum relative error of 32\%. This may appear large compared to the individual errors for each temperature regime. But when only using either one of the two reactions we would implicitly accept a very large error when the temperature is around 400-500~K, where both individual rates disagree by one to two orders of magnitude. Our simultaneous fit is a reasonable fit in an attempt to keep the experimental findings for both temperature regimes.  
\begin{table*}[ht]
\begin{center}
\caption{All reactions with multiple reaction rate coefficients for several temperature ranges. The last column gives the relative error brackets. For the fitting, we limited the upper temperature range to 20000~K, neglecting higher temperatures. The errors for higher temperatures might be significantly larger. The temperature limits for UDFA are given in Kelvin. }\label{tab2}
\begin{tiny}
\begin{tabular}{lllclllllrrrrrr}
\hline\hline
\multicolumn{3}{c}{Educts}&&\multicolumn{5}{c}{Products}&T$_\mathrm{min}$&T$_\mathrm{max}$ &{\footnotesize $\alpha_\mathrm{new}$}& {\footnotesize $\beta_\mathrm{new}$}& {\footnotesize $\gamma_\mathrm{new}$}&[error]\\ \hline
{H} & + & {NH}$_2$ & {$\longrightarrow$} & {NH} & + & H$_2$ & & {} & 73 & 3000 & 4.61(-12) & 1.02
& 2161 & -0.05, 0.32 \\
 {C} & + & O$_2$ & {$\longrightarrow$} & {CO} & + & {O} & & {} & 10 & 8000 & 5.56(-11) & 0.41
& -26.87 & -0.50, 2.89 \\
 {CH} & + & {O} & {$\longrightarrow$} & {CO} & + & {H} & & {} & 10 & 6000 & 6.02(-11) &
0.10 & -4.46 & -0.17, 0.13 \\
 {N} & + & {OH} & {$\longrightarrow$} & {NO} & + & {H} & & {} & 10 & 2500 & 6.65(-11) &
-0.23 & 0.24 & -0.12, 0.26 \\
 {O} & + & {OH} & {$\longrightarrow$} & O$_2$ & + & {H} & & {} & 10 & 5000 & 5.28(-11) & -0.46
& 4.14 & -0.22, 0.66 \\
 {O} & + & C$_2$ & {$\longrightarrow$} & {CO} & + & {C} & & {} & 10 & 8000 & 1.25(-10) & 0.47
& -16.13 & -0.27, 0.32 \\
 H$_2$ & + & {CH}$_4$$^+$ & {$\longrightarrow$} & {CH}$_5$$^+$ & + & {H} & & {} & 15 & 41000 & 4.89(-11)
& -0.14 & -36.14 & -0.47, 0.96 \\
 H$_2$ & + & {NH}$_3$$^+$ & {$\longrightarrow$} & {NH}$_4$$^+$ & + & {H} & & {} & 10 & 41000 & 3.09(-13)
& 1.08 & -50.87 & -0.80, 0.83 \\
 {He}$^+$ & + & {CO} & {$\longrightarrow$} & {O} & + & C$^+$ &+& {He} & 10 & 41000 & 2.27(-10) &
0.28 & -29.10 & -0.83, 1.51 \\
 {CH}$_3$$^+$ & + & {HCl} & {$\longrightarrow$} & H$_2${CCl}$^+$ & + & H$_2$ & & {} & 10 & 520 & 5.72(-11)
& -2.25 & 47.50 & -0.40, 0.36 \\
 O$^+$ & + & N$_2$ & {$\longrightarrow$} & {NO}$^+$ & + & {N} & & {} & 23 & 41000 & 2.42(-12) & -0.21 &
-43.99 & -0.19, 1.61 \\
 {H} & + & {He}$^+$ & {$\longrightarrow$} & {He} & + & H$^+$ & & {} & 10 & 41000 & 8.41(-16) & 0.43
& -9.59 & -0.36, 0.30 \\
 H$^+$ & + & {O} & {$\longrightarrow$} & O$^+$ & + & {H} & & {} & 10 & 41000 & 6.86(-10) & 0.26 & 224
& -0.08, 0.05 \\
 {NH}$_2$$^+$ & + & e$^-$ & {$\longrightarrow$} & {N} & + & {H} &+& {H} & 12 & 12400 & 1.78(-7) &
-0.80 & 17.14 & -0.23, 0.31 \\
 {NH}$_2$$^+$ & + & e$^-$ & {$\longrightarrow$} & {NH} & + & {H} & & {} & 12 & 12400 & 9.21(-8) &
-0.79 & 17.11 & -0.23, 0.31 \\
 {NH}$_4$$^+$ & + & e$^-$ & {$\longrightarrow$} & {NH}$_2$ & + & {H} &+& {H} & 12 & 34670 & 4.67(-7)
& -1.25 & 41.88 & -0.50, 4.02 \\
 {NH}$_4$$^+$ & + & e$^-$ & {$\longrightarrow$} & {NH}$_2$ & + & H$_2$ & & {} & 12 & 34670 & 2.22(-7) & -1.25
& 41.91 & -0.50, 4.02 \\
 {NH}$_4$$^+$ & + & e$^-$ & {$\longrightarrow$} & {NH}$_3$ & + & {H} & & {} & 12 & 34670 & 1.54(-6)
& -1.25 & 41.92 & -0.50, 4.02 \\
 {H} & + & e$^-$ & {$\longrightarrow$} & H$^-$ & + & {h$\nu $} & & {} & 10 & 41000 & 3.37(-16) & 0.64
& 9.17 & -0.20, 0.37 \\
 C$^+$ & + & e$^-$ & {$\longrightarrow$} & {C} & + & {h$\nu $} & & {} & 10 & 41000 & 2.36(-12) & -0.29
& -17.55 & -0.58, 0.41 \\
 N$^+$ & + & e$^-$ & {$\longrightarrow$} & {N} & + & {h$\nu $} & & {} & 10 & 41000 & 3.5(-12) &
-0.53 & -3.18 & -0.34, 0.08 \\
 H$^+$ & + & {H} & {$\longrightarrow$} & H$_2$$^+$ & + & {h$\nu $} & & {} & 200 & 32000 & 1.15(-18) & 1.49
& 228 & -0.18, 0.42 \\
 {C} & + & {N} & {$\longrightarrow$} & {CN} & + & {h$\nu $} & & {} & 300 & 14700 & 5.72(-19)
& 0.37 & 50.95 & -0.16, 0.46\\
 {C} & + & {O} & {$\longrightarrow$} & {CO} & + & {h$\nu $} & & {} & 10 & 14700 & 4.69(-19)
& 1.52 & -50.51 & -0.69, 3.10 \\
 C$^+$ & + & {O} & {$\longrightarrow$} & {CO}$^+$ & + & {h$\nu $} & & {} & 10 & 13900 & 2.44(-18)
& -0.07 & 2.77 & -0.09, 0.06 \\
 {O} & + & {Si}$^+$ & {$\longrightarrow$} & {SiO}$^+$ & + & {h$\nu $} & & {} & 10 & 14700 &
5.92(-18) & -0.67 & 19.37 & -0.45, 4.87 \\
 {H} & + & {HNC} & {$\longrightarrow$} & {HCN} & + & {H} & & {} & 10 & 1000 & 1.14(-13)
& 4.23 & -115 & -0.70, 5.10 \\ \hline
\end{tabular}
\end{tiny}
\end{center}
\end{table*}

The special case of carbon recombination is shown in Figure~\ref{tmaxinfluence}. Three different sets of fit parameters are given in UDFA for three distinct temperature ranges up to 41000~K. Especially the high temperature rates show a very different behavior compared to the low temperature range. A simultaneous fit gives very large errors because of the large deviations from the high temperature rates. However, for most applications, temperatures up to 8000~K are sufficient and we can neglect the errors for high temperatures. In that case, the new fit has much smaller errors. For many practical purposes, one could even completely discard the two higher temperature ranges and only use the low temperature rates. For the sake of completeness, we still list the new fit temperatures up to 20000~K. 

\subsection{Discussion}
For three reactions in Table~\ref{tab1} and \ref{tab2} one finds updated measurements and theoretical values of the reaction rate coefficients in KIDA $k^K(T)$. A detailed comparison of the respective reaction rate coefficient behavior is beyond the scope of this paper. We will only take a look at the updated recommended 10~K values for these reactions and compare them to our new values $k^f(10K)$ in Table~\ref{KIDA}.
\begin{table*}[ht]
\begin{center}
\caption{Comparison of the recommended 10~K reaction rate coefficients available in KIDA $k^K(10K)$ with $k^f(10K)$ from our new fits.}\label{KIDA}
\begin{tiny}
\begin{tabular}{lllcllllll}
\hline\hline
\multicolumn{3}{c}{Educts}&&\multicolumn{3}{c}{Products}&$k^K(10K)$&$k^f(10K)$&References\\
&&&&&&&cm$^3$s$^{-1}$&cm$^3$s$^{-1}$&\\\hline
N&+&OH&$\rightarrow$&NO&+&H&$4.5\times 10^{-11}$&$8.44\times 10^{-11}$&\citet{jorfi2009}\\
O&+&OH&$\rightarrow$&H&+&O$_2$&$4\times 10^{-11}$&$5.57\times 10^{-11}$&\citet{lique2009, quemener2009}\\
O&+&NH$_2$&$\rightarrow$&HNO&+&H&$1\times 10^{-10}$&$4.80\times 10^{-11}$&\citet{baulch2005}\\\hline
\end{tabular}
\end{tiny}
\end{center}
\end{table*}

For all three reactions, the 10~K rate coefficients achieved from our new fits are within a factor of two compared to the recommendations from the KIDA KInetic Database for Astrochemistry. For the reaction O~+~OH, our value is only 40\% higher than the recommended value. Given the overall uncertainties, the agreement between our new fits and the KIDA values is remarkable and an indication that our general fit approach is viable to achieve reasonable results.   
\section{Summary and conclusions}\label{sec-summary}
We presented an approach to overcome some numerical and practical difficulties in present-day databases of chemical reaction rates. One of these difficulties is the existence of fits to reaction rate coefficient that unphysically diverge at very low temperatures. This is especially problematic for numerical codes, which calculate chemical networks down to very low temperatures and thus have to use many of the reaction rate coefficients that are listed in chemical databases beyond their given temperature regime. We demonstrate that it is possible to find a new parametrization of these reactions that ensures a less erroneous behavior and allows calculations to temperatures of 10~K and lower. 

A second complication that occurs in chemical databases is that some of the listed reactions have multiple entries because the experimental data were obtained for different temperature ranges. Any user is instructed to use the listed reaction rate coefficients only in their valid temperature regime. Unfortunately, multiple entries in these databases very often display fairly inconsistent properties. This is particularly problematic for numerical codes that continuously change the temperatures when calculating chemical networks because it introduces artificial numerical jumps in the coefficient matrix of the equation system to be solved. One solution could be to restrict oneself to only use the one reaction rate that has a temperature regime closest to the computational regime. However, in many cases this is not possible because the parameter space relevant for application spans multiple temperature regimes. Additionally, this means that one implicitely accepts quite large errors when leaving the valid temperature regime. We suggest to replace these multiple entries with a single, new rate which is fitted simultaneously to the individual ones. This minimizes the error when moving from one temperature regime to the other and additionally overcomes the problem of discontinuous reaction rate jumps. 

We consider the new parametrization presented in this paper only a temporary solution until better experimental data for low temperature reaction rates are available. 

As a side effect we demonstrate that in some cases one might find an equally good parametrization of the reaction rate coefficients without any artificial numerical side effects just by taking care not to allow too low negative values of $\gamma$. The chemical databases try to express the possibly very complicated physics of the various chemical reactions with only three fitting parameters. Of course, this is an extremely strong constraint on any parameter fit to the complex experimental data. Although, under certain circumstances, chemical reactions listed in chemical databases like UDFA, OSU, and others might reasonably be applied even outside their given scope, it is important to ensure that numerical side effects from parametrization are minimized when possible. For many numerical applications this is much more important than a smaller total error of the reaction rate coefficients in a limited range in temperature. 

Any user should always take care not to blindly use reaction rate coefficients from chemical databases outside their temperature scope. This is also true for the new parametrizations listed in this paper. In each case, one has to balance advantages against disadvantages. If you are sure that you will keep within the given temperature regime, then there is no need to use our rate coefficients because they add an additional error to the computation. Under certain circumstances, this additional error might be negligible compared to the numerical problems that come with the original rate coefficients.

\begin{acknowledgements}
 This work was supported by the German
      \emph{Deut\-sche For\-schungs\-ge\-mein\-schaft, DFG\/} project
      number Os~177/1--1. We acknowledge the
use of UDFA (a.k.a. UMIST)
(http://www.udfa.net/) chemical reaction databases.
Some kinetic data we used were downloaded from the online 
database KIDA (KInetic Database for 
Astrochemistry, http://kida.obs.u-bordeaux1.fr).
The author thanks 
R. Simon and V. Ossenkopf for their helpful comments on the text. 
The author thanks the anonymous referee for his or her comments, which 
helped the final text a lot. 
\end{acknowledgements}

\bibliographystyle{aa}
\bibliography{14743}

\Online
\appendix
 \section{Plots of  all new fits.}
Here, we show the plots of all new fits to allow the user an assessment of the fit quality over the entire fitted temperature range. The plots are shown in the same order as in Table~\ref{tab1} and \ref{tab2}.
\begin{figure*}
 \centering
 \includegraphics[width=6cm]{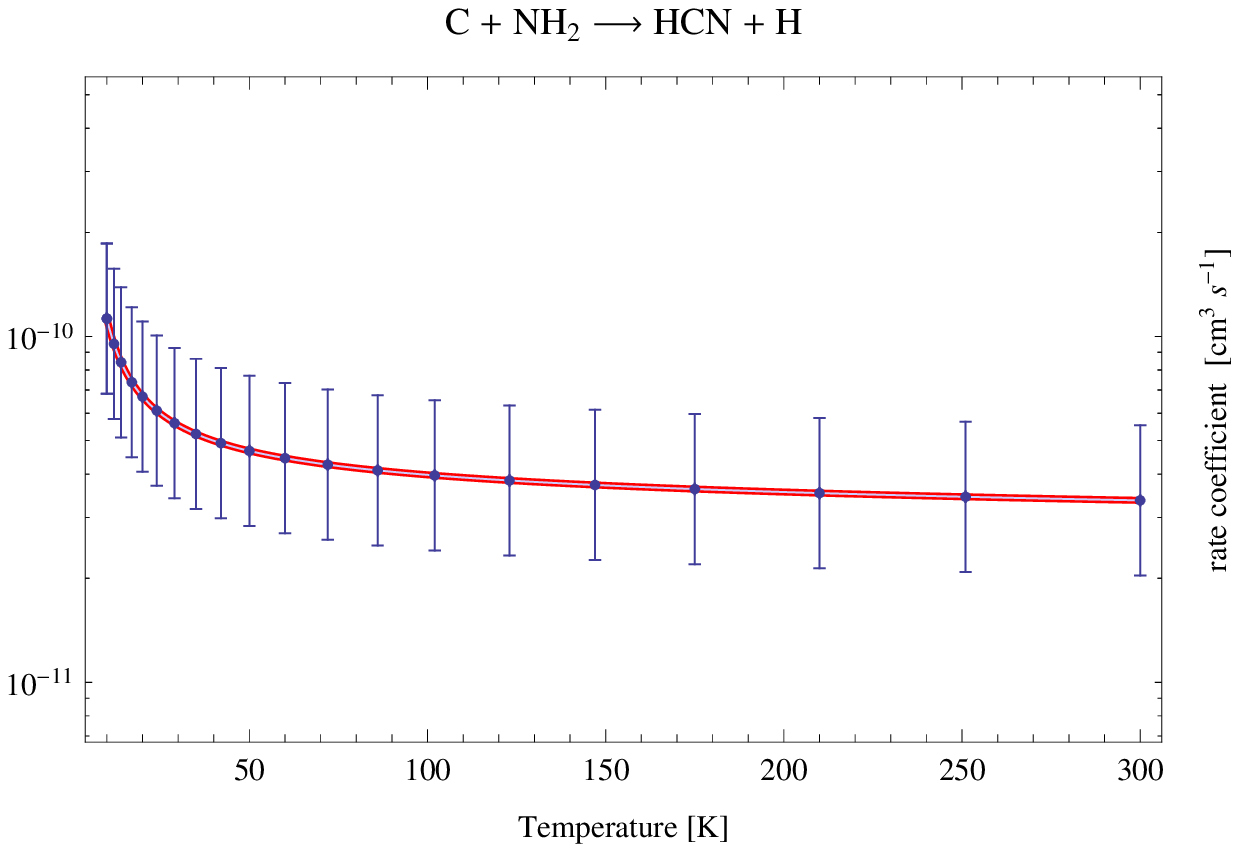}
 \hfill
 \includegraphics[width=6cm]{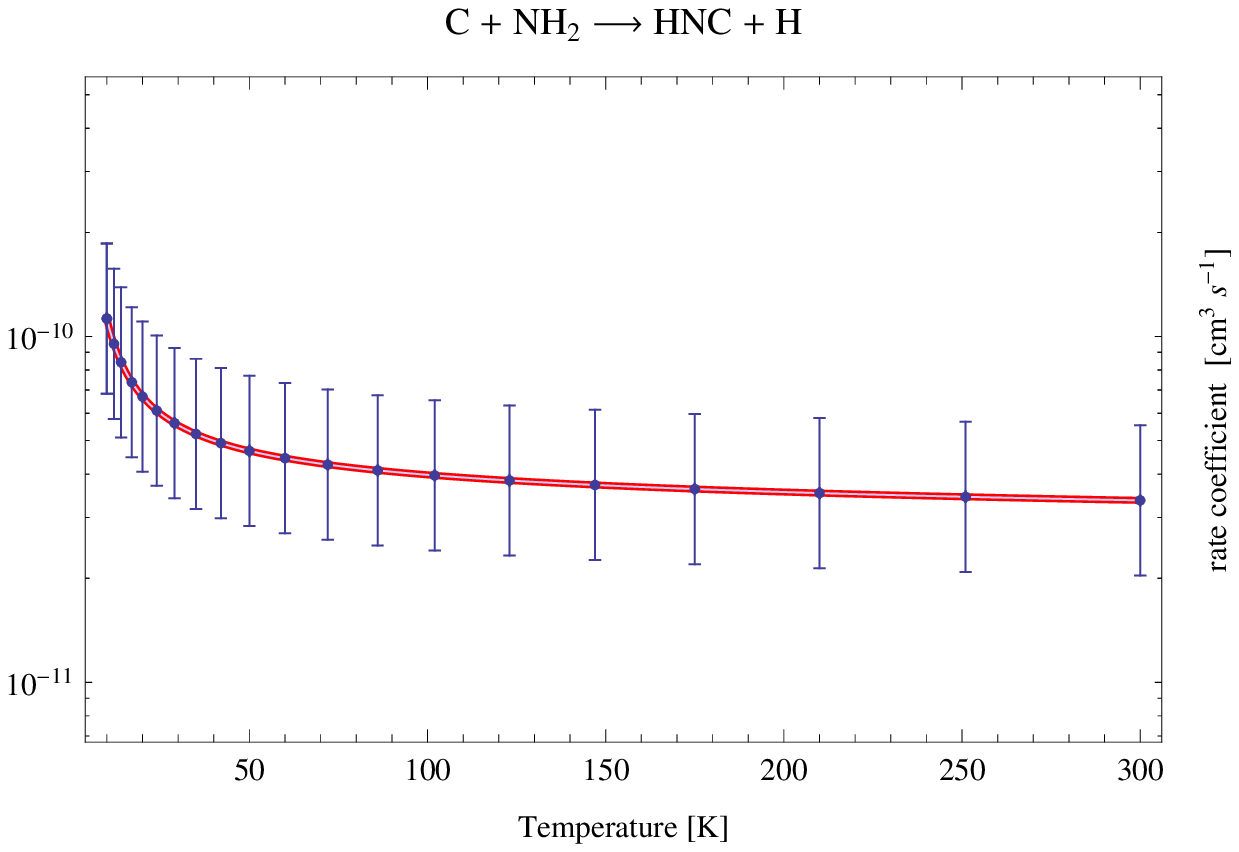}
 \hfill
 \includegraphics[width=6cm]{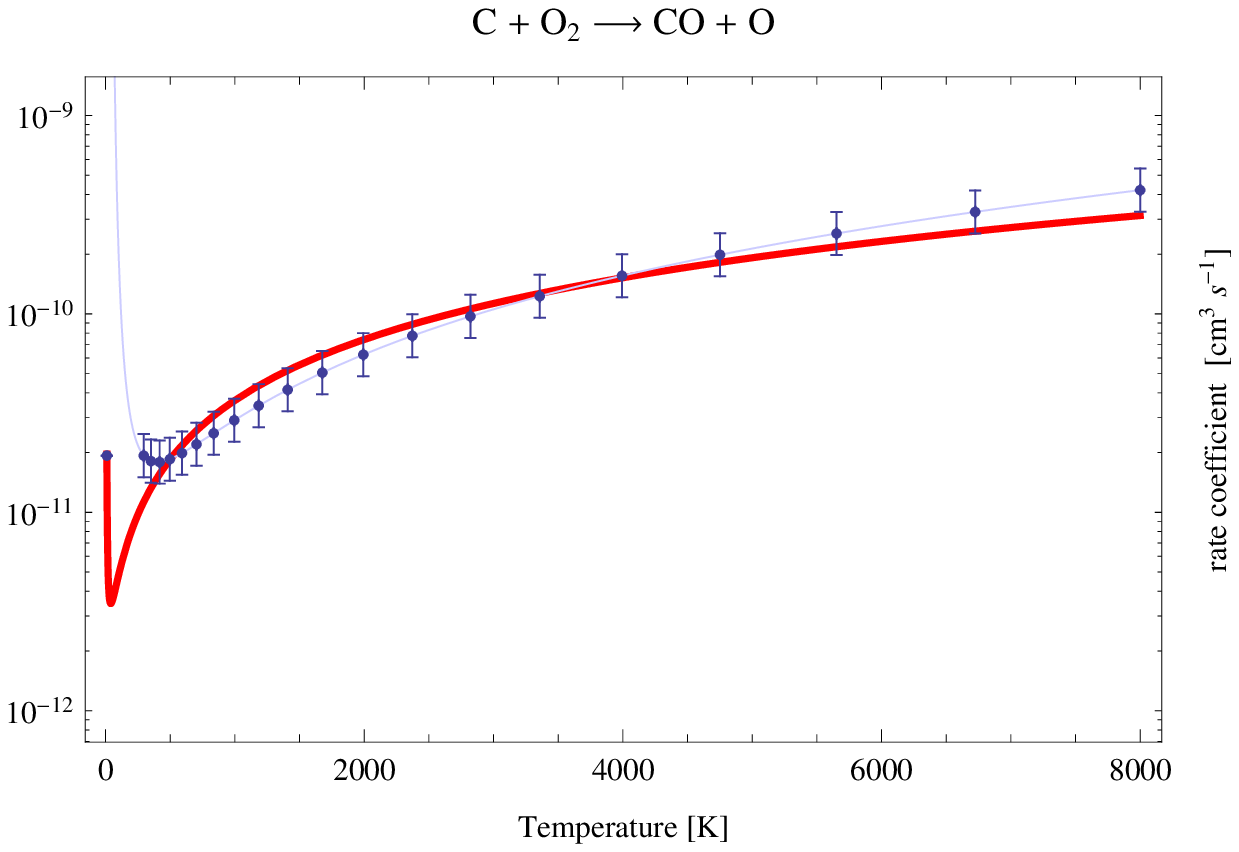}\\
 \includegraphics[width=6cm]{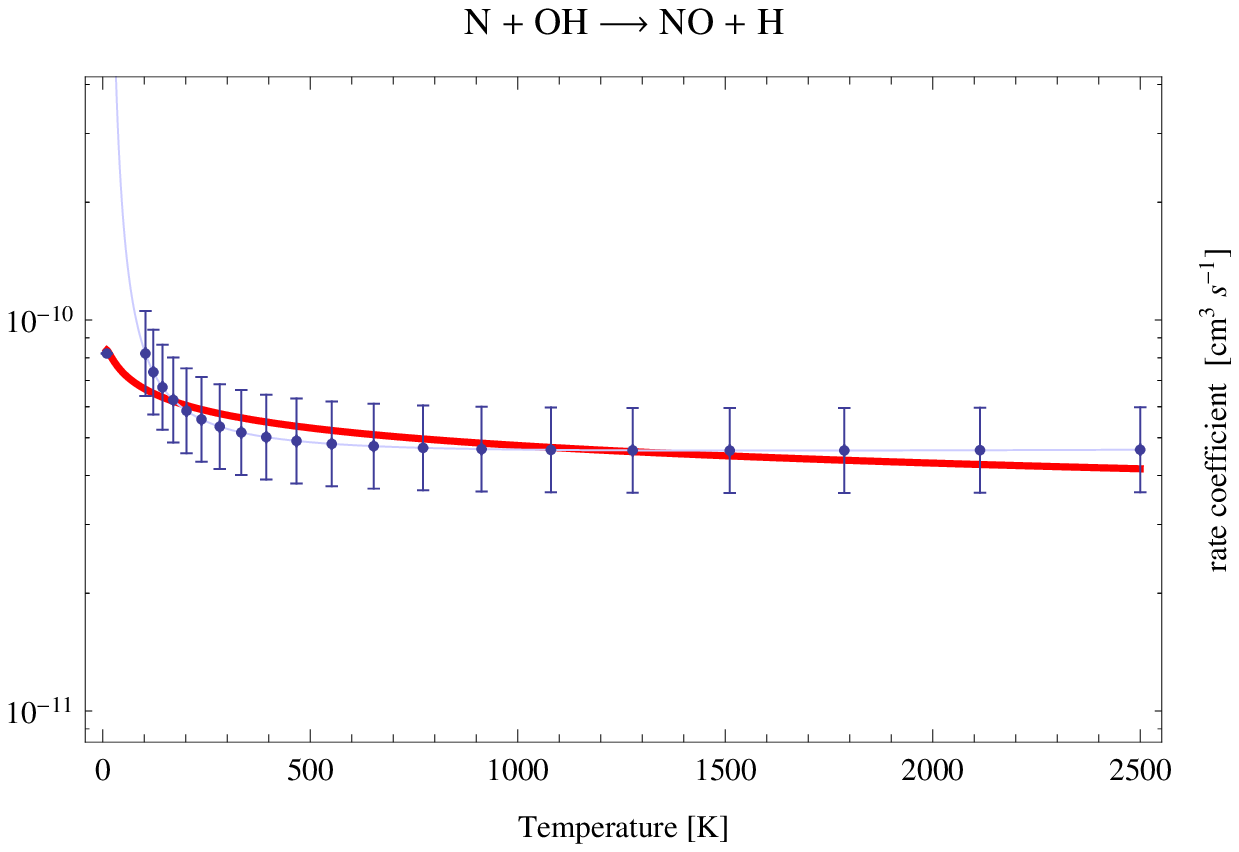}
 \hfill
 \includegraphics[width=6cm]{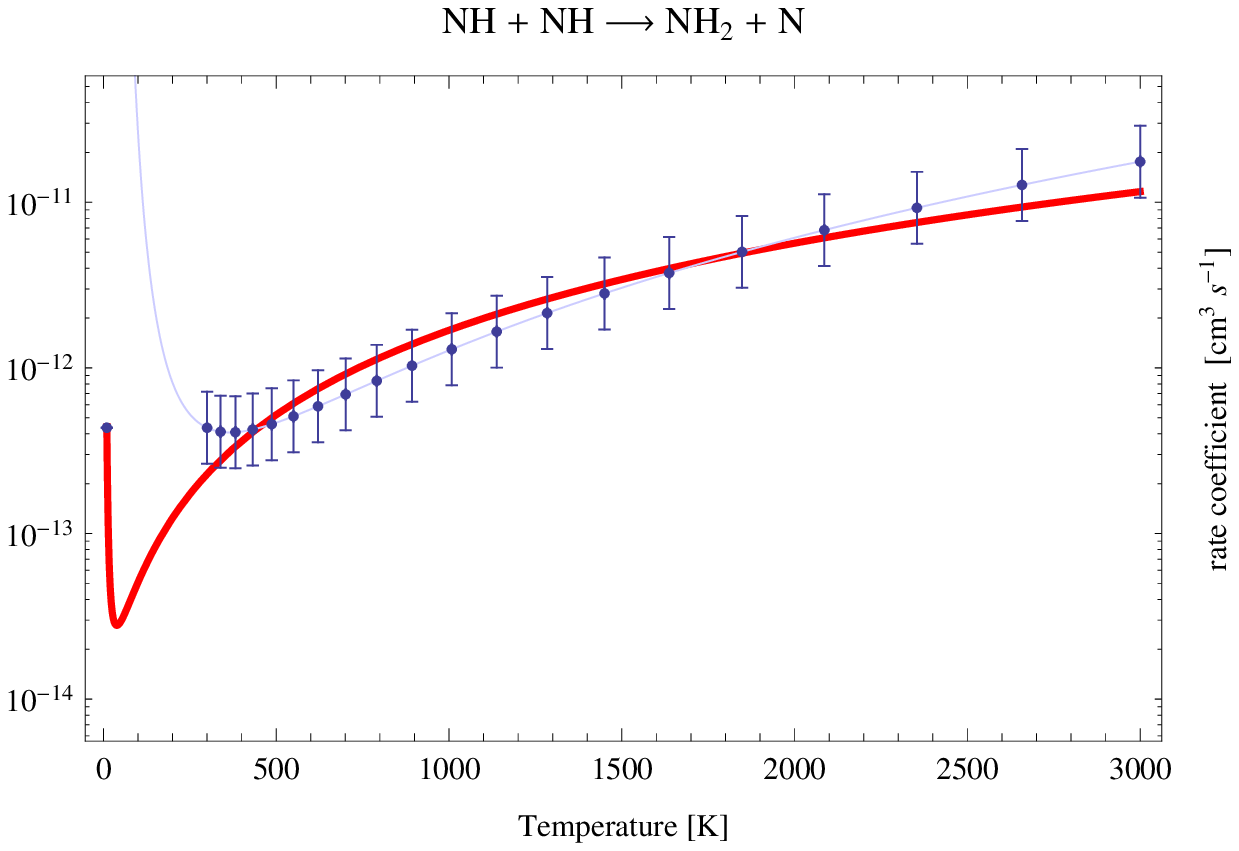}
 \hfill
 \includegraphics[width=6cm]{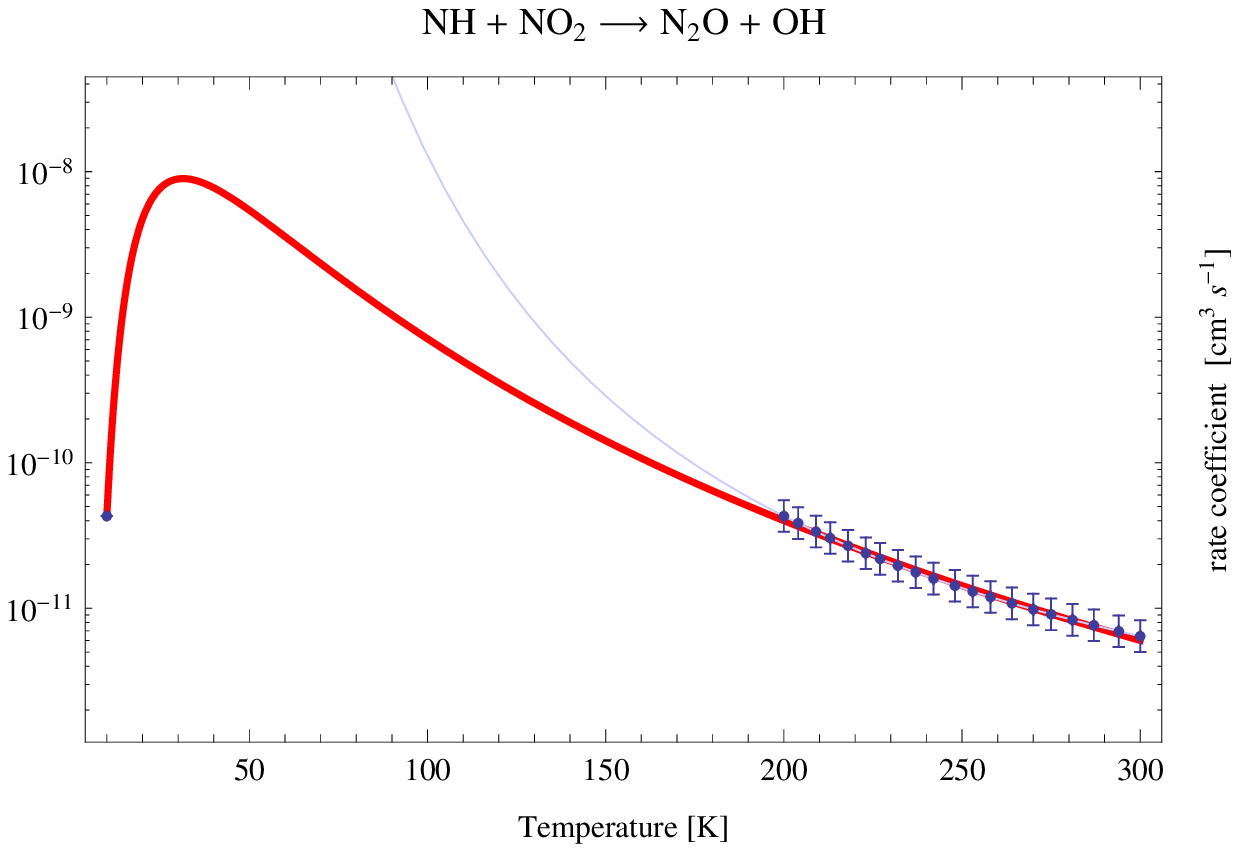}\\ 
\includegraphics[width=6cm]{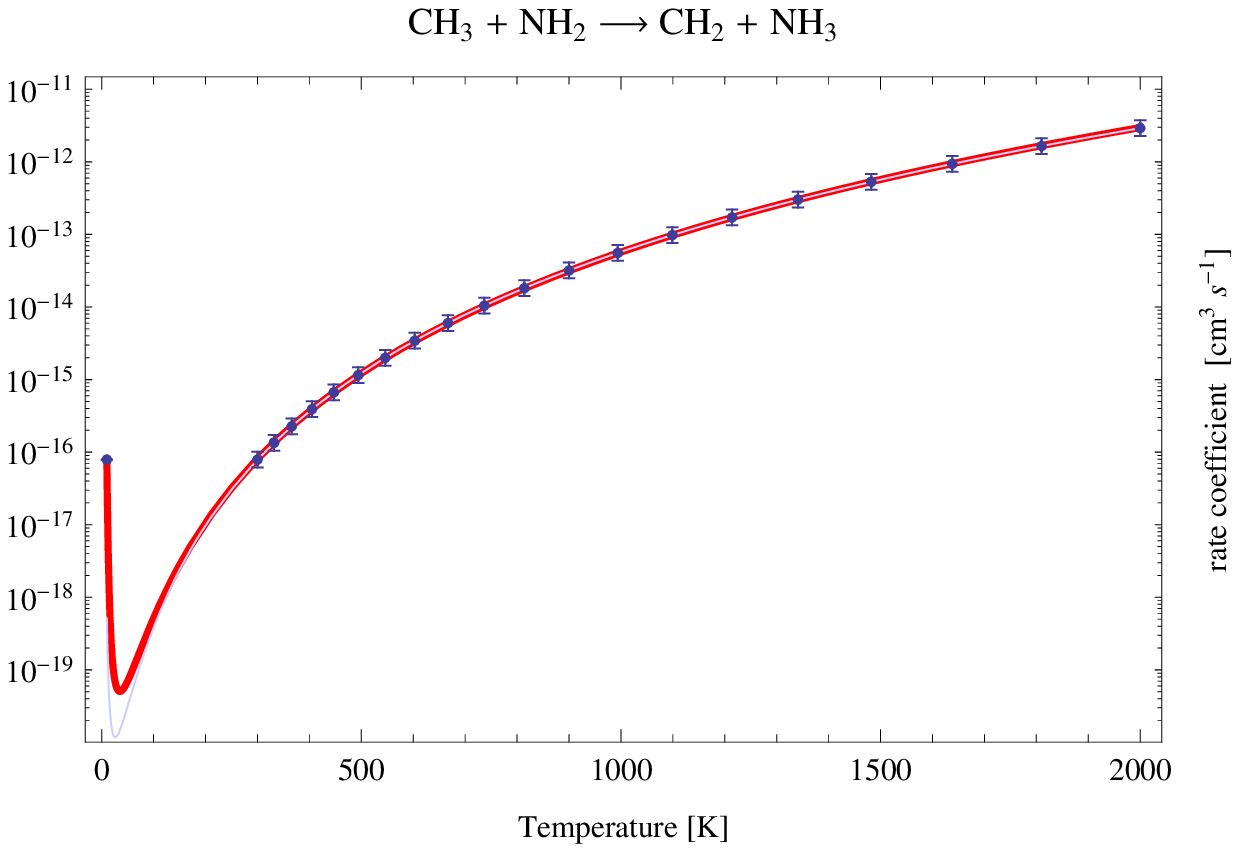}
 \hfill
 \includegraphics[width=6cm]{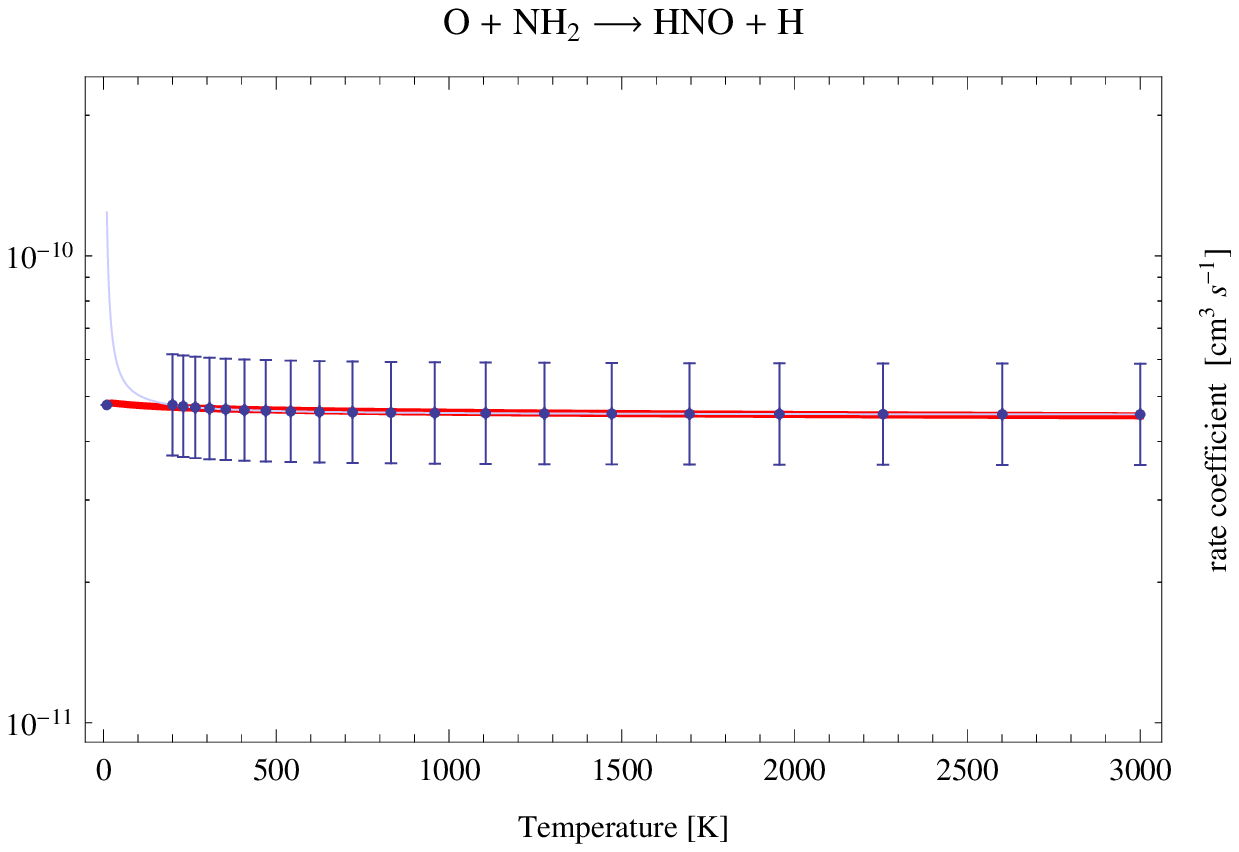}
 \hfill
 \includegraphics[width=6cm]{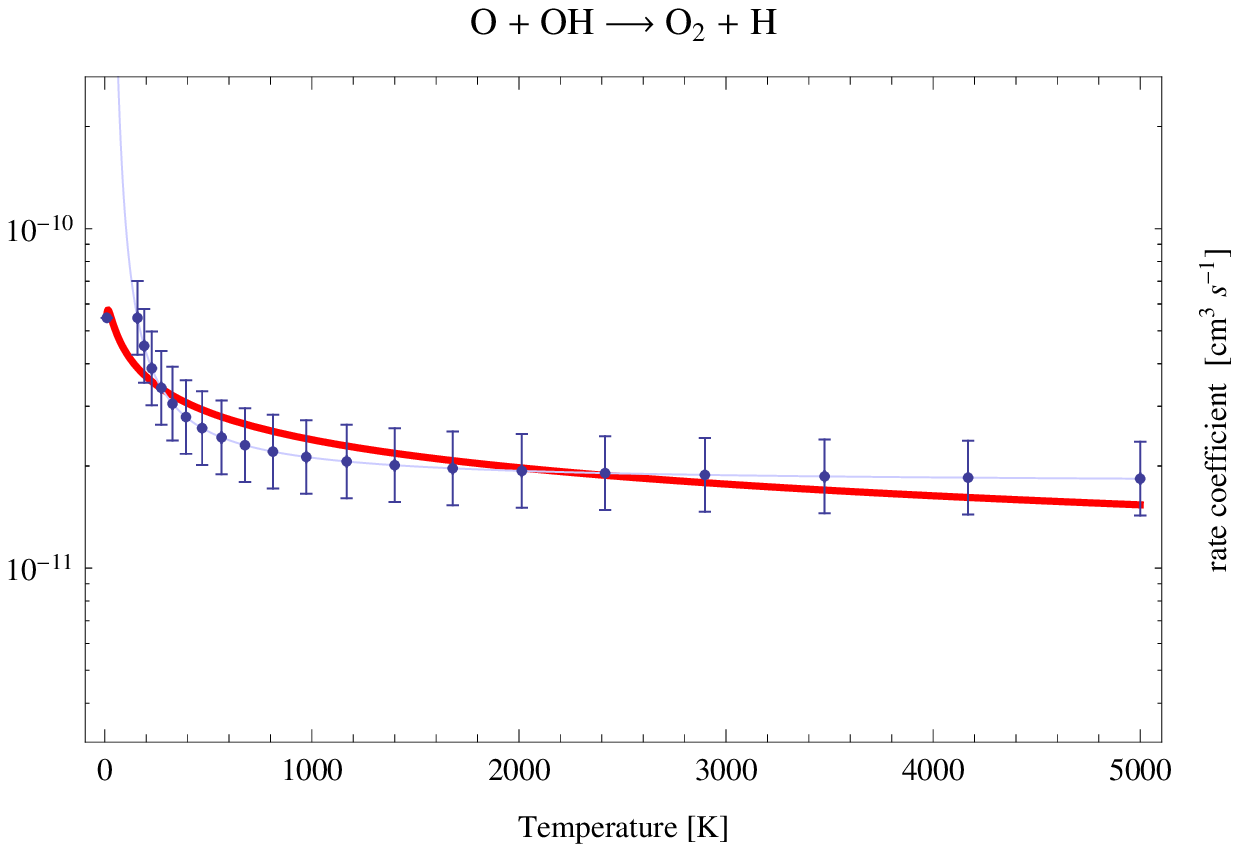}\\
\includegraphics[width=6cm]{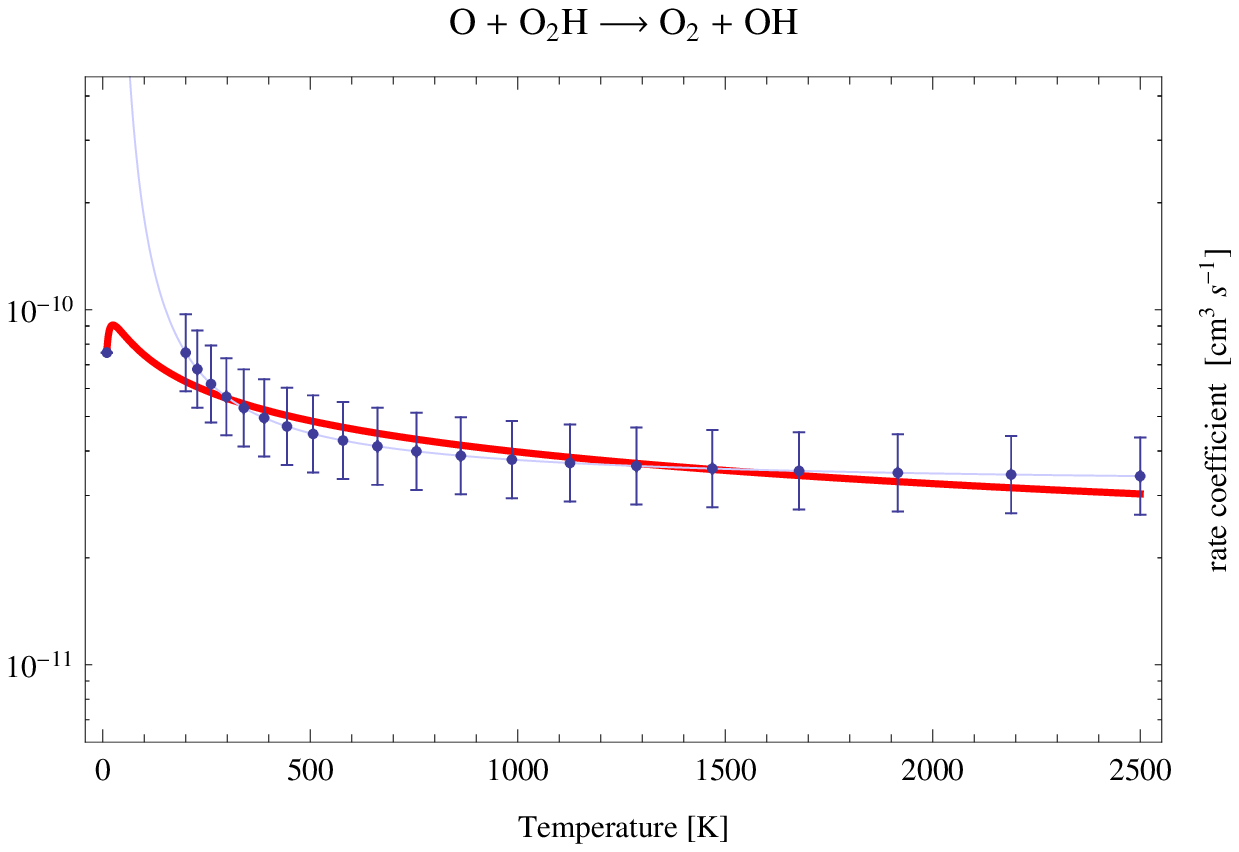}
 \hfill
 \includegraphics[width=6cm]{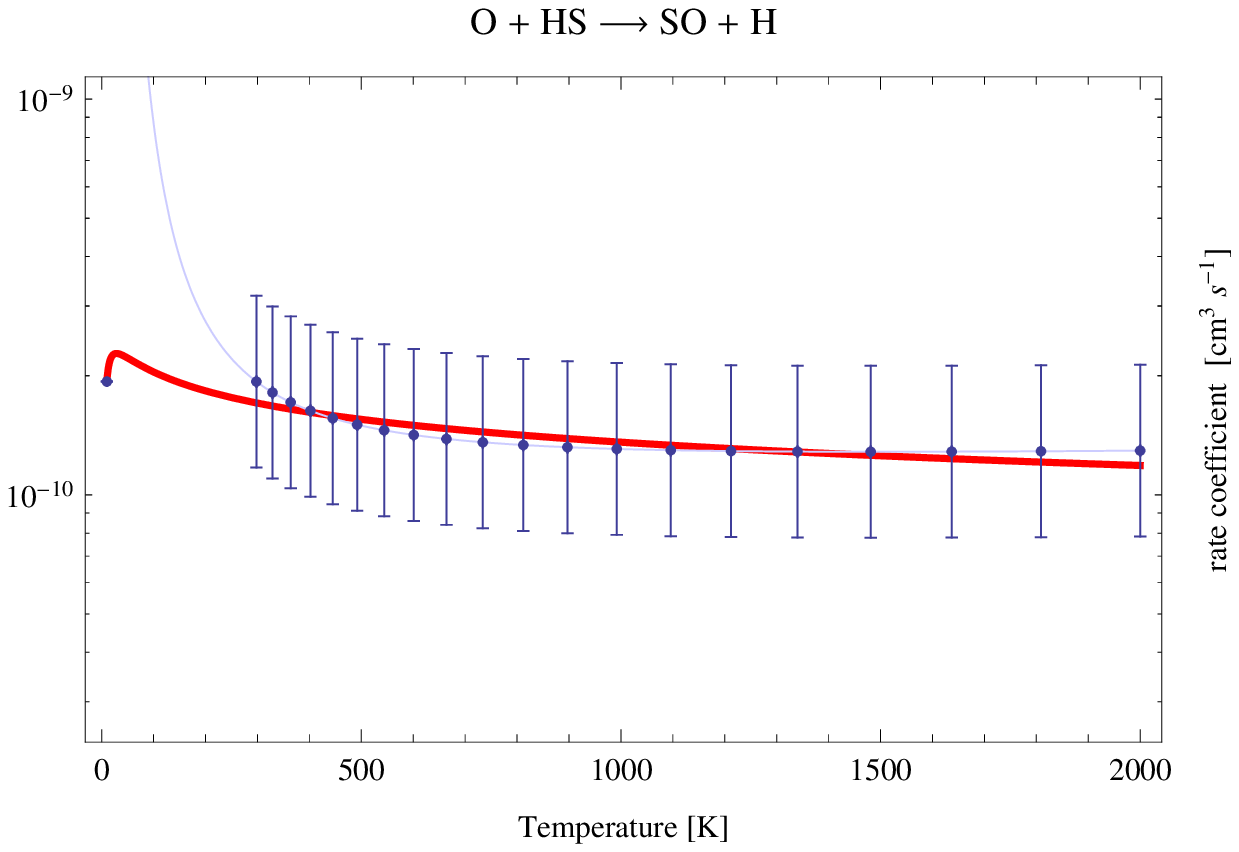}
 \hfill
 \includegraphics[width=6cm]{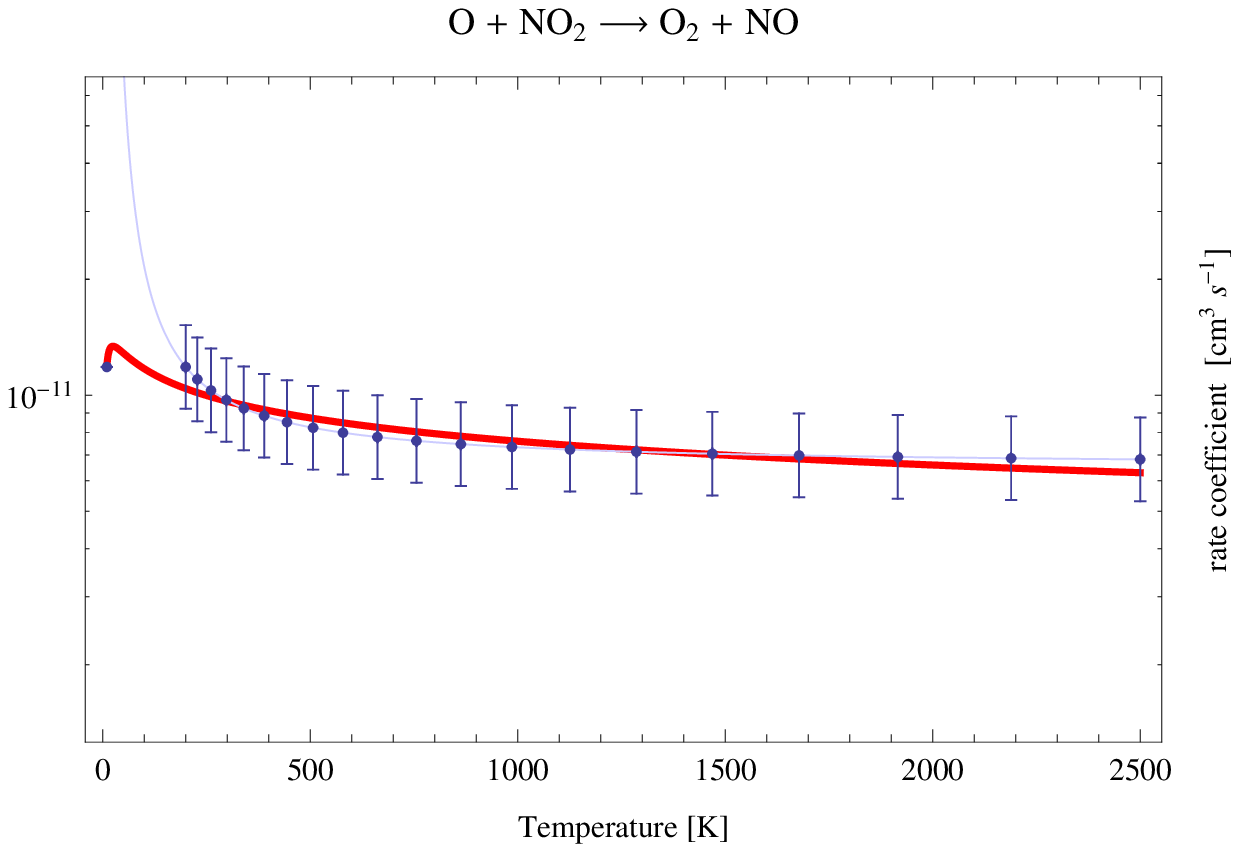}\\
\includegraphics[width=6cm]{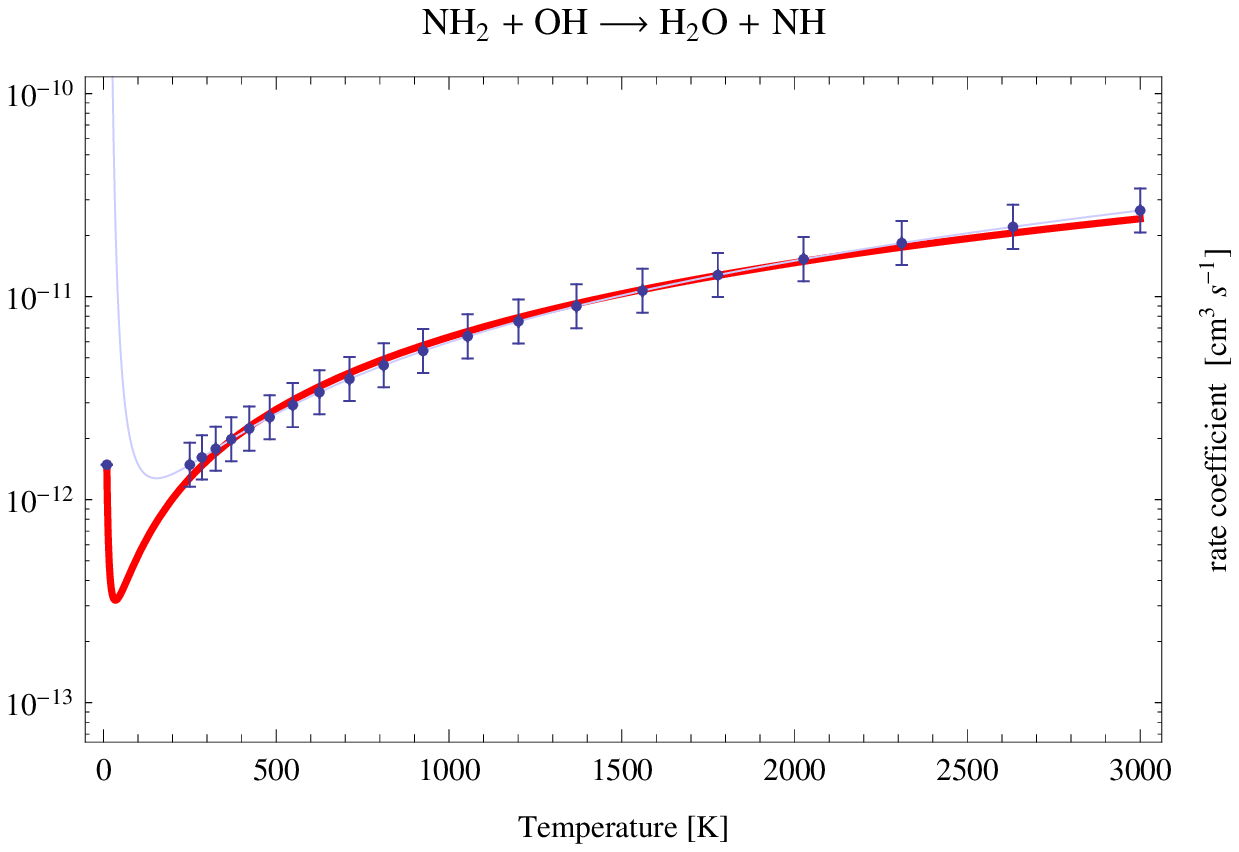}
 \hfill
 \includegraphics[width=6cm]{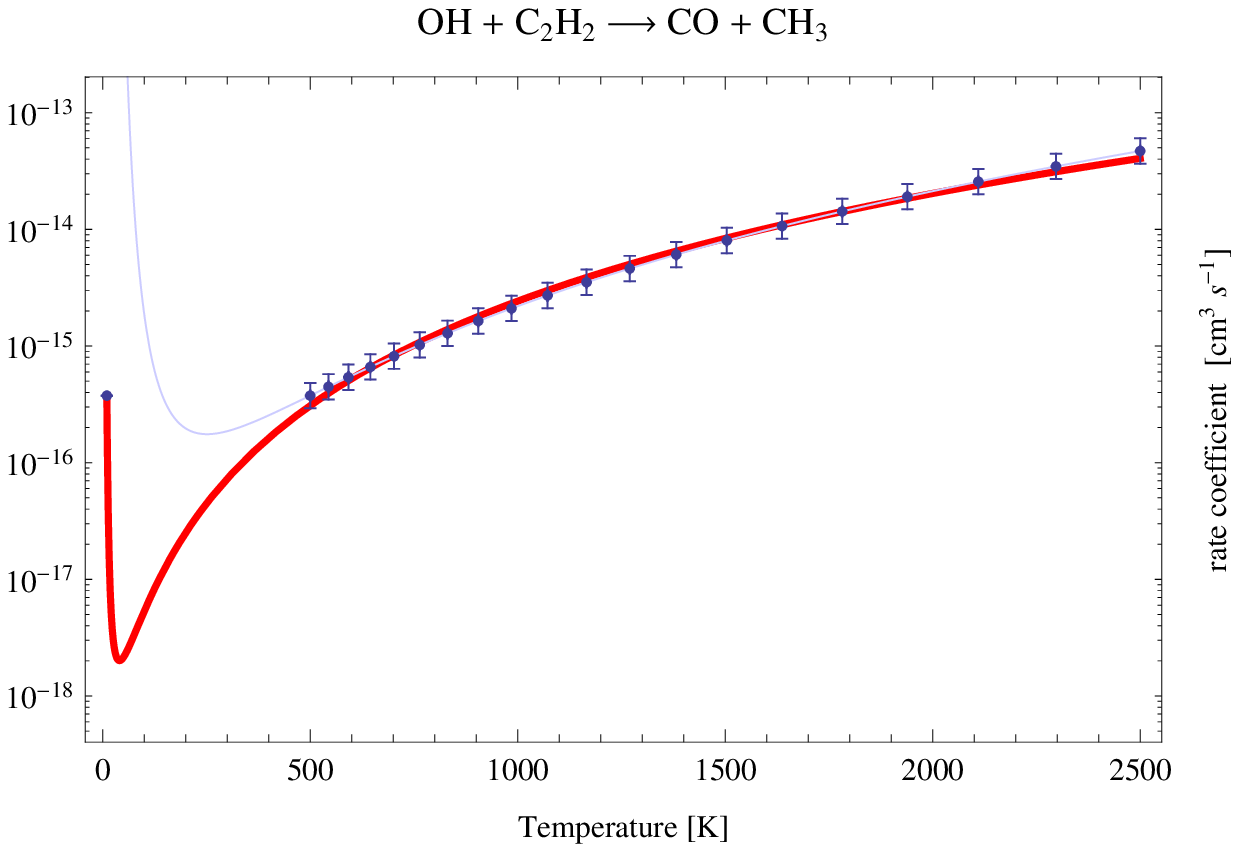}
 \hfill
 \includegraphics[width=6cm]{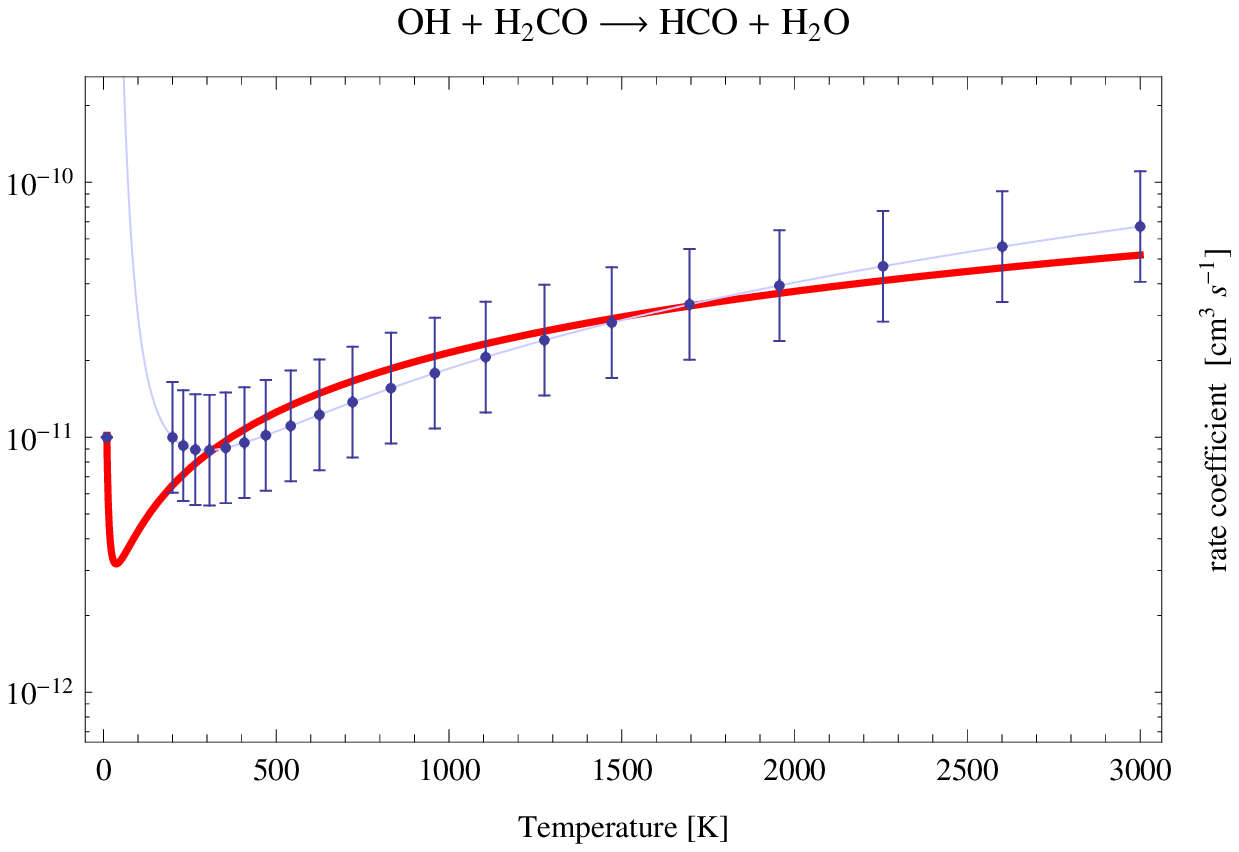}\\
 \includegraphics[width=6cm]{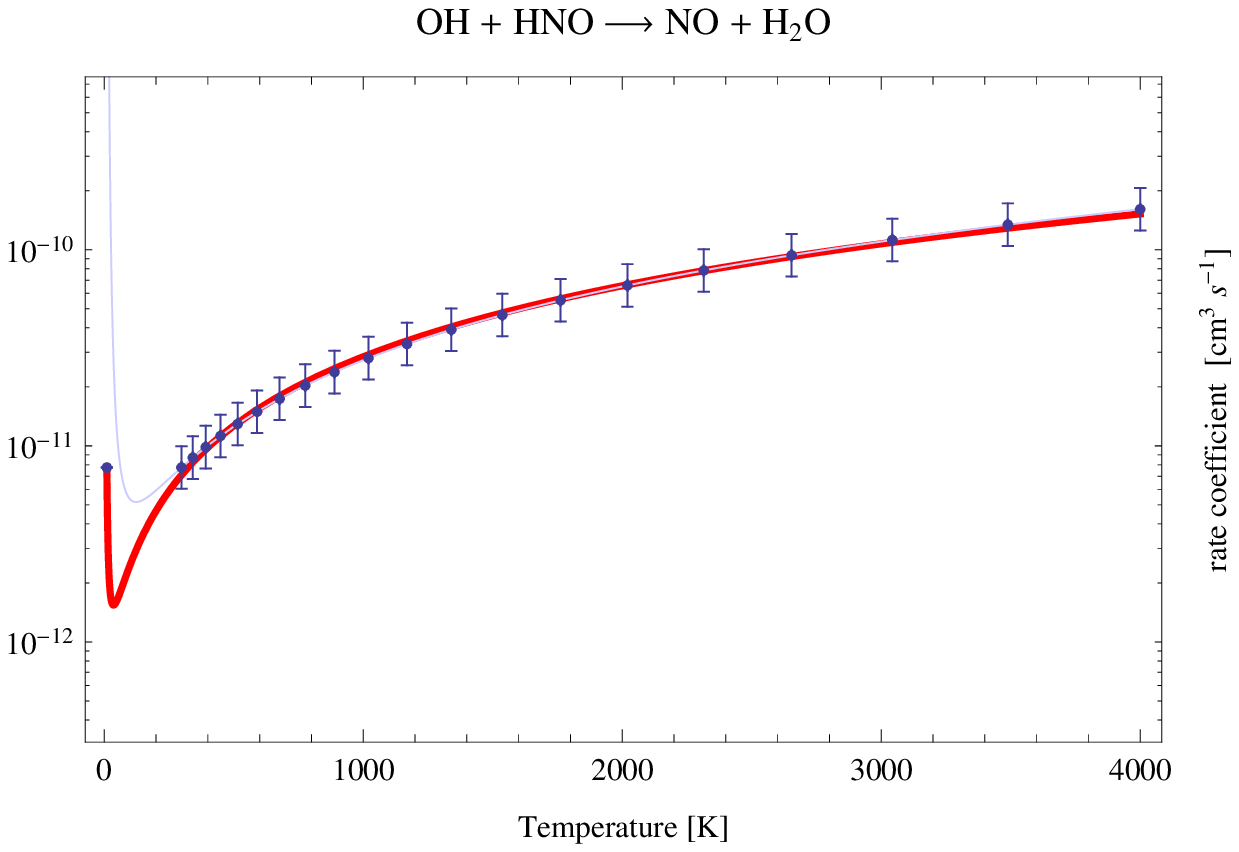}
 \hfill
 \includegraphics[width=6cm]{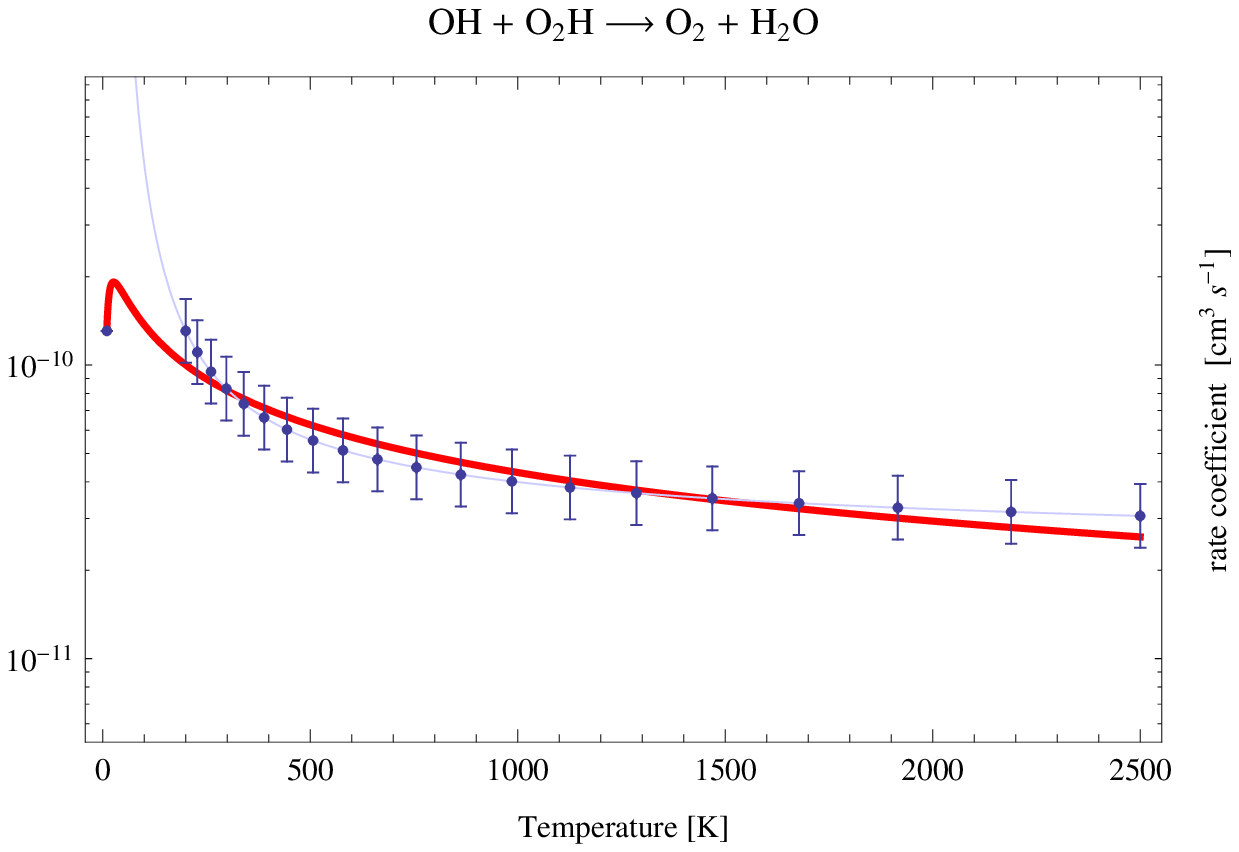}
 \hfill
 \includegraphics[width=6cm]{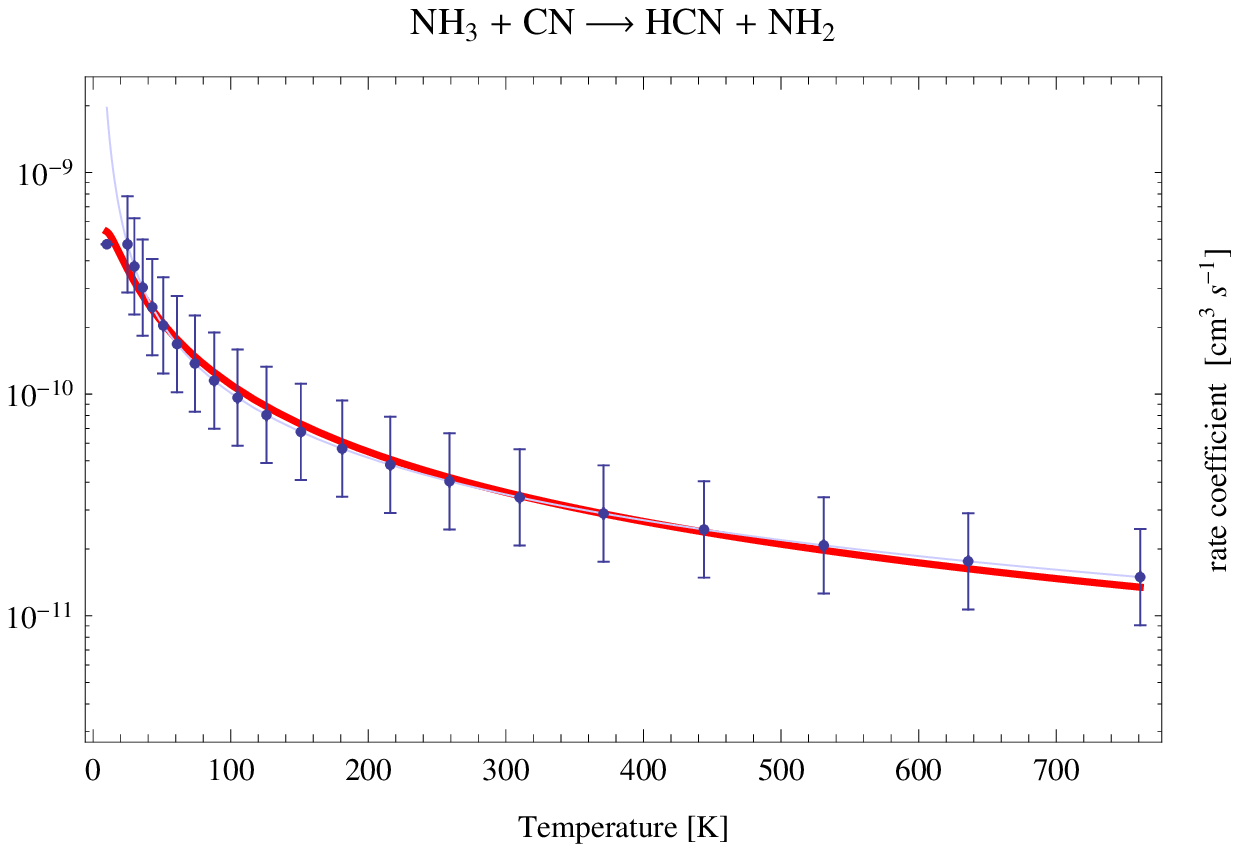}\\
 \label{appendix1}
 \caption{Fits to reactions with negative $\gamma$.}
 \end{figure*}
\begin{figure*}
 \centering
 \includegraphics[width=6cm]{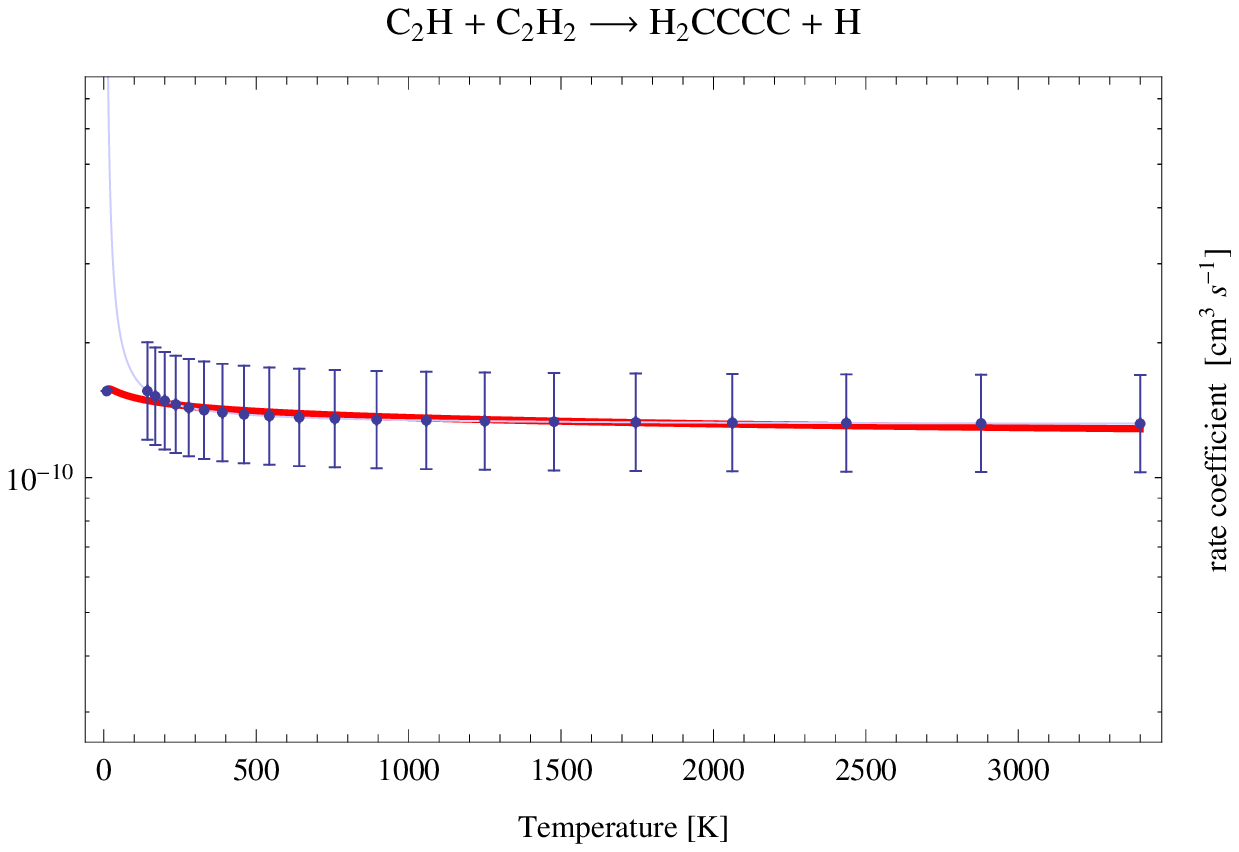}
 \hfill
 \includegraphics[width=6cm]{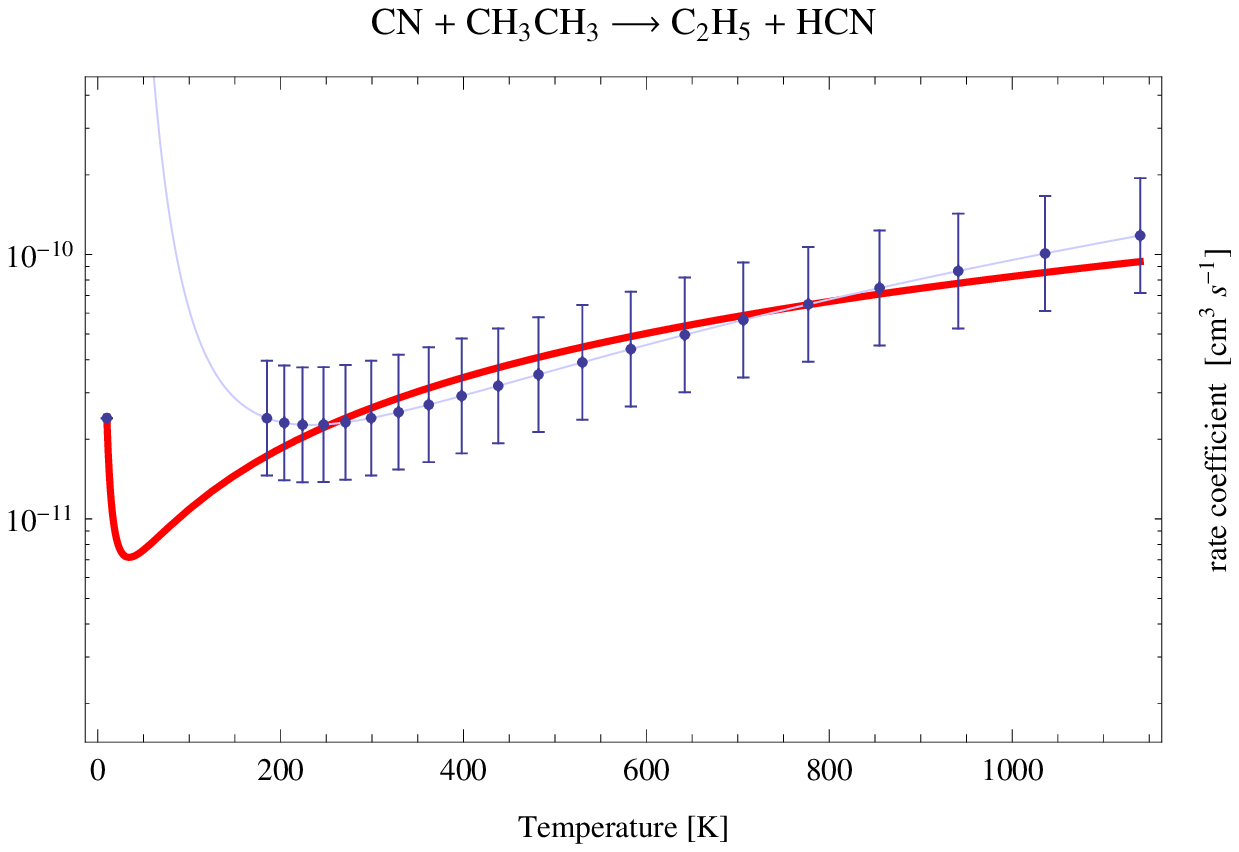}
 \hfill
 \includegraphics[width=6cm]{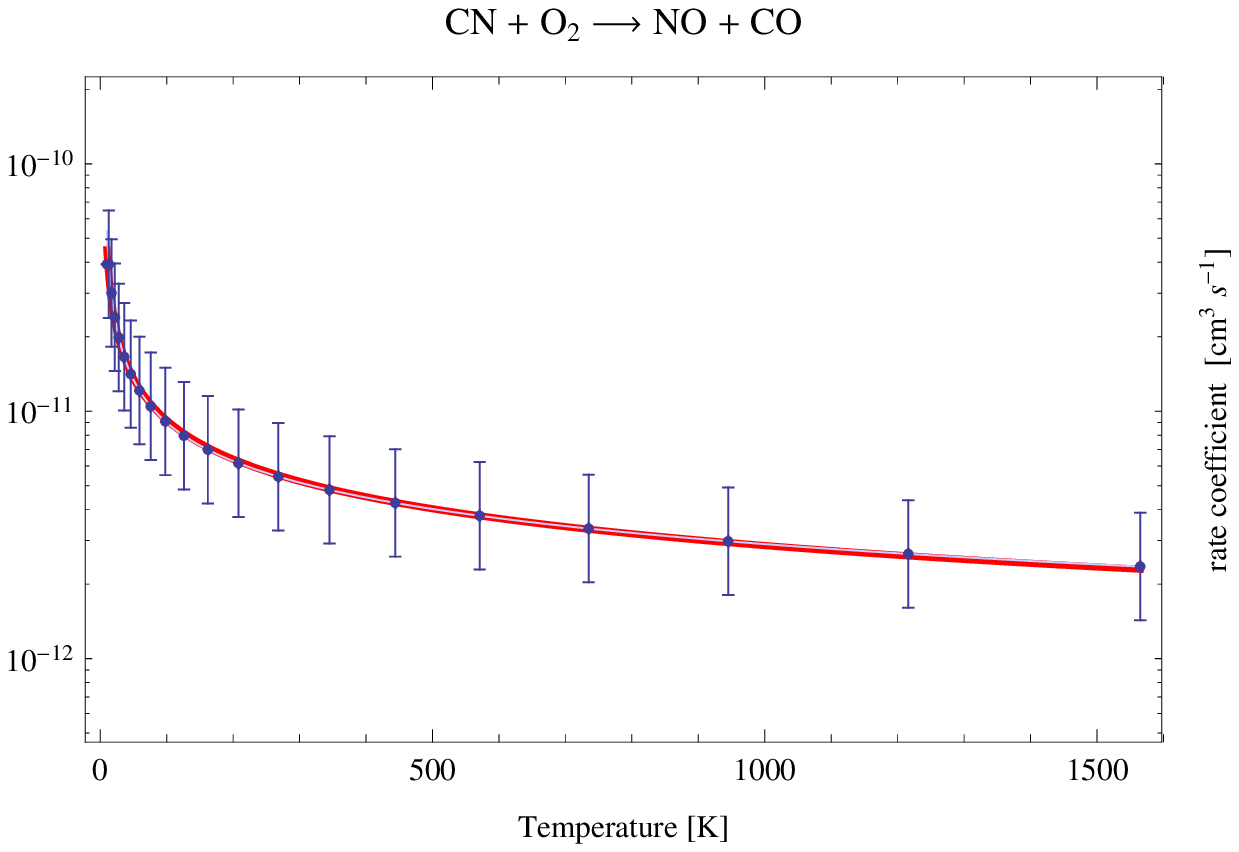}\\
 \includegraphics[width=6cm]{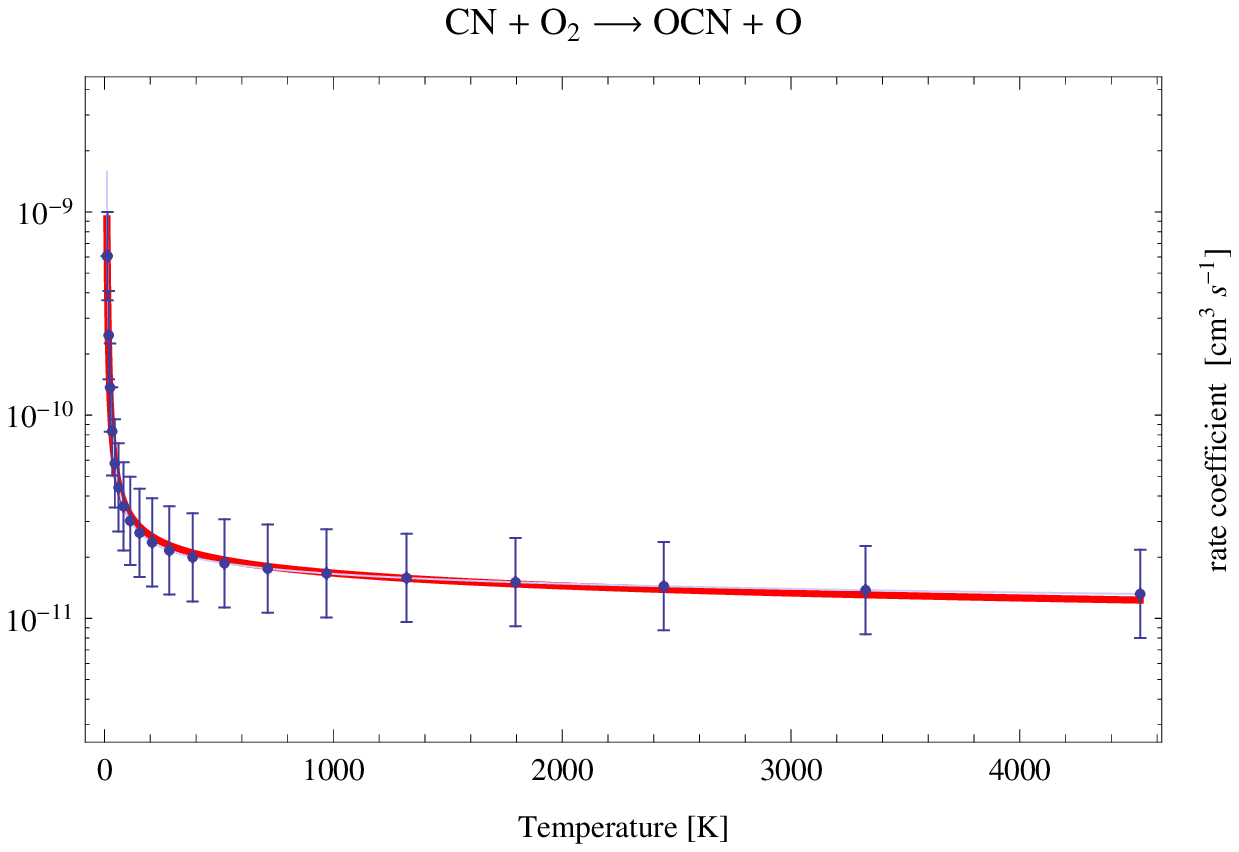}
 \hfill
 \includegraphics[width=6cm]{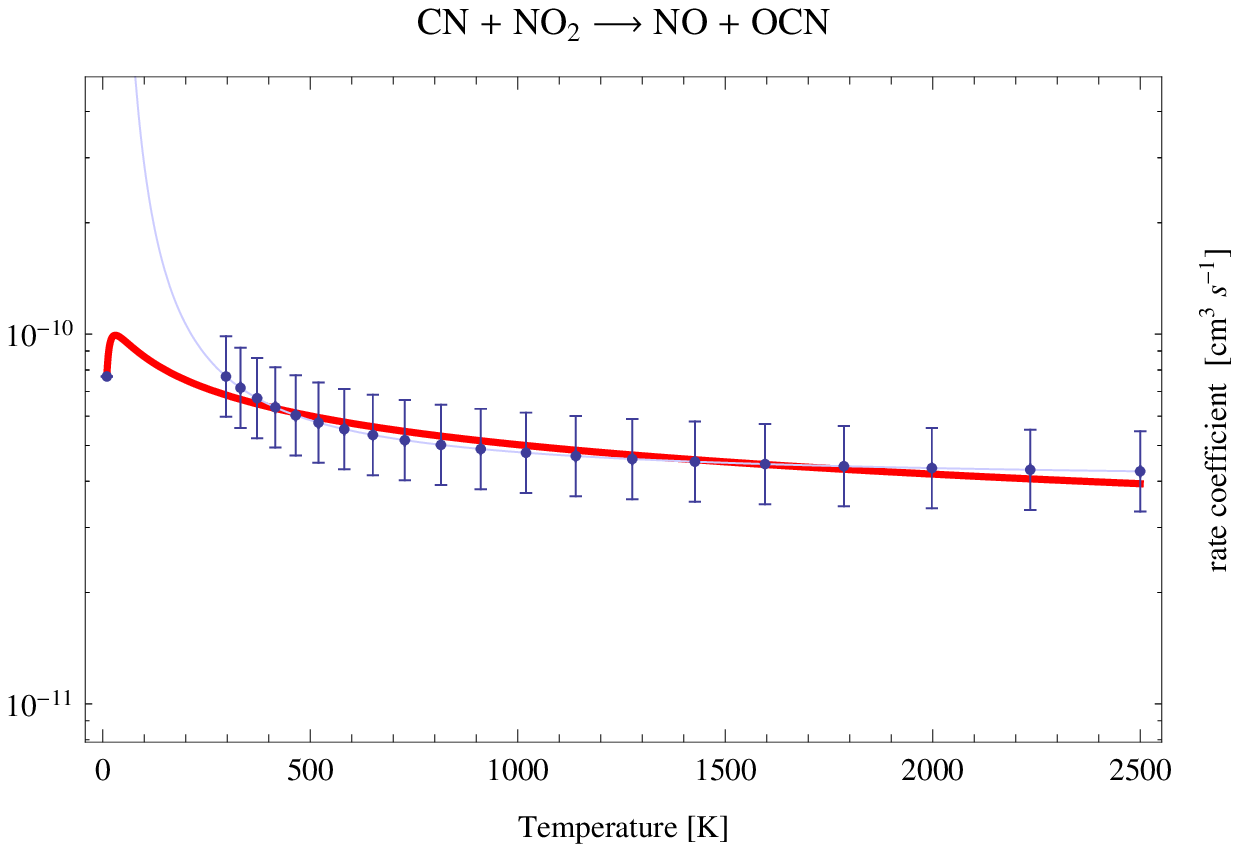}
 \hfill
 \includegraphics[width=6cm]{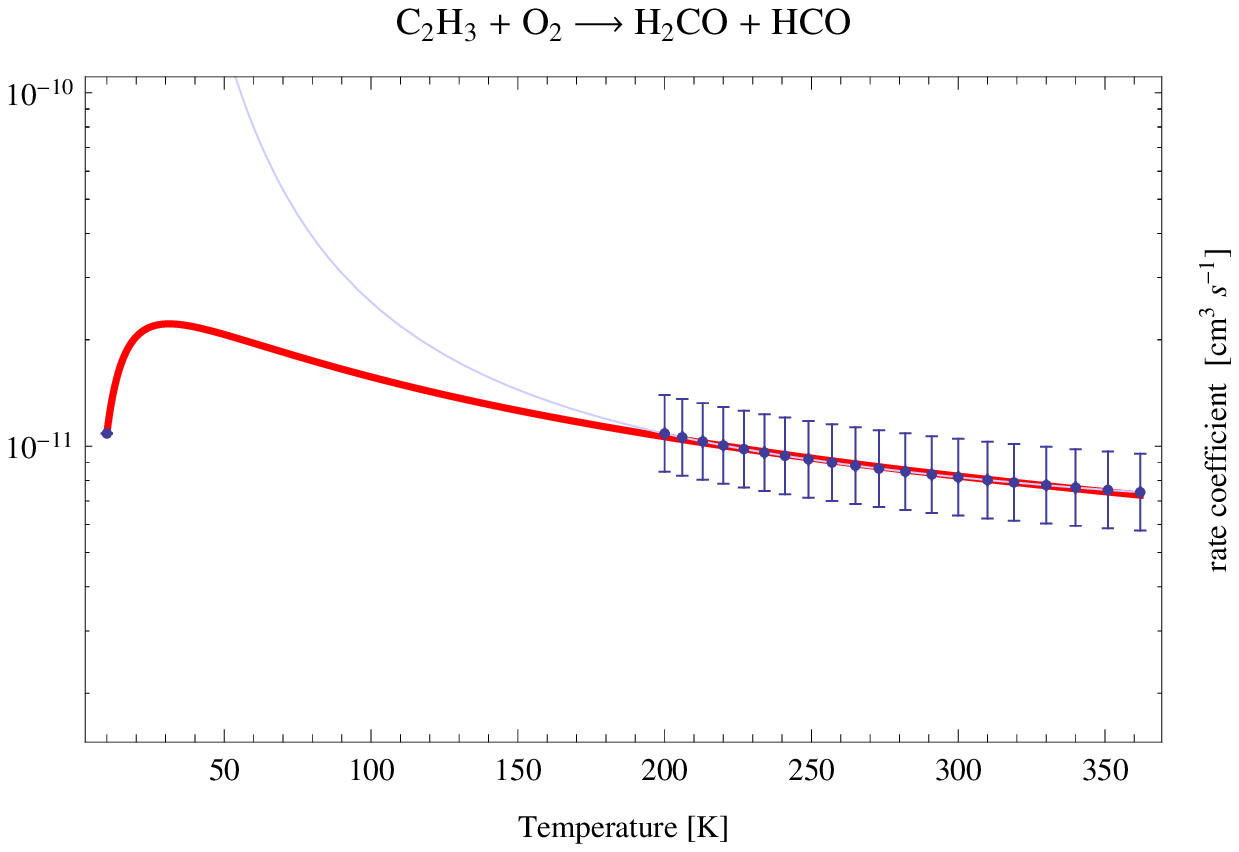}\\ 
\includegraphics[width=6cm]{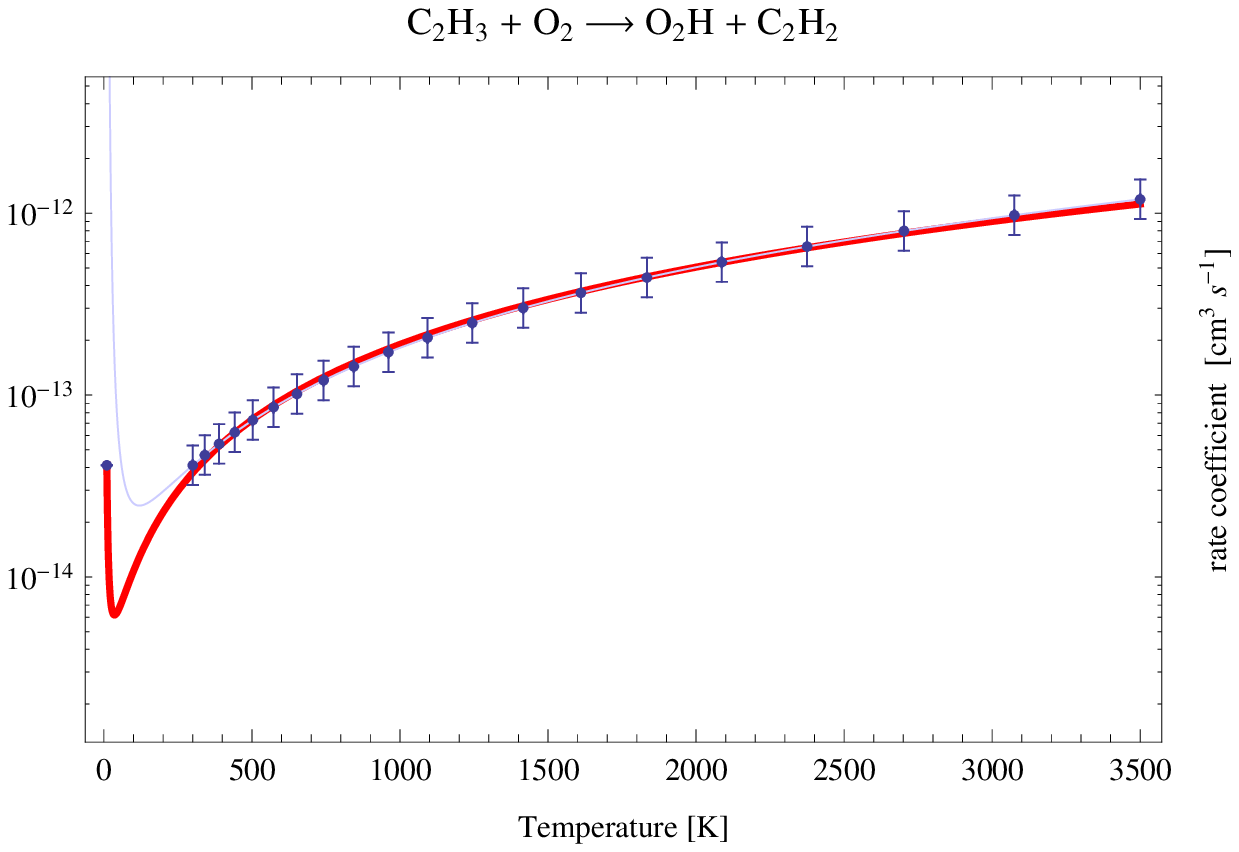}
 \hfill
 \includegraphics[width=6cm]{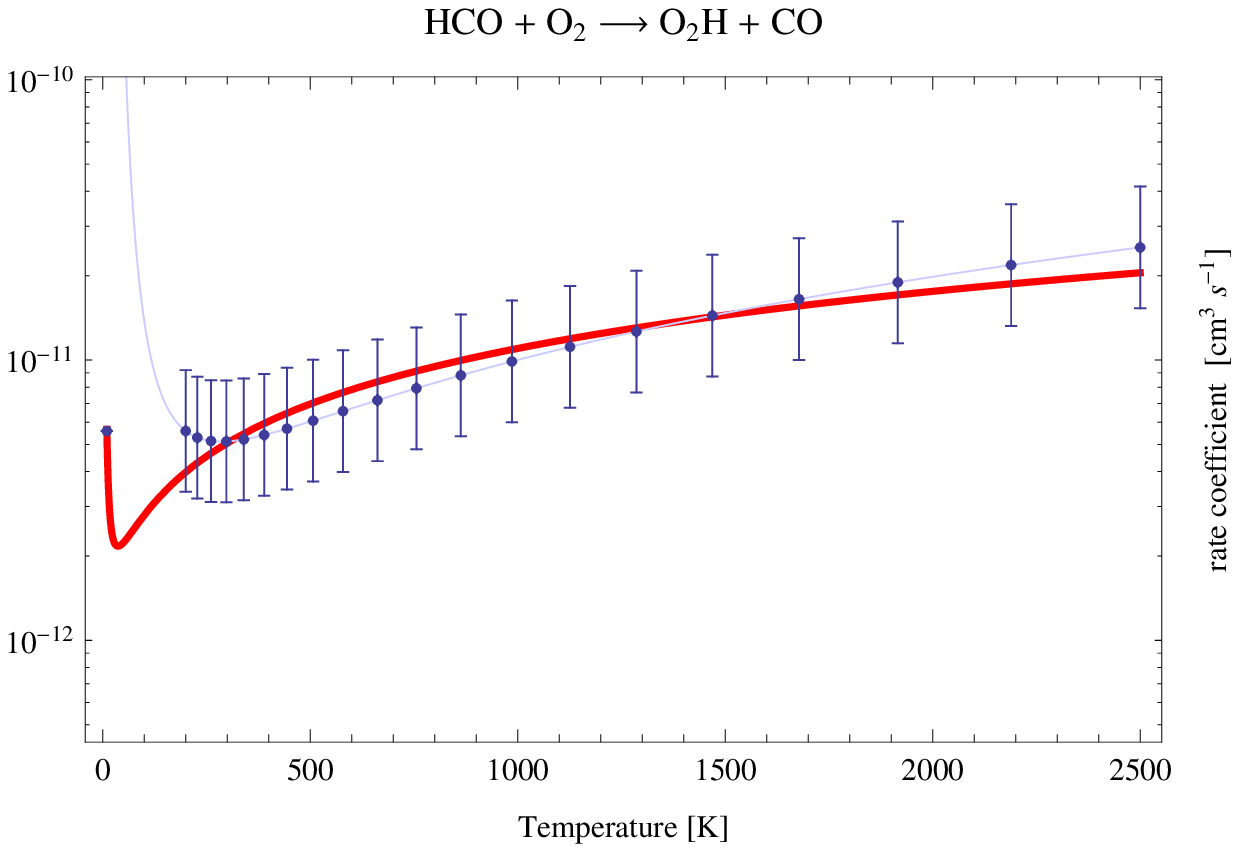}
 \hfill
 \includegraphics[width=6cm]{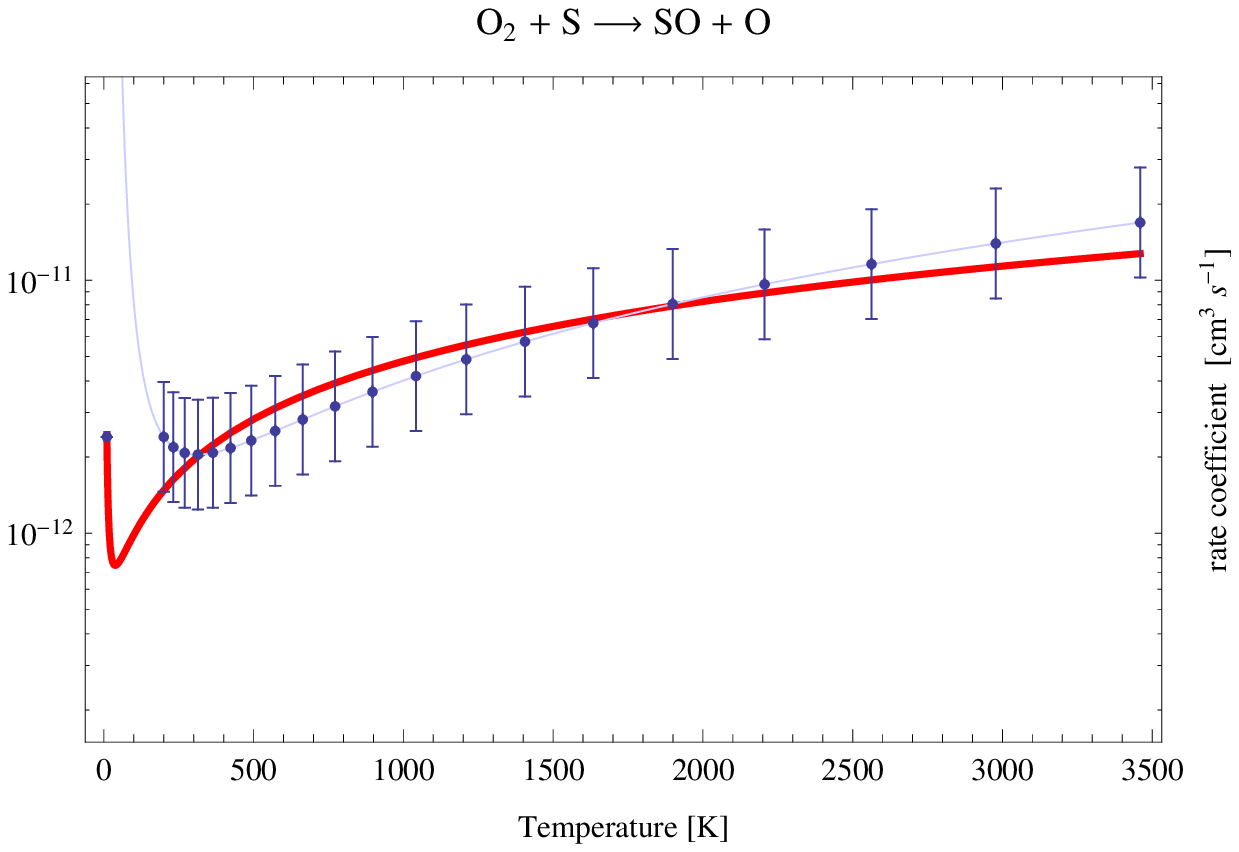}\\
 \caption{Fits to reactions with negative $\gamma$ (continued).}
 \end{figure*}
 
\begin{figure*}
 \centering
 \includegraphics[width=6cm]{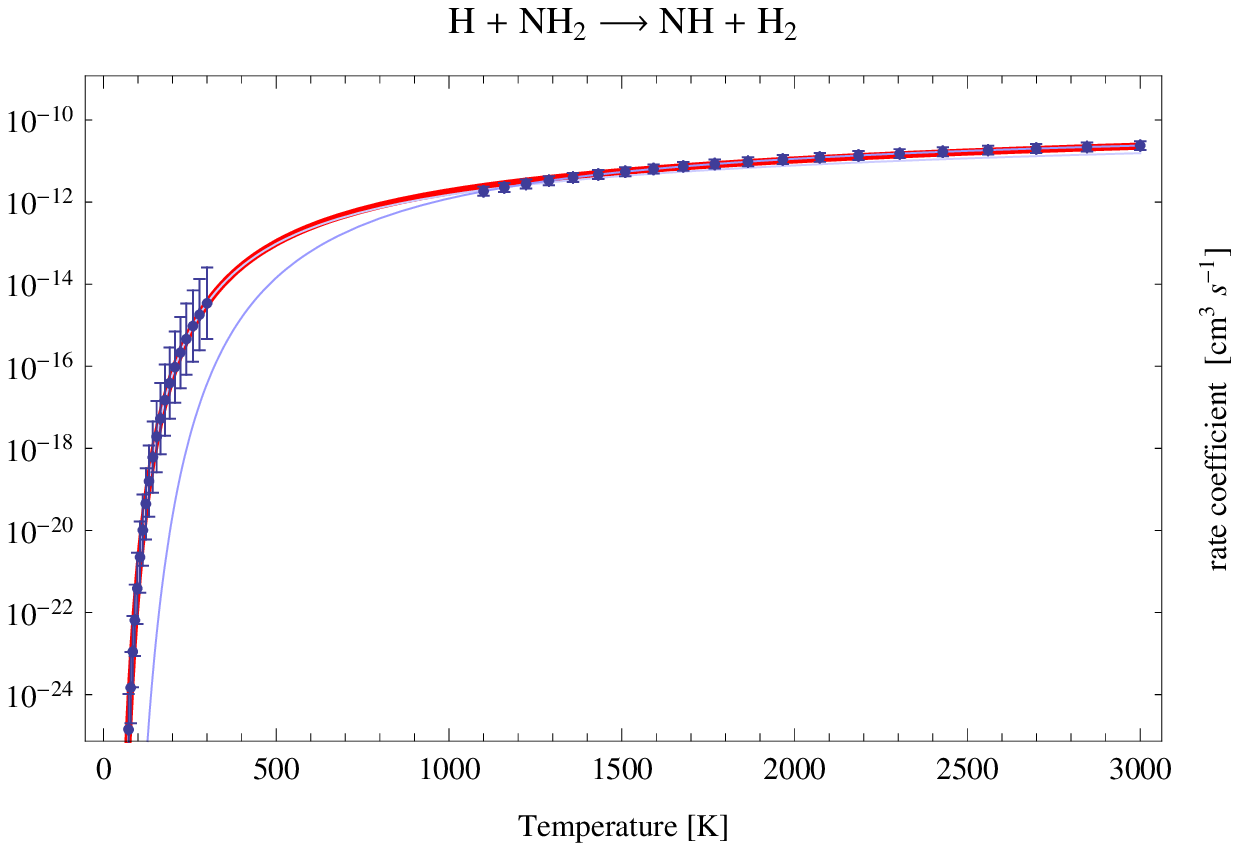}
 \hfill
 \includegraphics[width=6cm]{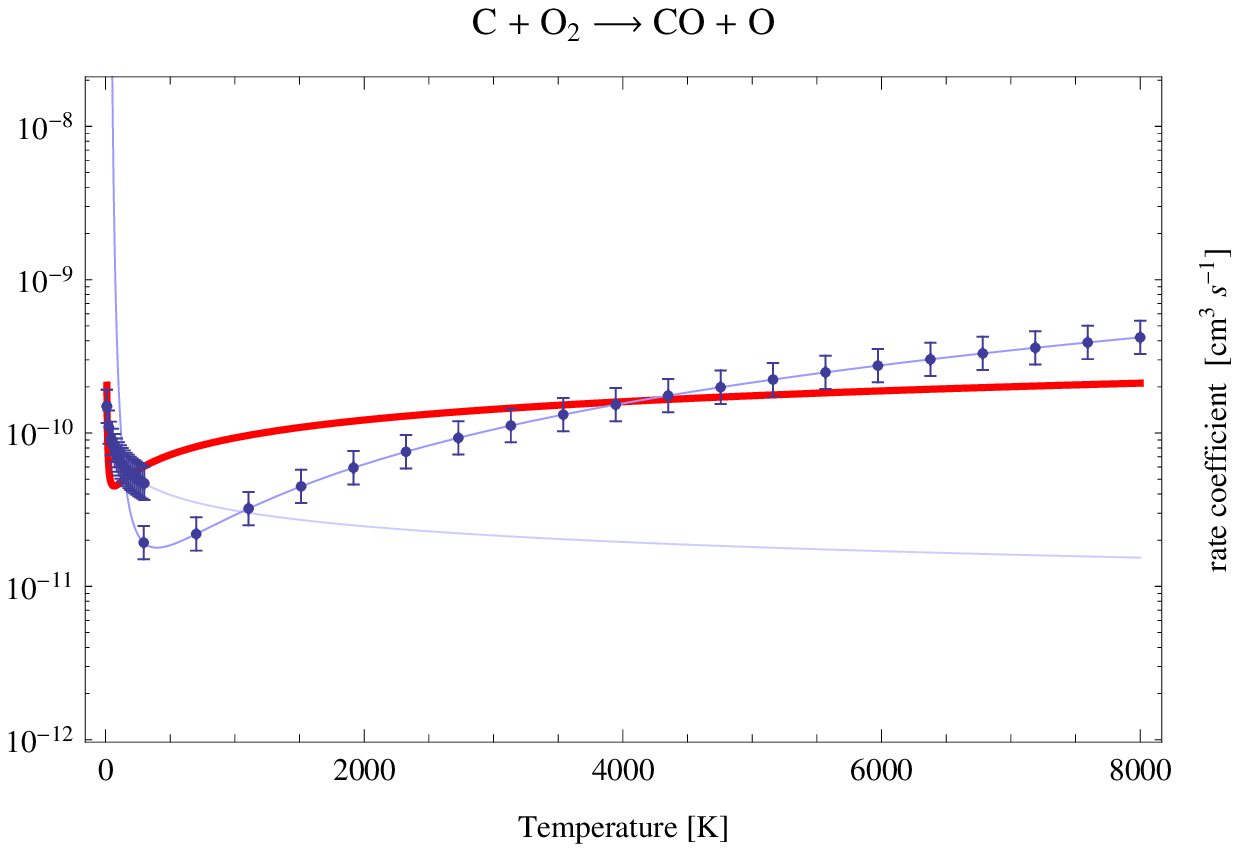}
 \hfill
 \includegraphics[width=6cm]{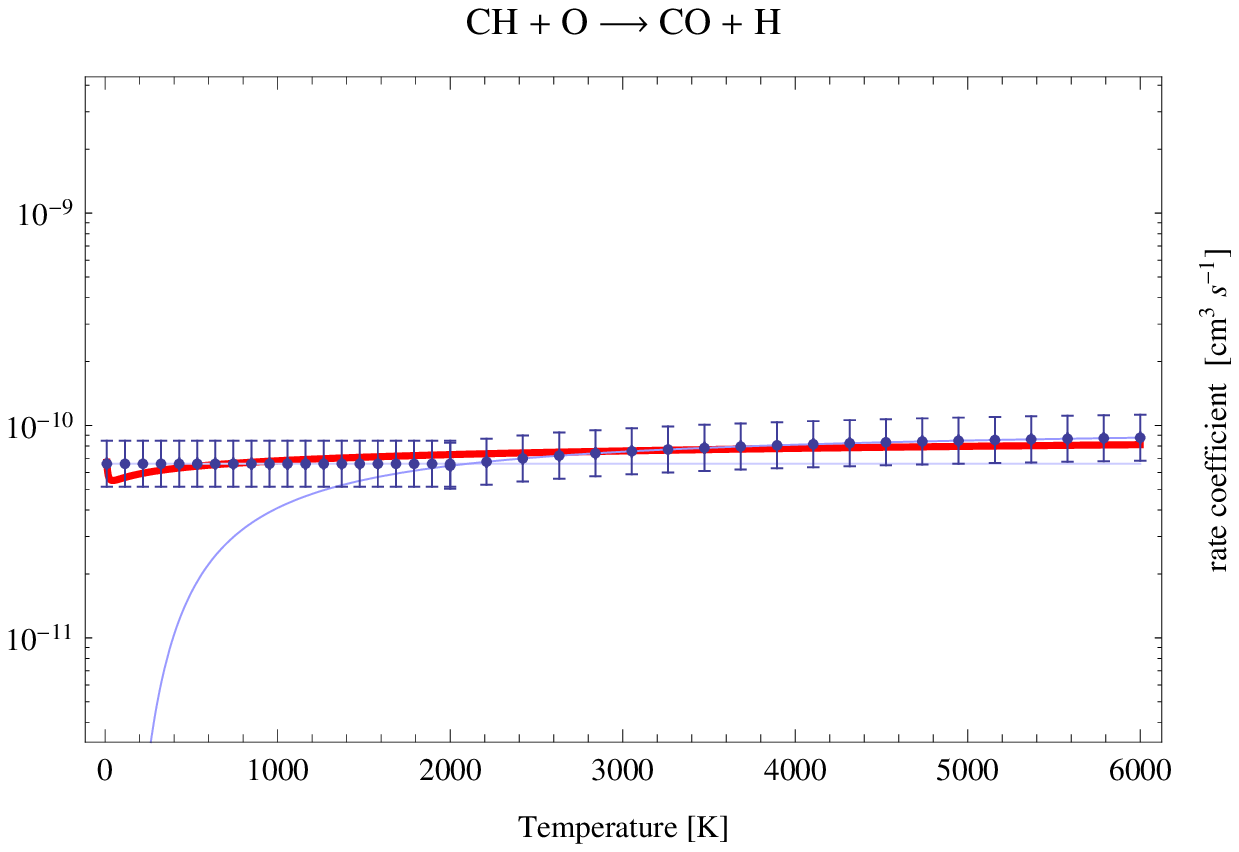}\\
 \includegraphics[width=6cm]{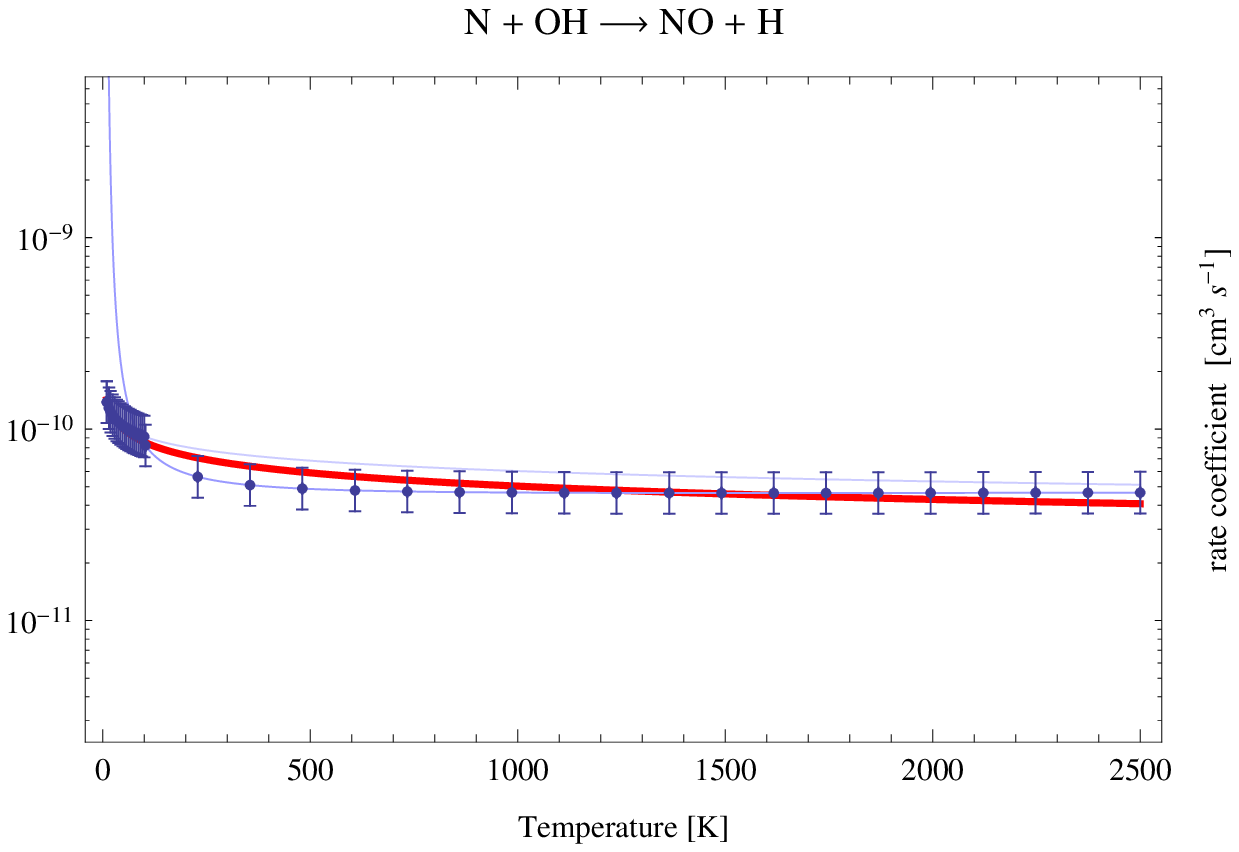}
 \hfill
 \includegraphics[width=6cm]{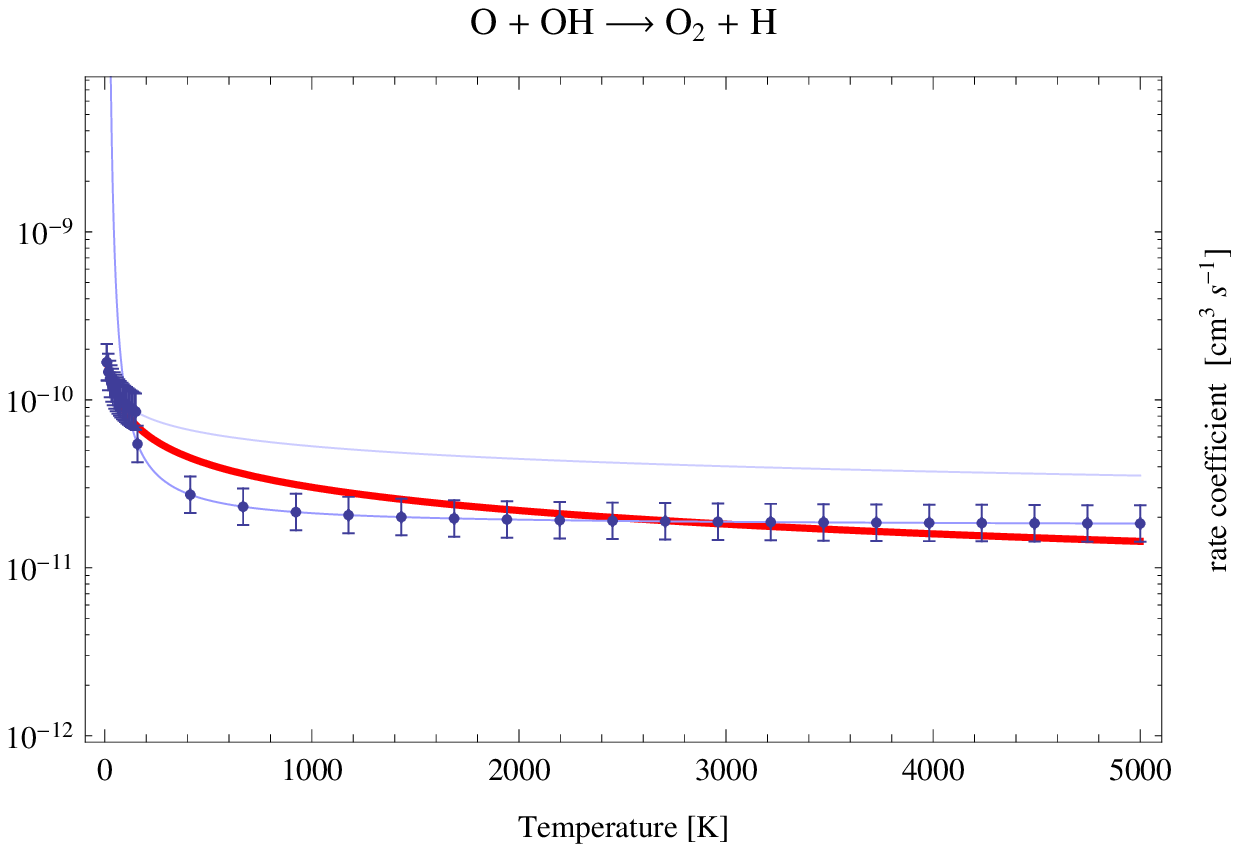}
 \hfill
 \includegraphics[width=6cm]{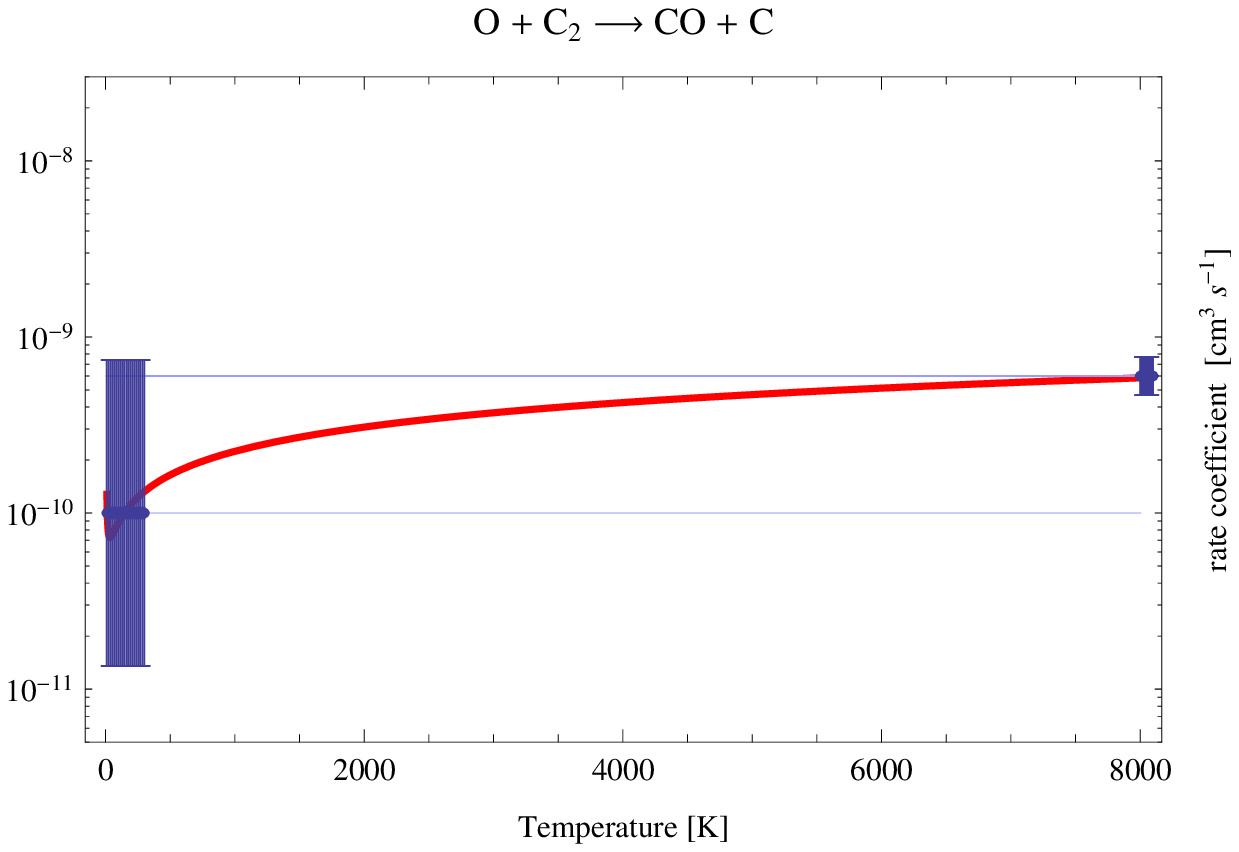}\\ 
\includegraphics[width=6cm]{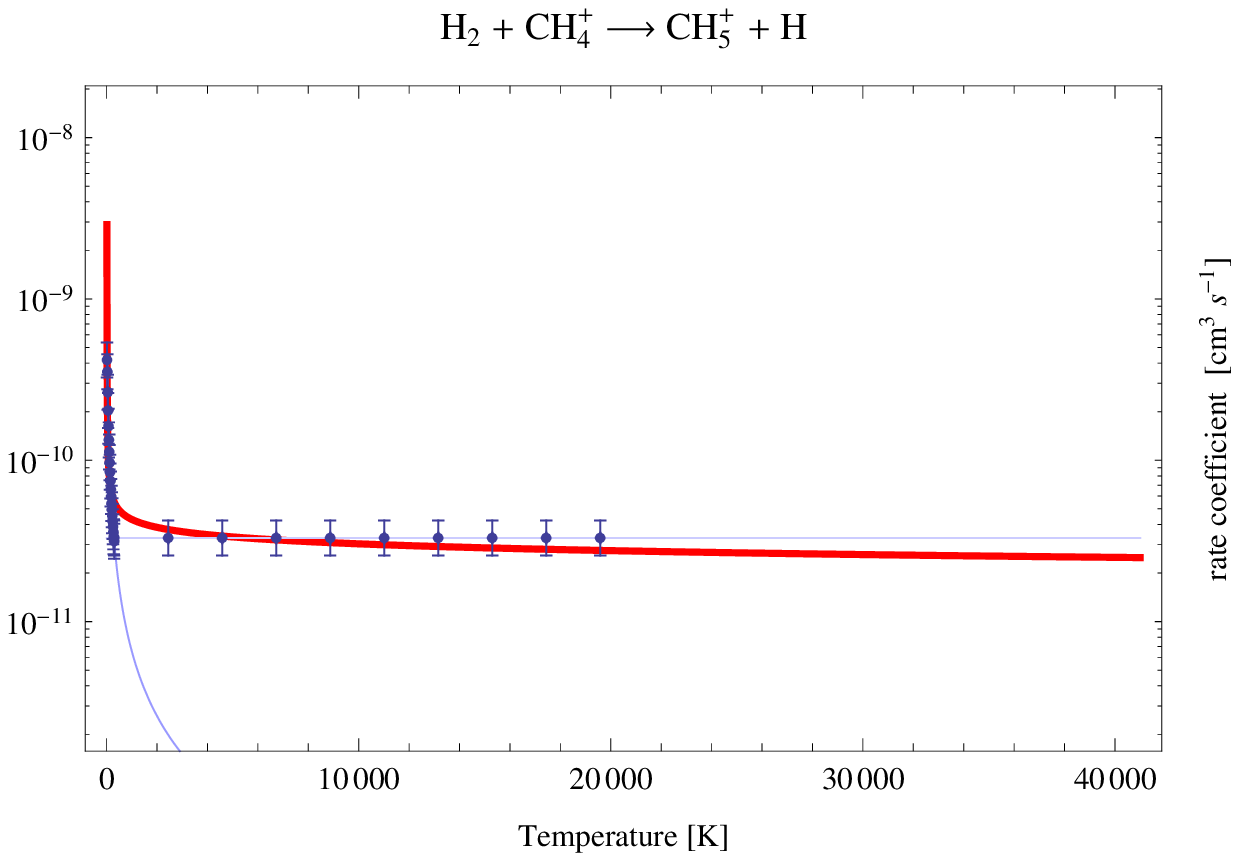}
 \hfill
 \includegraphics[width=6cm]{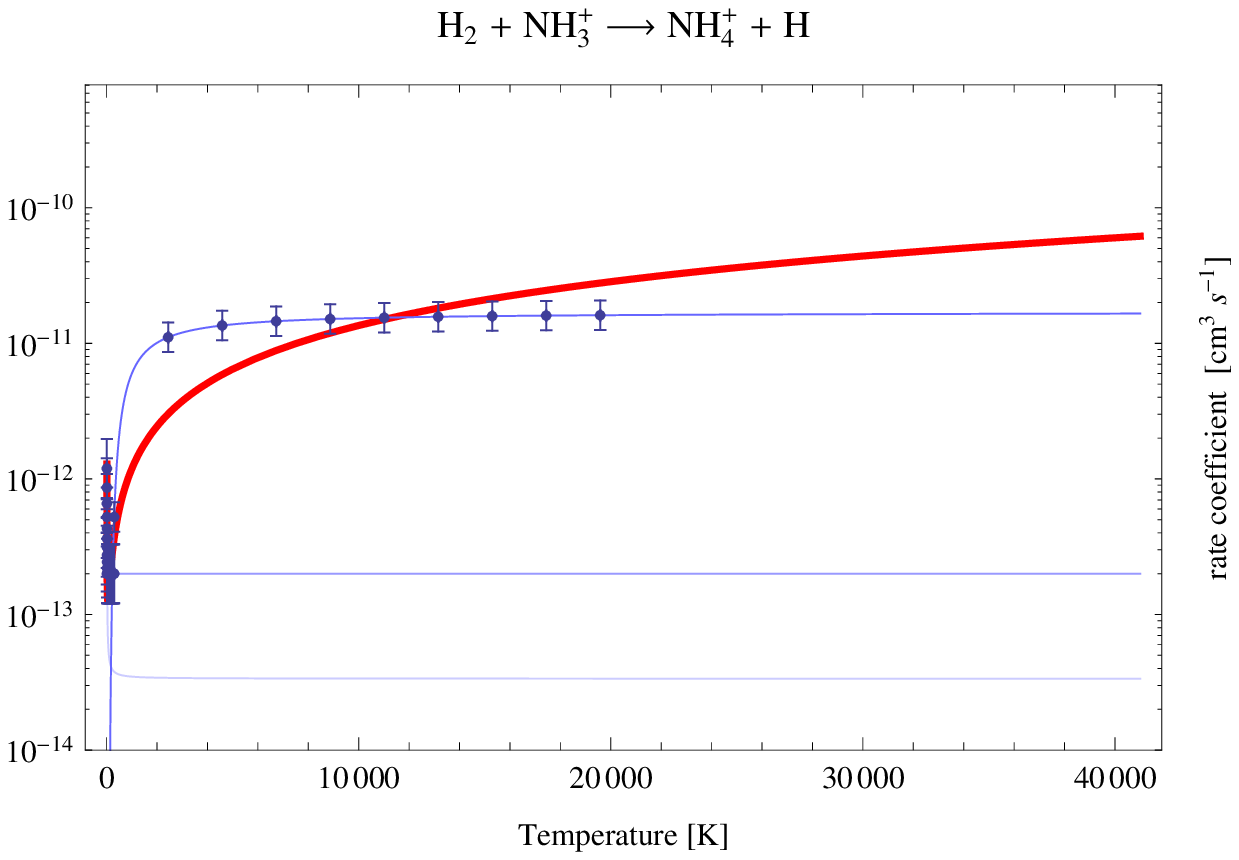}
 \hfill
 \includegraphics[width=6cm]{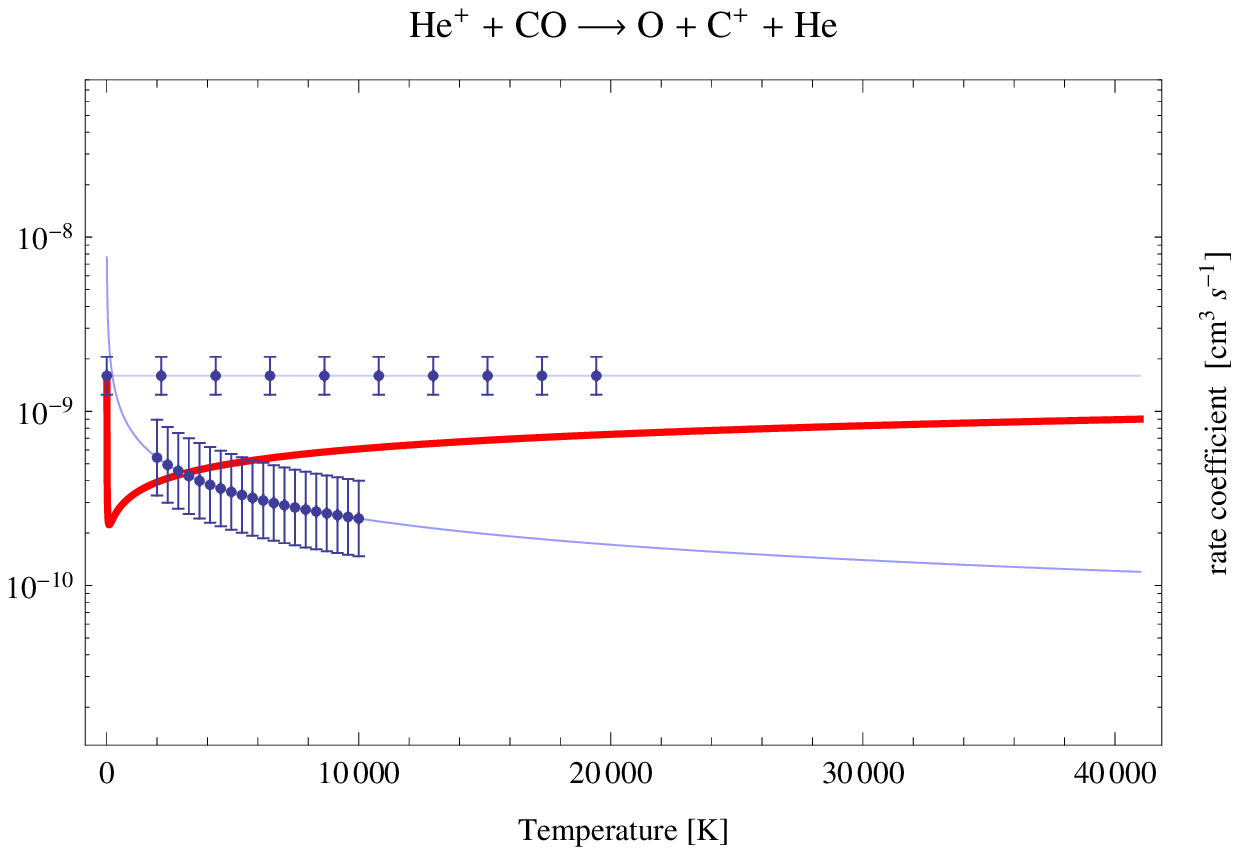}\\
\includegraphics[width=6cm]{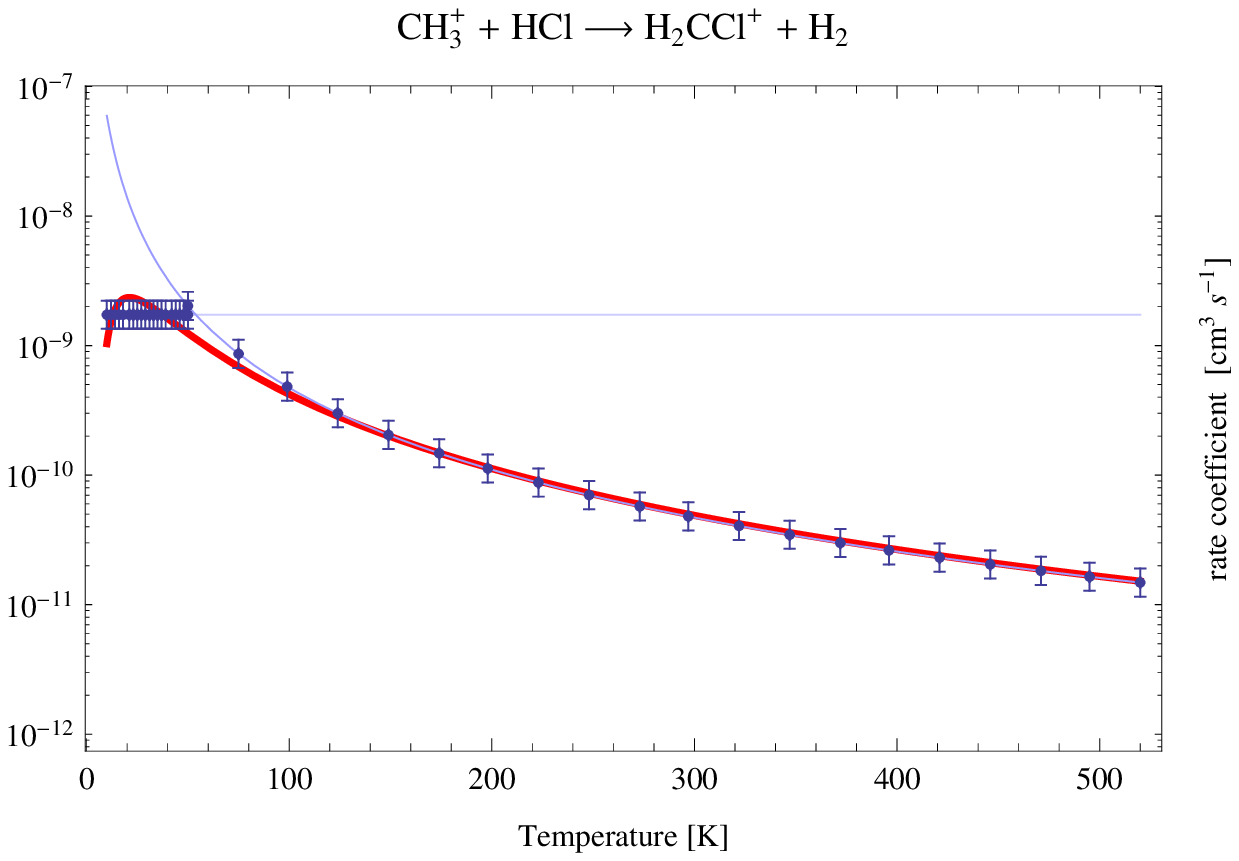}
 \hfill
 \includegraphics[width=6cm]{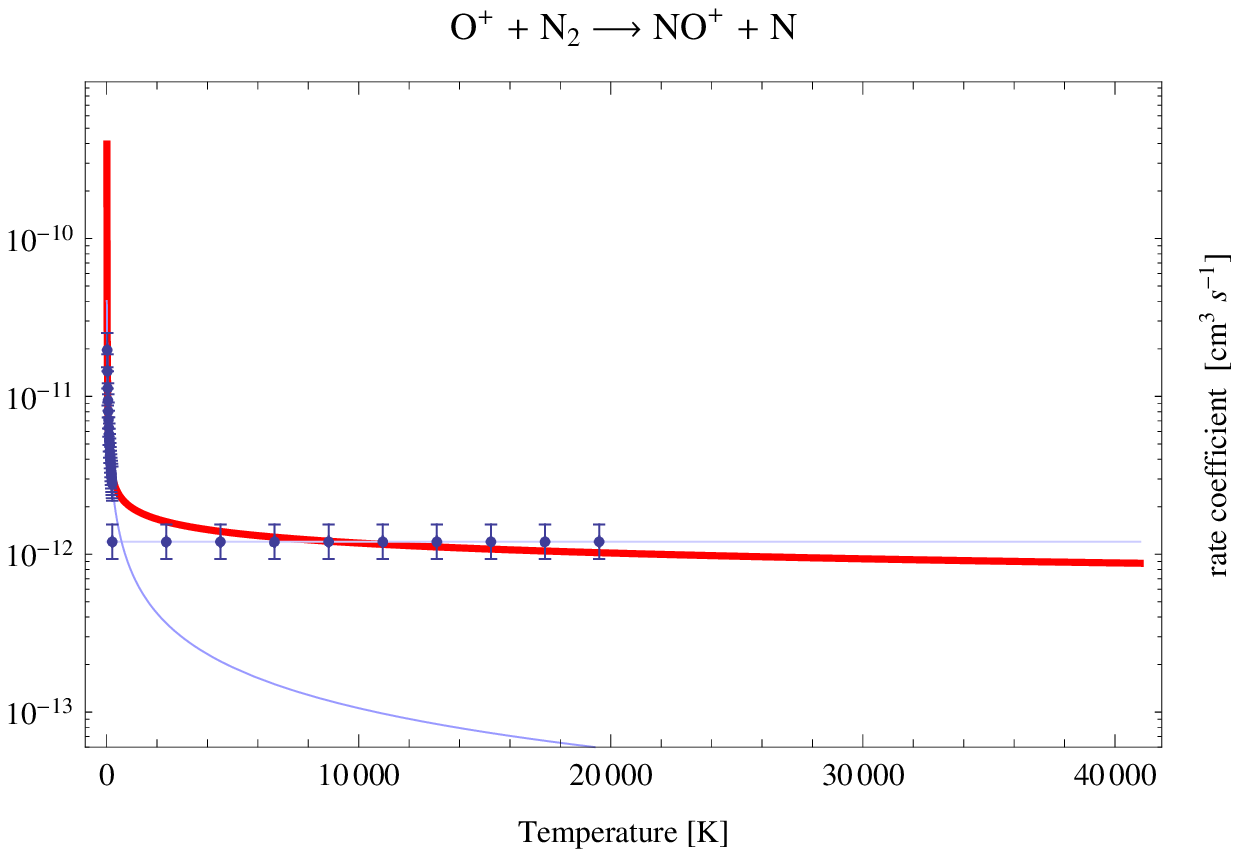}
 \hfill
 \includegraphics[width=6cm]{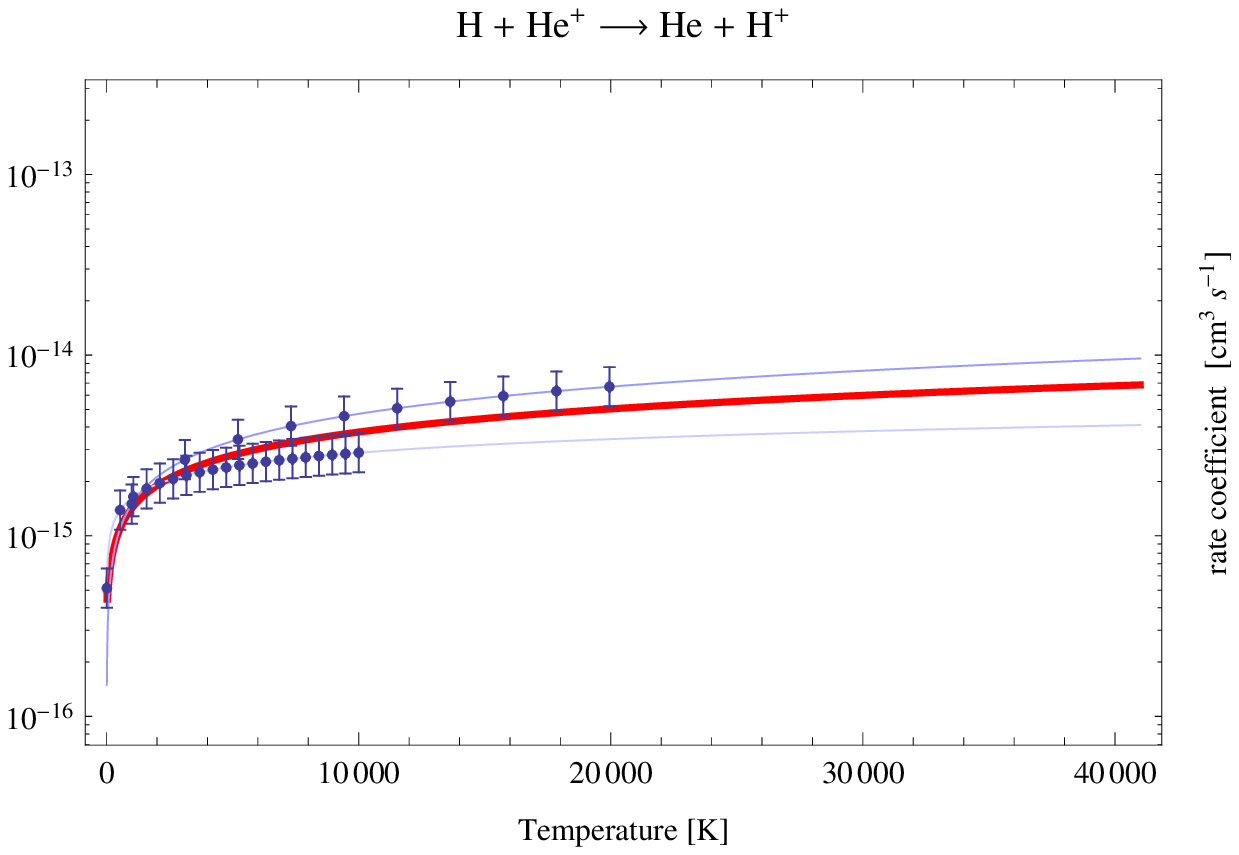}\\
\includegraphics[width=6cm]{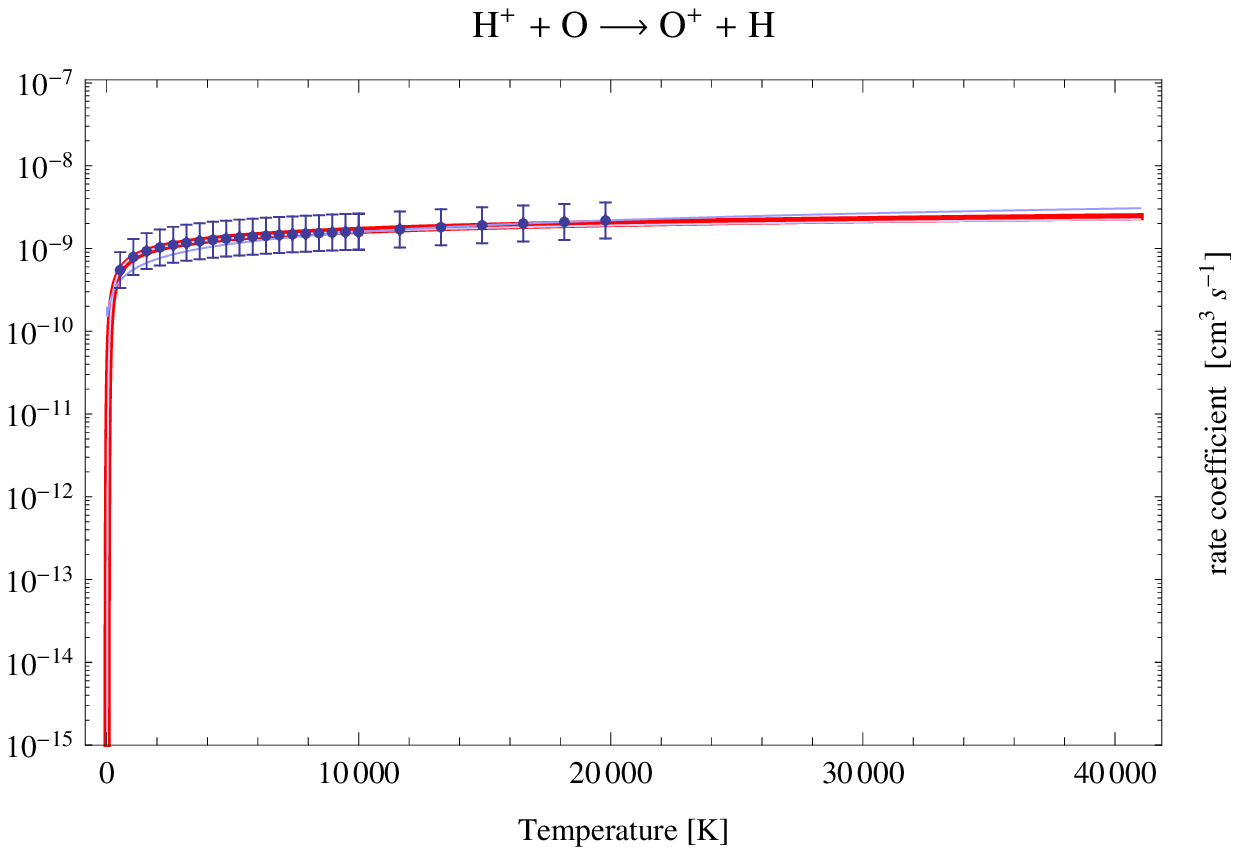}
 \hfill
 \includegraphics[width=6cm]{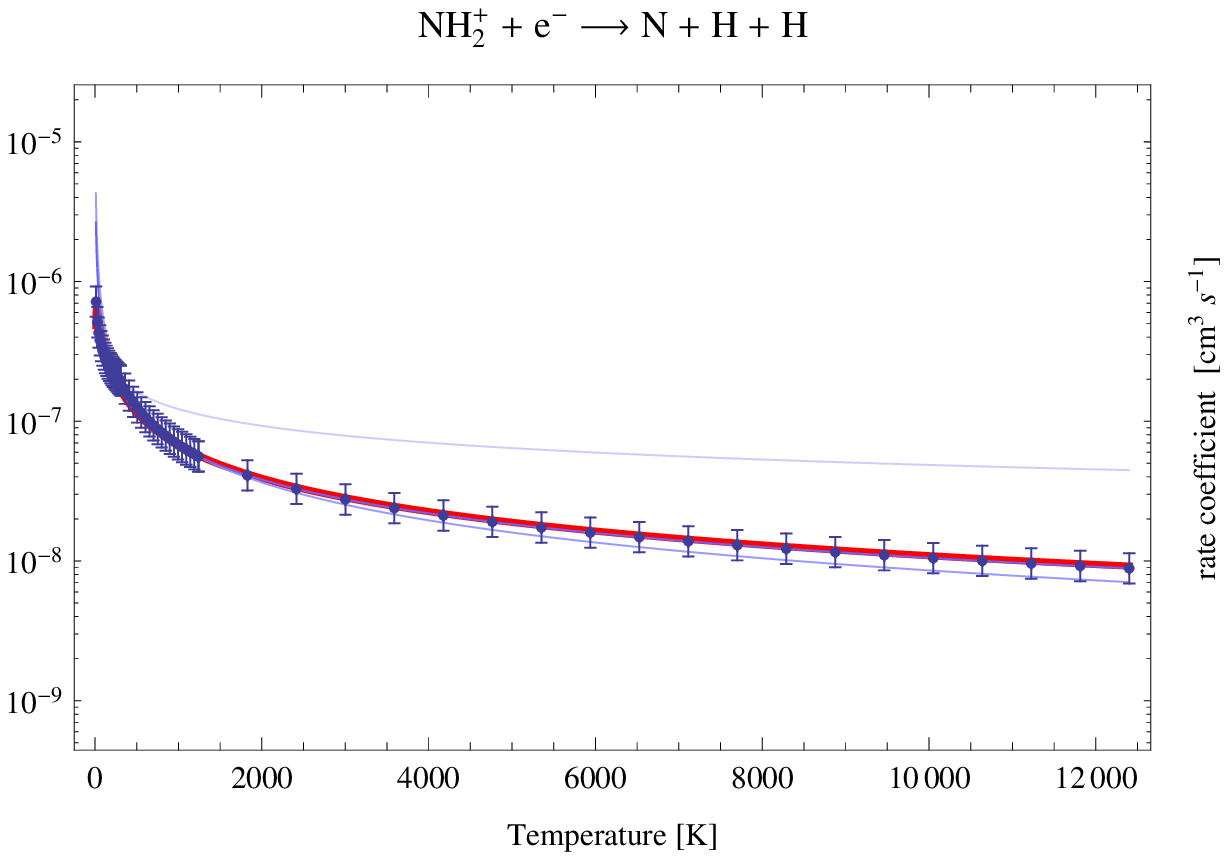}
 \hfill
 \includegraphics[width=6cm]{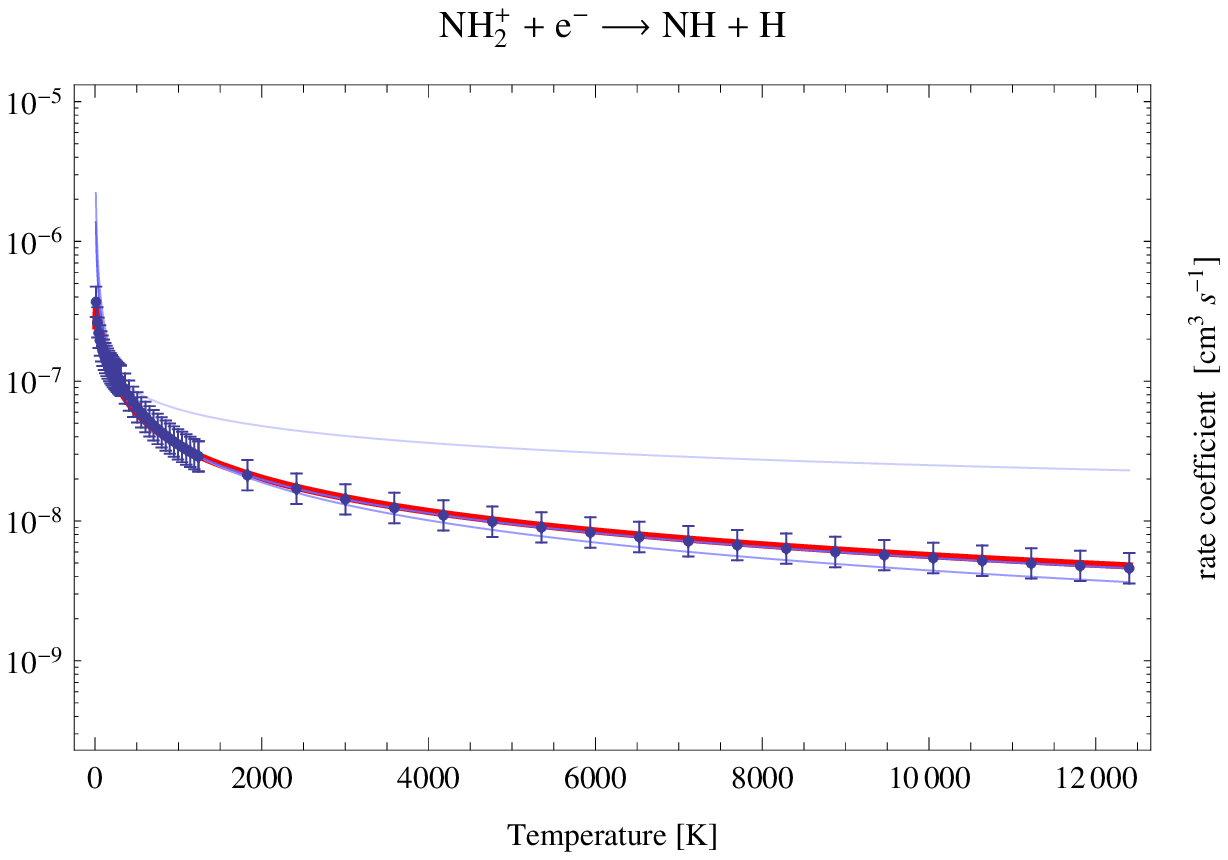}\\
 \includegraphics[width=6cm]{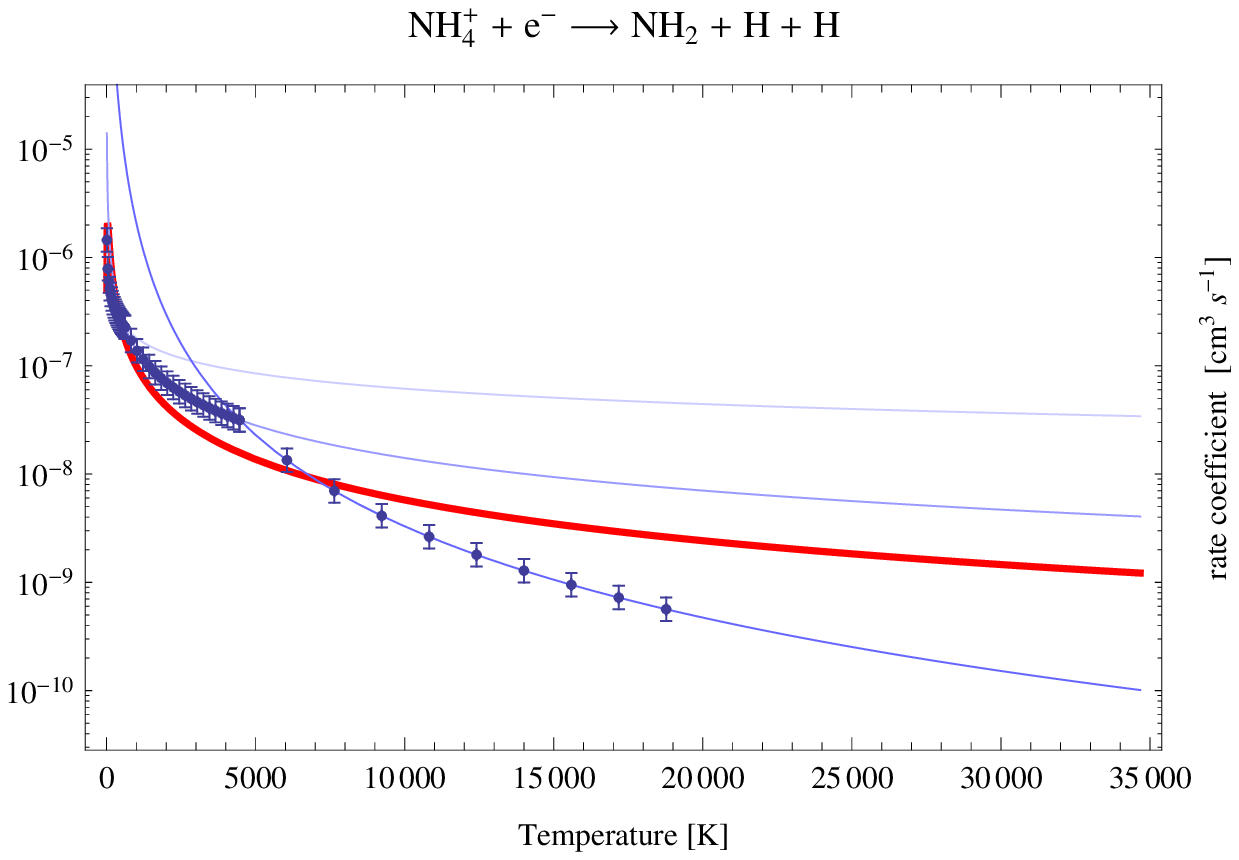}
 \hfill
 \includegraphics[width=6cm]{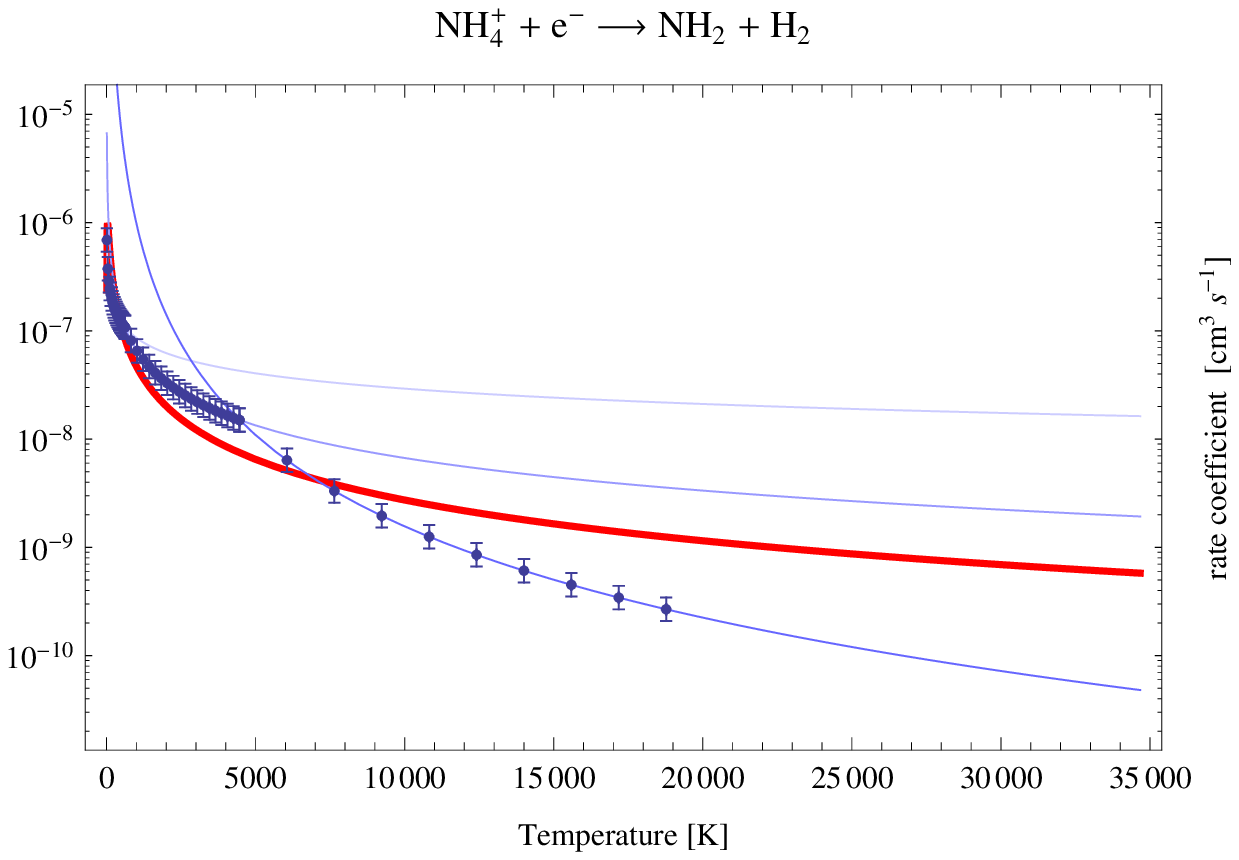}
 \hfill
 \includegraphics[width=6cm]{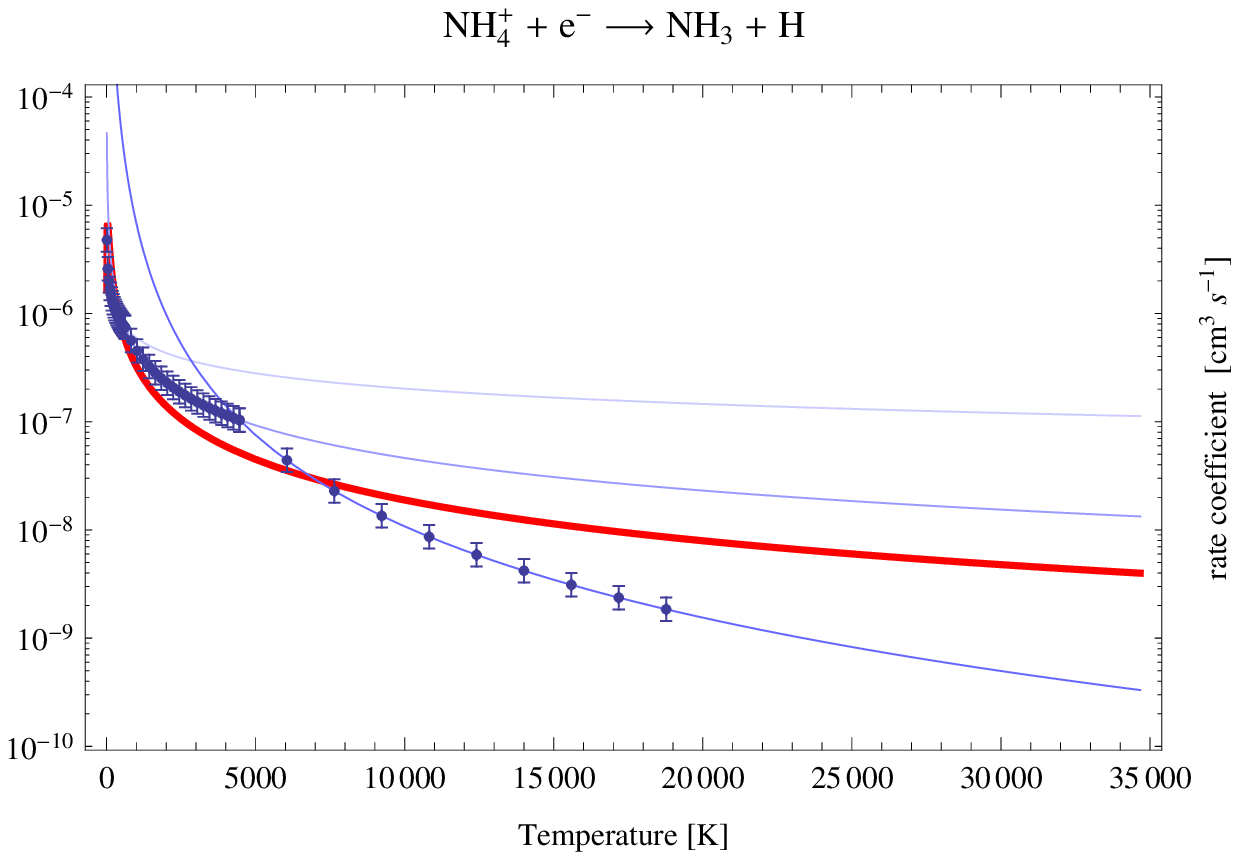}\\
 \label{appendix2}
 \caption{Fits to reactions with multiple entries.}
 \end{figure*}
\begin{figure*}
 \centering
 \includegraphics[width=6cm]{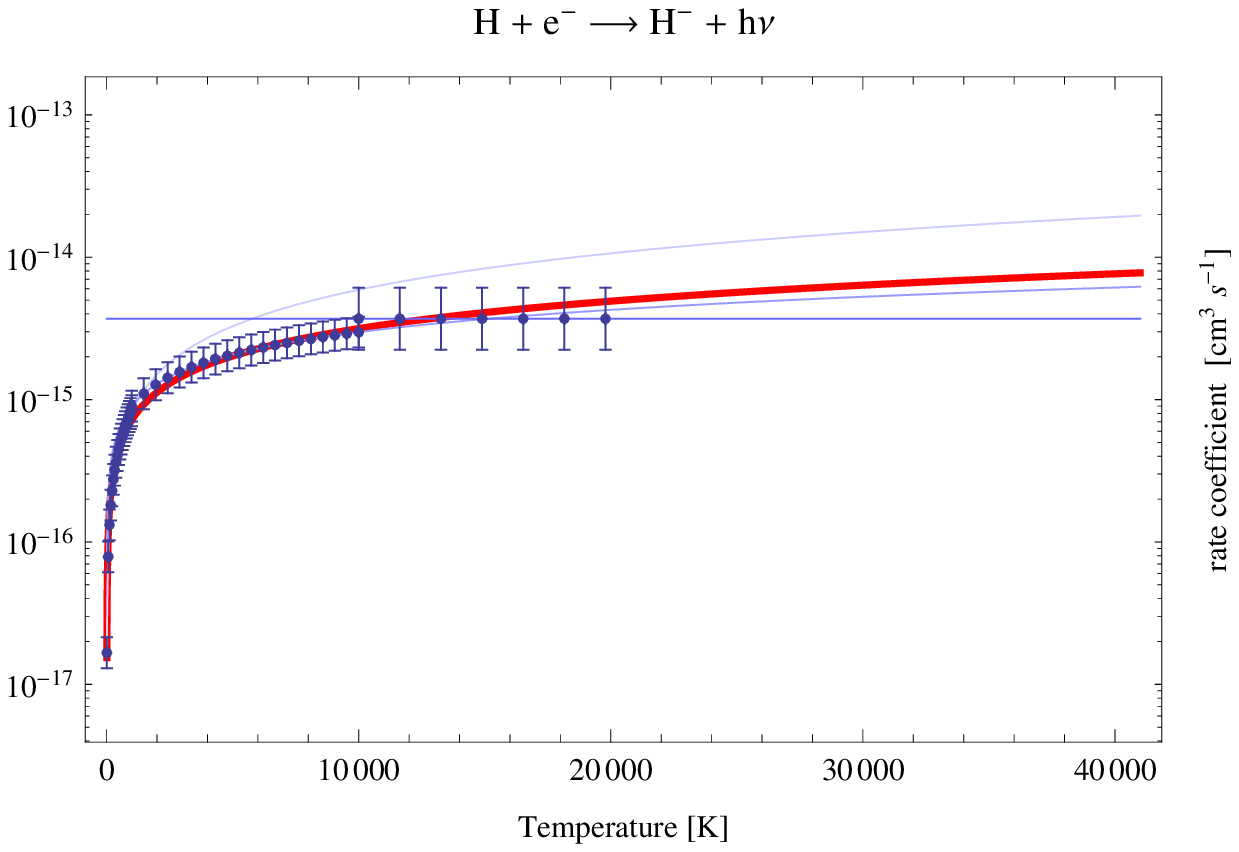}
 \hfill
 \includegraphics[width=6cm]{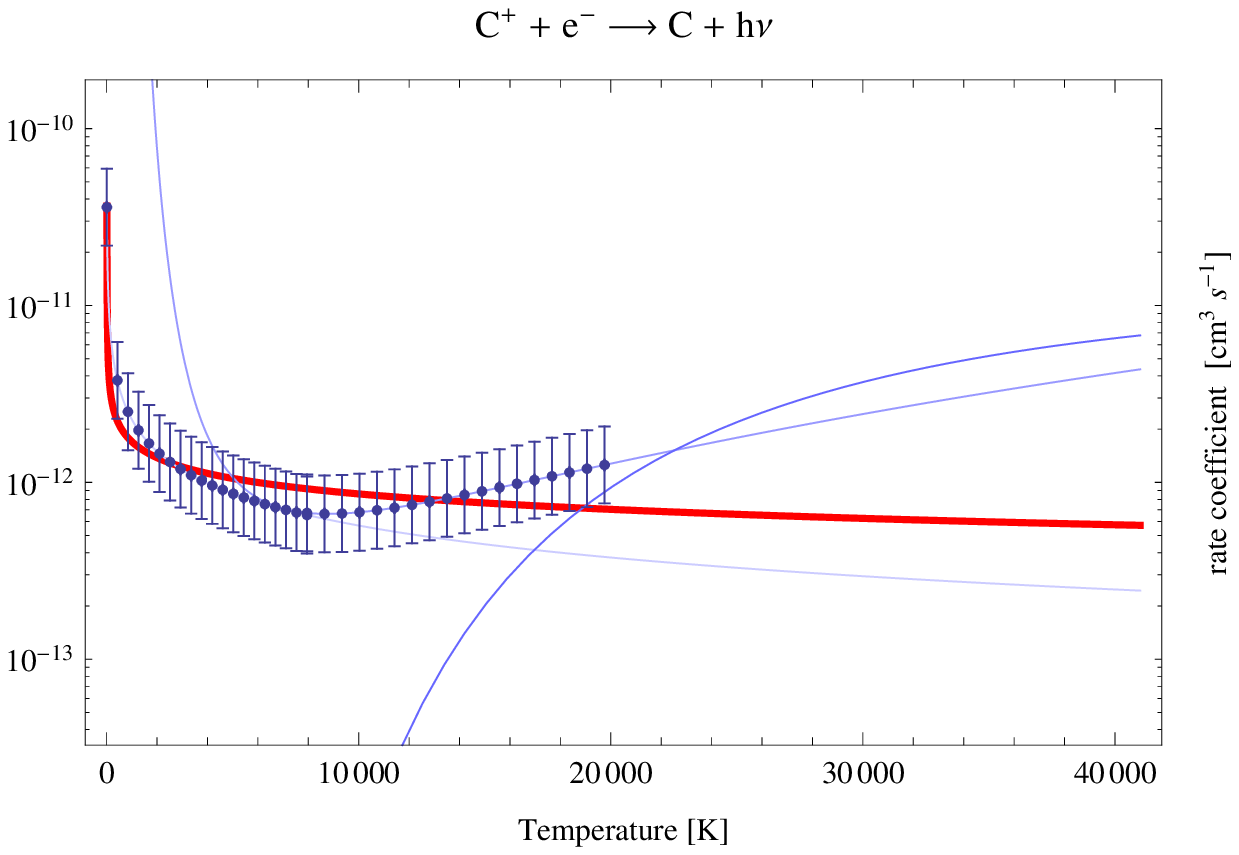}
 \hfill
 \includegraphics[width=6cm]{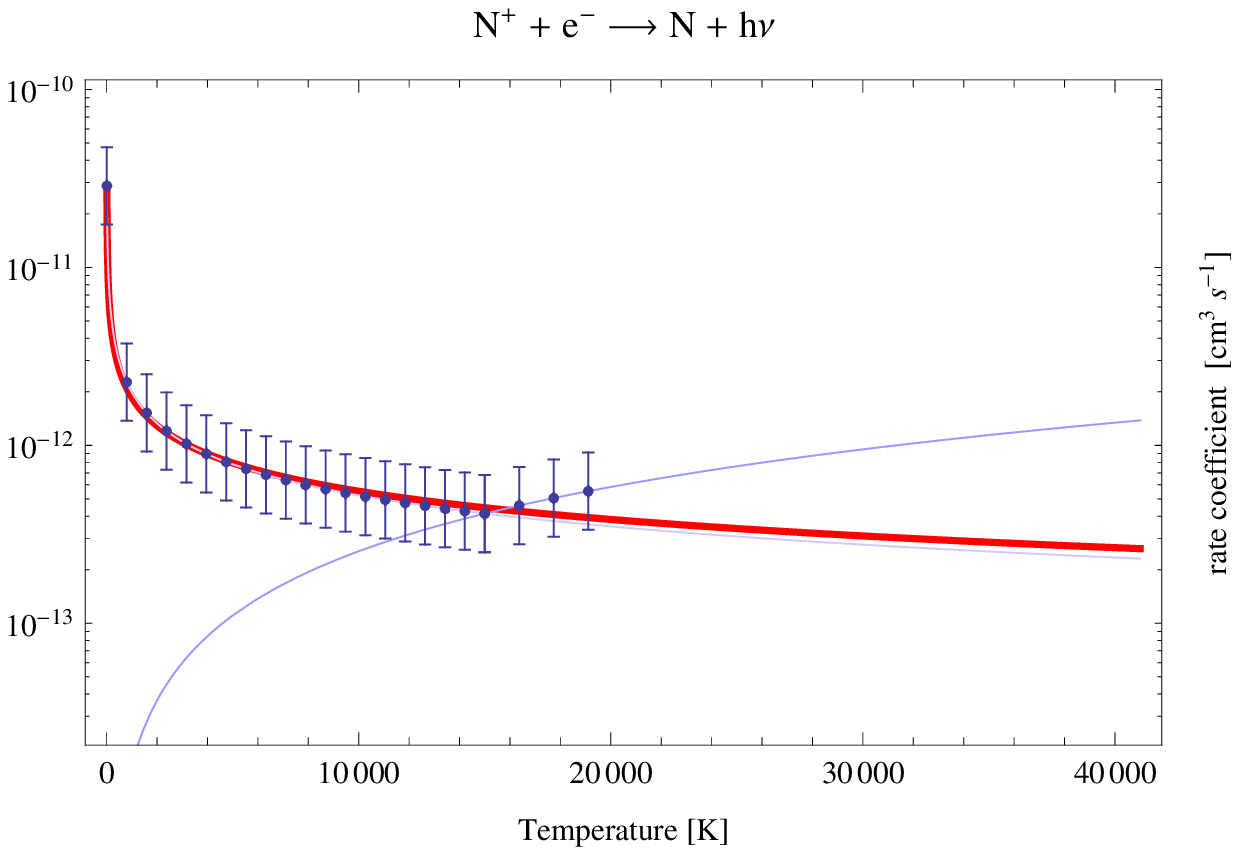}\\
 \includegraphics[width=6cm]{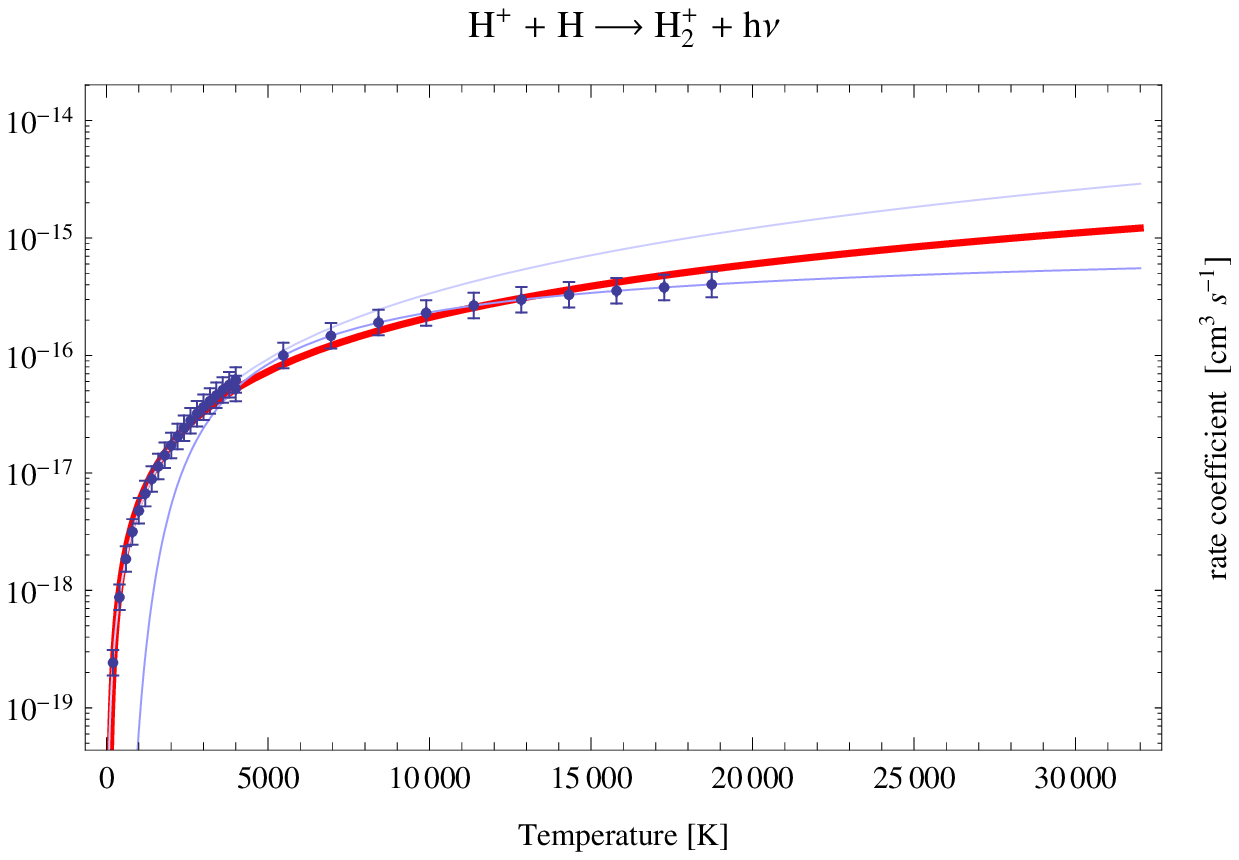}
 \hfill
 \includegraphics[width=6cm]{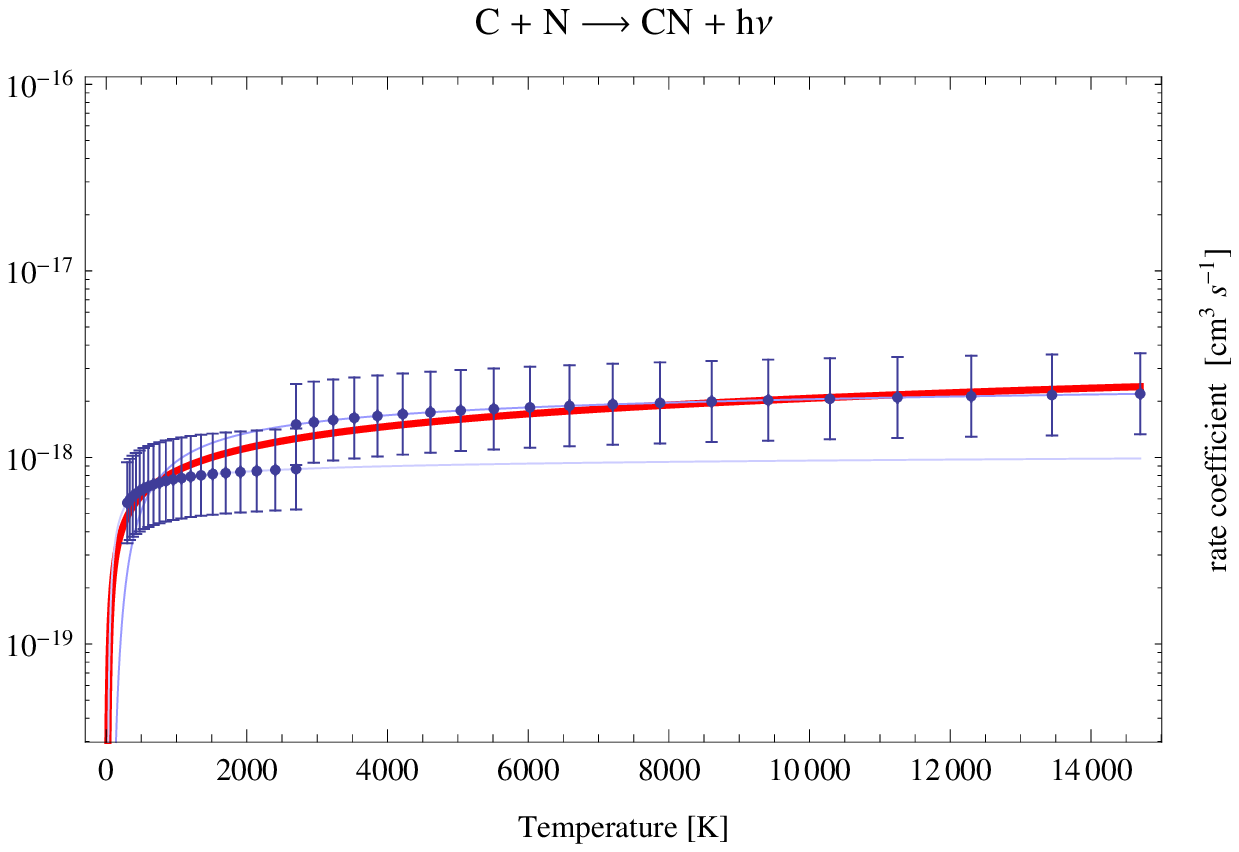}
 \hfill
 \includegraphics[width=6cm]{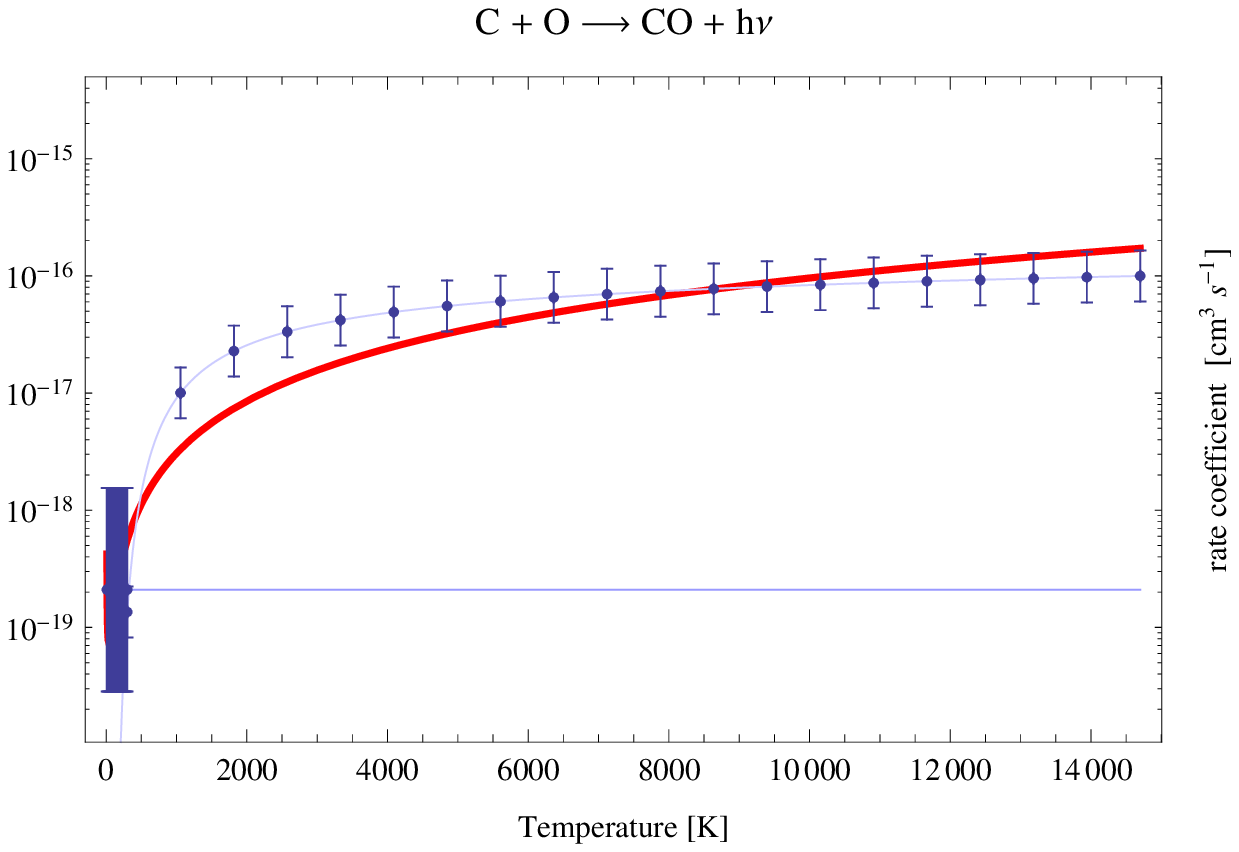}\\ 
\includegraphics[width=6cm]{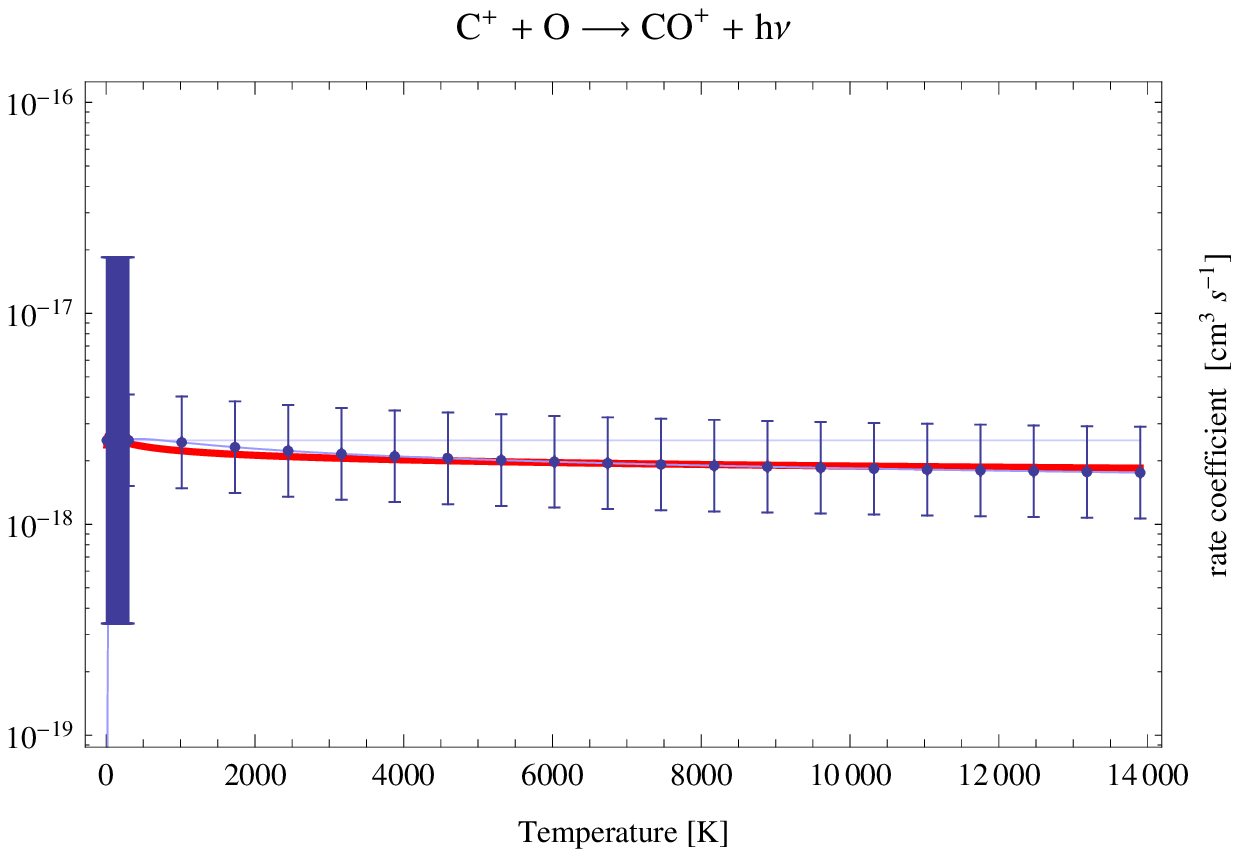}
 \hfill
 \includegraphics[width=6cm]{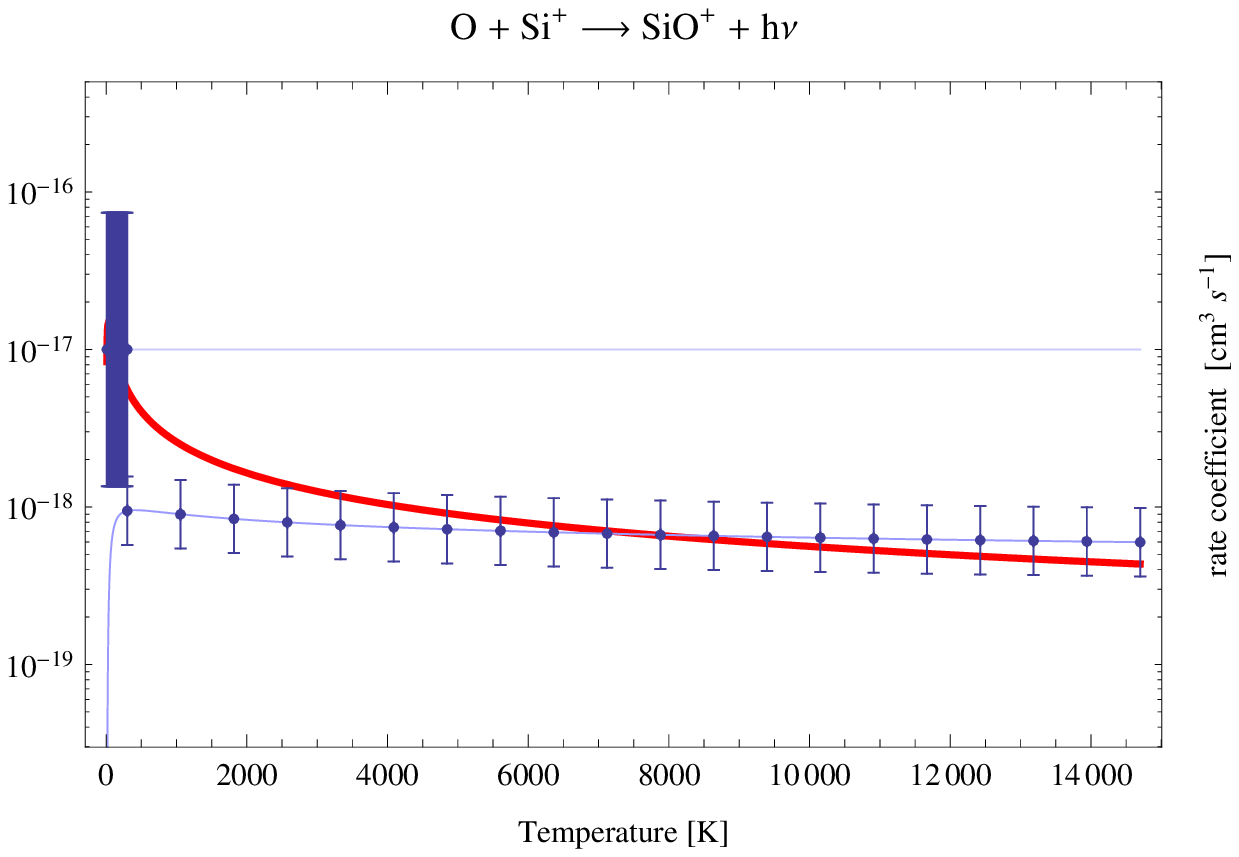}
 \hfill
 \includegraphics[width=6cm]{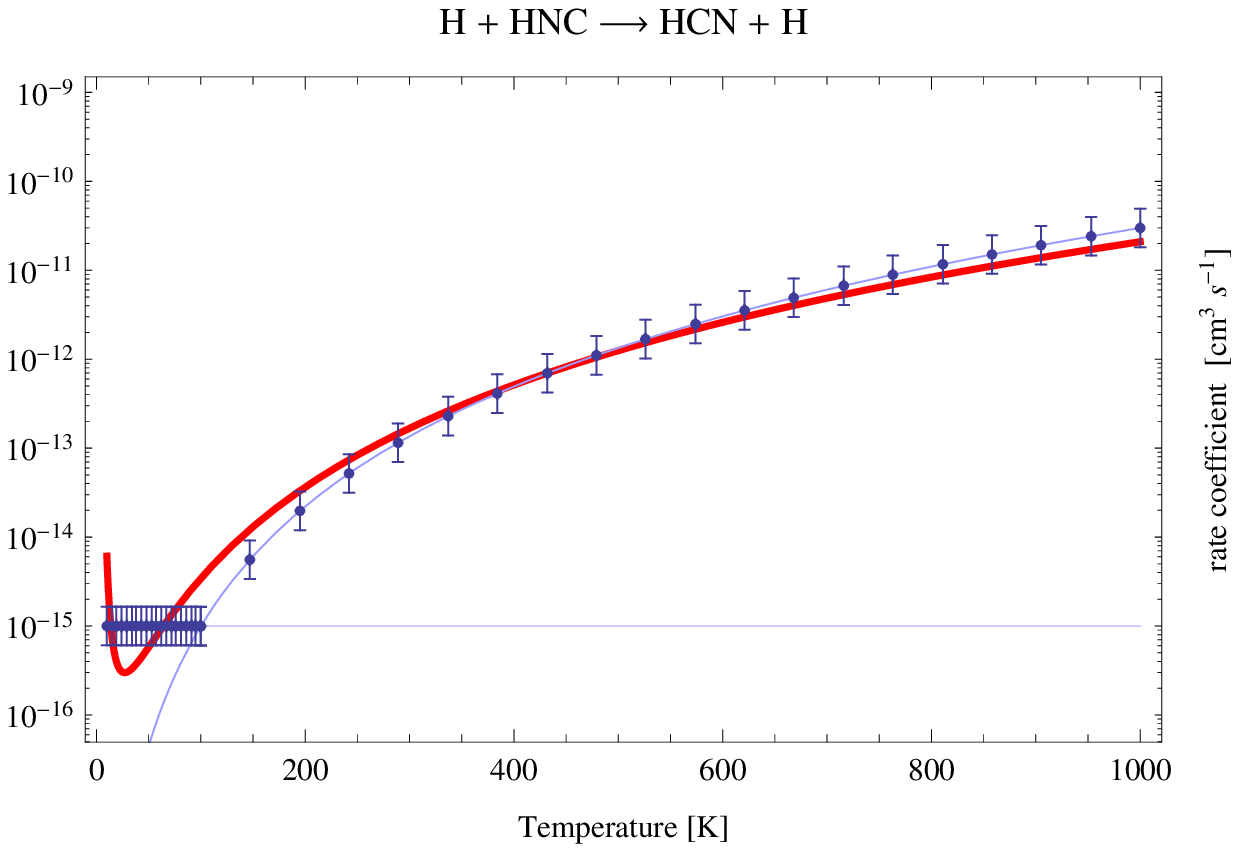}\\
 \includegraphics[width=6cm]{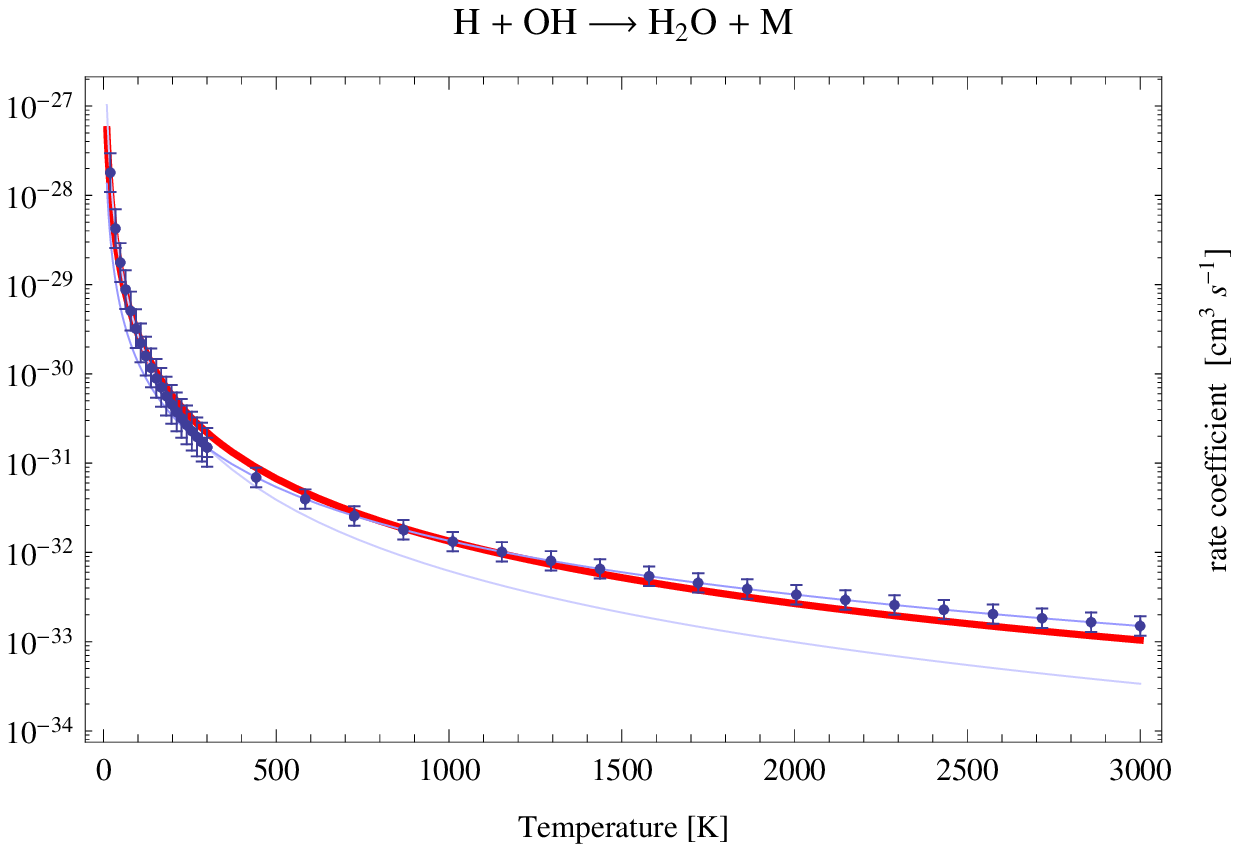}\hfill\hfill
 \caption{Fits to reactions with multiple entries (continued).}
 \end{figure*}
\end{document}

%% file: journals.tex
\def\aj{AJ}%
\def\actaa{Acta Astron.}%
\def\araa{ARA\&A}%
\def\apj{ApJ}%
\def\apjl{ApJ}%
\def\apjs{ApJS}%
\def\ao{Appl.~Opt.}%
\def\apss{Ap\&SS}%
\def\aap{A\&A}%
\def\aapr{A\&A~Rev.}%
\def\aaps{A\&AS}%
\def\azh{AZh}%
\def\baas{BAAS}%
\def\bac{Bull. astr. Inst. Czechosl.}%
\def\caa{Chinese Astron. Astrophys.}%
\def\cjaa{Chinese J. Astron. Astrophys.}%
\def\icarus{Icarus}%
\def\jcap{J. Cosmology Astropart. Phys.}%
\def\jrasc{JRASC}%
\def\mnras{MNRAS}%
\def\memras{MmRAS}%
\def\na{New A}%
\def\nar{New A Rev.}%
\def\pasa{PASA}%
\def\pra{Phys.~Rev.~A}%
\def\prb{Phys.~Rev.~B}%
\def\prc{Phys.~Rev.~C}%
\def\prd{Phys.~Rev.~D}%
\def\pre{Phys.~Rev.~E}%
\def\prl{Phys.~Rev.~Lett.}%
\def\pasp{PASP}%
\def\pasj{PASJ}%
\def\qjras{QJRAS}%
\def\rmxaa{Rev. Mexicana Astron. Astrofis.}%
\def\skytel{S\&T}%
\def\solphys{Sol.~Phys.}%
\def\sovast{Soviet~Ast.}%
\def\ssr{Space~Sci.~Rev.}%
\def\zap{ZAp}%
\def\nat{Nature}%
\def\iaucirc{IAU~Circ.}%
\def\aplett{Astrophys.~Lett.}%
\def\apspr{Astrophys.~Space~Phys.~Res.}%
\def\bain{Bull.~Astron.~Inst.~Netherlands}%
\def\fcp{Fund.~Cosmic~Phys.}%
\def\gca{Geochim.~Cosmochim.~Acta}%
\def\grl{Geophys.~Res.~Lett.}%
\def\jcp{J.~Chem.~Phys.}%
\def\jgr{J.~Geophys.~Res.}%
\def\jqsrt{J.~Quant.~Spec.~Radiat.~Transf.}%
\def\memsai{Mem.~Soc.~Astron.~Italiana}%
\def\nphysa{Nucl.~Phys.~A}%
\def\physrep{Phys.~Rep.}%
\def\physscr{Phys.~Scr}%
\def\planss{Planet.~Space~Sci.}%
\def\procspie{Proc.~SPIE}%
\let\astap=\aap
\let\apjlett=\apjl
\let\apjsupp=\apjs
\let\applopt=\ao